\documentclass[12pt]{article}
\usepackage{cite}
\usepackage{amsmath, amsthm, amssymb,slashed,url}
\usepackage{ifpdf}
\ifpdf
  \usepackage[pdftex]{graphicx}
  \usepackage{epstopdf}
\else  
  \usepackage[dvips]{graphicx}
\fi
\parskip=\baselineskip
\def\Bbb{\mathbb}
\def\Tr{{\rm Tr}}
\def\16{{\bf 16}}
\def\1{{\bf 1}}
\def\2{{\bf 2}}
\def\3{{\bf 3}}
\def\4{{\bf 4}}
\def\c{{\sf c}}
\def\x{{\sf x}}
\def\y{{\sf y}}
\def\I{{\mathrm I}}
\def\II{{\mathrm {II}}}
\def\III{{\mathrm{ III}}}
\def\K{{\mathcal K}}
\def\CPT{\sf{CPT}}
\def\a{{\sf a}}
\def\b{{\sf b}}
\def\nn{{[n]}}
\def\kk{{[k]}}
\def\oone{{[1]}}
\def\ttwo{{[2]}}
\def\dn{{\la n\ra}}

\def\tr{{\mathrm{tr}}}
\def\CRT{{\sf{CRT}}}
\def\h{\widehat}
\def\bar{\overline}

\def\ra{\rangle}

\def\bp{\begin{pmatrix}}
\def\ep{\end{pmatrix}}
\def\la{\langle}
\def\R{{\Bbb{R}}}\def\Z{{\Bbb{Z}}}

\def\RR{{\sf R}}
\def\S{{\mathcal S}}
\def\hat{\widehat}
\font\teneurm=eurm10 \font\seveneurm=eurm7 \font\fiveeurm=eurm5
\newfam\eurmfam
\textfont\eurmfam=\teneurm \scriptfont\eurmfam=\seveneurm
\scriptscriptfont\eurmfam=\fiveeurm

\font\teneusm=eusm10 \font\seveneusm=eusm7 \font\fiveeusm=eusm5
\newfam\eusmfam
\textfont\eusmfam=\teneusm \scriptfont\eusmfam=\seveneusm
\scriptscriptfont\eusmfam=\fiveeusm

\font\tencmmib=cmmib10 \skewchar\tencmmib='177
\font\sevencmmib=cmmib7 \skewchar\sevencmmib='177
\font\fivecmmib=cmmib5 \skewchar\fivecmmib='177
\newfam\cmmibfam
\textfont\cmmibfam=\tencmmib \scriptfont\cmmibfam=\sevencmmib
\scriptscriptfont\cmmibfam=\fivecmmib

\numberwithin{equation}{section}

\def\d{\mathrm d}
\def\u{{\sf u}}
\def\v{{\sf v}}
\def\C{{\Bbb C}}
\def\Z{{\Bbb Z}}
\def\bar{\overline}
\def\A{{\mathcal A}}
\def\bar{\overline}
\begin{document}
\begin{titlepage}
\begin{flushright}
\end{flushright}
\vskip 1.5in
\begin{center}
{\bf\Large{Notes On Some Entanglement Properties \\ \vskip.5cm Of Quantum Field Theory}}
\vskip
0.5cm {Edward Witten} \vskip 0.05in {\small{ \textit{School of
Natural Sciences, Institute for Advanced Study}\vskip -.4cm
{\textit{Einstein Drive, Princeton, NJ 08540 USA}}}
}
\end{center}
\vskip 0.5in
\baselineskip 16pt
\abstract{ These are notes on some entanglement properties of quantum field theory, aiming to make accessible a variety of ideas
that are known in the literature.  The main goal is to explain how to deal with entanglement when -- as in quantum field theory -- it is a property
of the algebra of observables and not just of the states. }
\date{January, 2018}
\end{titlepage}
\def\Hom{\mathrm{Hom}}
\def\H{{\mathcal H}}
\def\d{{\mathrm d}}
\def\t{\widetilde}
\def\U{{\mathcal U}}
\def\UU{{\mathrm U}}
\def\V{{\mathcal V}}
\def\st{{\sf t}}
\def\O{{\mathcal O}}
\def\i{{\mathrm i}}
\def\A{{\mathcal A}}
\def\be{\begin{equation}}
\def\ee{\end{equation}}

\tableofcontents

\section{Introduction}

Ideas of quantum information theory and entanglement have played an increasingly important role in quantum field theory
and string theory in recent years. Unfortunately, it is really not possible in a short space to give references to the many developments in this general area that
have occurred in the last decade.  Many important developments are cited and summarized in the recent review article
\cite{Nishioka}.  

The present notes are not an overall introduction to this subject.  The goal here is more narrow: to make accessible some of the mathematical
ideas that underlie some of these developments, and which are present in the existing literature but not always so easy to extract.  
In the process, we will also make contact with some of the older literature on axiomatic and algebraic quantum field theory.

In section \ref{rst}, we describe the Reeh-Schlieder theorem \cite{RS}, which demonstrates that in quantum field theory,
all field variables in any one region of
spacetime are entangled with variables in other regions.   Actually,  the entanglement of spatially adjacent field modes
is so strong that entanglement entropy between adjoining
spacetime regions in quantum field theory  is not just large but ultraviolet divergent.   (Early references on this ultraviolet
divergence include \cite{BKLS, Sred, SU,McG,CW,HLW}.)   This ultraviolet divergence means that the entanglement is not just a property of the states but of the algebras
of observables.  Explaining  this statement  and how to deal with it in the context of local quantum
field theory is a primary goal in what follows.   (We do not consider the implications of quantum gravity.)

An important tool in dealing with entanglement when it is a property of the algebras and not just the states is provided by Tomita-Takesaki
theory, which we introduce in section \ref{relmood}.    It has been used in a number of recent developments, 
including an attempt to see behind the horizon of a black hole \cite{PapaRaju}, a proof of the quantum null energy 
condition \cite{Faulkner}, and too many others to properly cite here.   
As an inducement for the reader who is not sure this mathematical tool is worthwhile,
we describe in section \ref{relmood} a rigorous definition -- due to Araki \cite{Araki,Araki2} -- of relative entropy in quantum field theory, with a surprisingly
simple proof of its main properties, including its monotonicity when one enlarges the region in which measurements are made.

In section \ref{fte}, we explain what Tomita-Takesaki theory means for a quantum system with a finite-dimensional Hilbert space.
This motivates the statement of some of the subtler properties of Tomita-Takesaki theory.  It also leads -- following Araki's work and  later
developments by Petz \cite{Petz} and Petz and Nielsen \cite{PetzNielsen} -- to a natural proof of monotonicity of quantum relative
entropy for a finite-dimensional quantum system.  Monotonicity of relative entropy and its close cousin, strong subadditivity of quantum 
entropy, were first proved by Lieb and Ruskai \cite{LiebRuskai}, using a lemma by Lieb \cite{Lieb}.   These results underlie many of the
deeper statements in quantum information theory.

In section \ref{more}, we describe a fundamental -- and fairly well-known -- example of entanglement in quantum field theory.
This is the case, first analyzed by Bisognano and Wichmann \cite{BiWi} and Unruh \cite{Unruh}, of two complementary  ``wedges'' or
Rindler regions in Minkowski spacetime.   In Unruh's formulation, the question is what is seen by an accelerating observer in Minkowski
spacetime.   We approach this problem both from a path integral point of view -- which is important in
black hole physics \cite{GibbonsHawking}  -- and following the  rigorous approach of Bisognano and Wichman, which was based on analyticity rather than path integrals.  

In section \ref{algex}, we explain, following von Neumann and others \cite{vonN,Powers,ArakiWoods}, a short direct construction of algebras -- such as the algebra of quantum field theory observables
in a given spacetime region -- with the property that a divergent entanglement entropy is built into the structure of the algebra.

Finally, in section \ref{indo}, we give some examples of the use of Tomita-Takesaki theory to prove statements in quantum field
theory that would be more obvious if one could assume a simple factorization of the Hilbert space between degrees of freedom
localized in different spacetime regions.   All of these statements have been analyzed in previous rigorous papers, in some cases
before the relevance of Tomita-Takesaski theory was understood.

The topics discussed in these notes can be treated rigorously, but the presentation here is certainly not rigorous.   More complete
treatments of most of the points about quantum field theory can be found in the article of Borchers 
\cite{Borchers} and the book of Haag \cite{Haag}. 
 Quantitative measures of entanglement in quantum field theory such as Bell's inequalities have been discussed by
Summers and Werner \cite{SW} and from a different standpoint by Narnhofer and Thirring \cite{NaT}.
See also  a recent
article of Hollands and Sanders \cite{HSa} for another point of view on entanglement measures in quantum field theory and much interesting detail.  
For general mathematical background on von Neumann algebras, a convenient reference is
the lecture notes of Jones \cite{Jones}.   

\section{The Reeh-Schlieder Theorem}\label{rst}

\subsection{Statement}

Our starting point will be the Reeh-Schlieder Theorem \cite{RS}, which back in 1961 came as a ``surprise'' 
according to Streater and Wightman 
\cite{StW}.

We consider a quantum field theory in Minkowski spacetime $M_D$ of dimension $D$ with
spacetime coordinates $x^\mu=(t,\vec x)$ and 
 metric \be\label{tehmetric}\d s^2=\sum_{\mu,\nu=0}^{D-1}\eta_{\mu\nu}\d x^\mu\d x^\nu=
-\d t^2+\d \vec x^2.\ee  We write $\Omega$ for the vacuum state
and $\H_0$ for the vacuum sector of Hilbert space, which consists of all states that can be created from the vacuum by local field     
operators.  ($\H_0$ is not necessarily the full Hilbert space $\H$ of the given theory, since there may be ``superselection sectors'';
see section \ref{bounded}.)
For simplicity of notation, we assume that the algebra of local fields of the theory under discussion
is generated by a hermitian scalar field $\phi(x^\mu)$; otherwise, additional generators are included in what
follows.   Whether $\phi(x^\mu)$ is an ``elementary field'' is not relevant.
For any smooth function $f$, we write $\phi_f $ for the smeared field $\int \d^Dx\,f(\vec x,t) \phi(\vec x,t)$.
Then states of the form
\be\label{vanq}|\Psi_{\vec f}\,\rangle =\phi_{f_1}\phi_{f_2}\cdots \phi_{f_n}|\Omega\rangle\ee  
are sufficient to generate $\H_0$ in the Hilbert space sense.    (The purpose of smearing is to make sure that these states
have finite norm and thus really are Hilbert space states.)
In other words, any state in $\H_0$ can be approximated arbitrarily well by a linear combination of states $\Psi_{\vec f}\,$.  This is 
the definition of the vacuum sector $\H_0$. 

An initial value hypersurface (or Cauchy hypersurface) $\Sigma$ is a complete spacelike hypersurface on which, classically, one could
formulate initial data for the theory.  For example, $\Sigma$  could be the hypersurface $t=0$.   In eqn. (\ref{vanq}), we can require
that the functions $f_i$ are supported in any given open neighborhood $\U$ of $\Sigma$ (for example, in the open set $|t|<\epsilon$ for some
$\epsilon>0$ if $\Sigma$ is defined by $t=0$), and it is reasonable to hope that such states will still be enough to generate
the Hilbert space $\H_0$.    This statement is a quantum version of the fact that, classically, a solution of the field
equations is determined by initial data (fields and their time derivatives) on $\Sigma$.   Quantum mechanically, one
 may view this statement as part of what we mean by quantum field theory; it is Postulate 8(a) in \cite{HS}.    But actually, we will prove a stronger statement that is
 known as the Reeh-Schlieder theorem.

The Reeh-Schlieder theorem states that one can further restrict to an arbitrary small open set $\V\subset \Sigma$, and a corresponding
small neighborhood $\U_\V$ of $\V$ in spacetime.   Thus, even if we restrict the functions $f_1,\dots,f_n$ to be supported in $\U_\V$,
the states $\Psi_{\vec f}$ still suffice to generate $\H_0$.

If this were false, there would be some state $|\chi\rangle$ orthogonal to all states $|\Psi_{\vec f}\rangle$ such that the $f_i$
are supported in $\U_\V$:
\be\label{zeldox}0=\langle \chi|\Psi_{f_1f_2\cdots f_n}\rangle.\ee 
This is true  for all functions $f_1,\dots,f_n$ if and only if it is true without smearing, in other words if and only if
\be\label{meldox}\langle\chi|\phi(x_1)\phi(x_2)\cdots \phi(x_n)|\Omega\rangle=0,~~x_1,\dots,x_n\in \U_\V. \ee 
There is not really much difference between the two statements, since the matrix element of a product of local fields,
as in (\ref{meldox}), has singularities as a function of the $x_i$ and must be interpreted as a distribution.  So a precise interpretation of
eqn. (\ref{meldox}) involves a slightly smeared version, as in (\ref{zeldox}).

\subsection{Proof}\label{proof}

To prove the Reeh-Schlieder theorem, we will show that if, for some $\chi$, the left hand side of (\ref{meldox}) vanishes
for all $x_1,\dots,x_n\in \U_\V$, then it actually vanishes for all $x_1,\dots,x_n$ in Minkowski spacetime $M_D$.  This then
implies that $\chi$ must vanish, by the definition of the vacuum sector.  So only the zero vector is orthogonal to all states
created from the vacuum by local operators supported in $\U_\V$; in other words, such states are dense in $\H_0$.

First let us show\footnote{The following
argument is along the lines of that in \cite{StW}.  However, to avoid invoking the multi-dimensional edge of the wedge theorem,
we consider one variable at a time, as suggested by R. Longo.}  that
\be\label{goodness}\varphi(x_1,x_2,\cdots,x_n)=\langle\chi|\phi(x_1)\phi(x_2)\cdots \phi(x_n)|\Omega\rangle\ee
  continues to vanish if $x_n$ is moved outside of $\U_\V$, keeping the other variables in $\U_\V$.  We write
$\st$ for the time-like vector $(1,0,\dots,0)$ and examine the effect of shifting $x_n$ to $x_n+u\st$ for some real $u$.  In other words,
we shift $x_n$ by $u$ in the time direction, leaving its spatial coordinates unchanged.
Consider the function
\be\label{zono}g(u)=\langle\chi|\phi(x_1)\phi(x_2)\cdots \phi(x_{n-1})\phi(x_n+u\st)|\Omega\rangle=
\langle\chi|\phi(x_1)\phi(x_2)\cdots \exp(\i H u)\phi(x_n)\exp(-\i H u)|\Omega\rangle,\ee
where $H$ is the Hamiltonian.  We are given that $g(u)=0$ for sufficiently small real $u$ (since for small enough $u$, $x_n+u\st\in \U_\V$)
and we want to prove that it is identically 0.  Because $H|\Omega\rangle=0$, we can drop the last factor of $\exp(-\i H u)$ in eqn.
(\ref{zono}):
\be\label{zongo}g(u)=
\langle\chi|\phi(x_1)\phi(x_2)\cdots \exp(\i H u)\phi(x_n)|\Omega\rangle.\ee
Because $H$ is bounded below by 0, the operator $\exp(\i H u)$ is holomorphic for $u$ in the upper half plane.\footnote{The rigorous
proof of this sort of statement in \cite{StW} uses some smearing with respect to $x_n$ to first replace $\phi(x_n)|\Omega\rangle$
with a normalizable vector.  So although it is true that the smeared and unsmeared statements (\ref{zeldox}) and (\ref{meldox})
are equivalent, the smeared version is convenient in the rigorous proof.}  Thus the function
$g(u)$ is holomorphic in the upper half plane, continuous as one approaches the real axis, and vanishes on a segment $I=[-\epsilon,\epsilon]$ of the real axis.

\begin{figure}
 \begin{center}
   \includegraphics[width=5.5in]{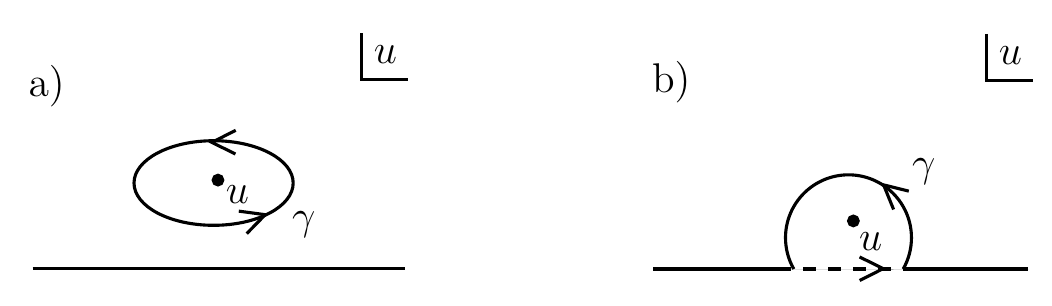}
 \end{center}
\caption{\small (a)  A function $g(u)$ holomorphic in the upper half $u$ plane can be computed by a Cauchy integral formula: any contour
$\gamma$ in the upper half-plane can be used to compute $g(u)$ for $u$ in the interior of $\gamma$.   (b) If $g(u)$ is continuous
on the boundary of the upper half-plane, one can take $\gamma$ to run partly along the boundary.  If in addition $g(u)=0$ along
part of the boundary -- indicated here by dashed lines -- then that part of the contour can be dropped.  In this case, the Cauchy integral
formula remains holomorphic as $u$ is moved through the gap and into the lower half-plane, implying that $g(u)$ is holomorphic
on that part of the real axis and  is identically zero. \label{Fig1}}
\end{figure}

If $g(u)$ were known to be holomorphic along the segment $I$, its vanishing along $I$ would imply that a Taylor series of $g(u)$
around, say, $u=0$ must be identically 0 and therefore that $g(u)$ is identically 0.  As it is, to begin with, we only have continuity
along the real axis and holomorphy in the upper half-plane.  However, using the fact that $g(u)$ vanishes in a segment of the
real axis (and imitating the proof of the Schwarz reflection principle), we can argue as follows.   For
$u$ in the upper half-plane, $g(u)$ can be represented by a Cauchy integral formula 
\be\label{tefo} g(u)=\frac{1}{2\pi \i}\oint_\gamma \d u' \frac{g(u')}{u'-u}. \ee
Here $\gamma$ is any contour that wraps counterclockwise once around $u$ (fig. \ref{Fig1}(a)).  For fixed $\gamma$, the formula is only valid for $u$ inside
the contour, since if we move $u$ across the contour, we meet the pole of the integrand.  However, if it is known
that $g(u)$ is identically 0 in a segment $I$ of the real axis, we can choose $\gamma$ to include that segment 
and then we can drop that part of the integral since $g(u')$ vanishes for $u'\in I$.  Once we do this, we are free to
move $u$ through the segment $I$ and into the lower half-plane (fig. \ref{Fig1}(b)); in particular, we learn that $g(u)$ is holomorphic along $I$.
As already explained, it follows that $g(u)$ is identically 0.  

In this argument, we could replace $\st$ by any other time-like vector.\footnote{In the case of a past-pointing timelike
vector, we make the same argument as before using holomorphy in the lower half $u$-plane.} Using some other 
timelike vector instead,\ we learn that 
$\langle\chi|\phi(x_1)\phi(x_2)\cdots \phi(x_n')|\Omega\rangle=0$ if $x_n'-x_n$ is any timelike vector and $x_1,\dots, x_n\in\U_V$.
But now we repeat the process with $x_n'$ replaced by $x_n''=x_n'+v \st'$ for  real $v$ and with some possibly different timelike vector $\st'$.
Analyzing the dependence on $v$ in exactly the same way, we learn that $\langle\chi|\phi(x_1)\phi(x_2)\cdots \phi(x_n'')|\Omega\rangle=0$
for any $x_n''$ of this form.  But since every point in Minkowski spacetime can be reached by starting with $\U_\V$ and zigzagging back
and forth in different timelike directions,  we learn that if, for some $x_1,\dots,x_{n-1}$,
$\varphi(x_1,\dots,x_{n-1},x_n)$
vanishes for all $x_n\in \U_\V$, then it actually vanishes for all $x_n$, without the restriction $x_n\in \U_\V$.

The next step is to remove the restriction $x_{n-1}\in\U_\V$.   We do this in exactly the same way, now shifting the last two coordinates
in a timelike direction.  Thus  we look now at
\be\label{urfu}g(u)=\langle\chi|\phi(x_1)\phi(x_2)\dots \phi(x_{n-2})\phi(x_{n-1}+u\st)\phi(x_n+u\st)|\Omega\rangle.\ee
Using again the fact that $H|\Omega\rangle=0$, we have
\be\label{rfu}g(u)=\langle\chi|\phi(x_1)\phi(x_2)\dots \exp(\i H u)\phi(x_{n-1}) \phi(x_n)|\Omega\rangle .\ee
Just as before, the function $g(u)$ is holomorphic in the upper half plane and vanishes along a segment of the real axis, so it is
identically zero.  Repeating this with a second timelike vector, we learn that we can make an arbitrary shift $x_{n-1},x_n
\to x_{n-1}+w,x_n+w$ without  affecting the vanishing of $\varphi(x_1,\dots,x_n)$.  Since we are also free to shift $x_n$ in an arbitrary
fashion, we learn that for $x_1,\dots,x_{n-2}\in \U_\V$, $\varphi(x_1,\dots,x_{n})$ is identically zero, with no restriction on $x_{n-1}$ and $x_n$.

The rest of the argument is hopefully clear at this point.  At the $k^{th}$ step, we make a timelike shift of the last
$k$ points, adding $u\st$ to each of them, and show as above that this does not affect the vanishing of $\varphi(x_1,x_2,\dots,x_n)$.    Repeating this with a shift by $v\st'$ and combining with the results of previous steps, we learn that vanishing of $\varphi(x_1,x_2,\cdots,x_n)$ is
not affected by moving the last $k$ points.
At the $n^{th}$ step, we finally learn that $\varphi(x_1,x_2,\dots,x_n)$ is identically zero for all $x_1,x_2,\cdots,x_n$.

For future reference, a systematic holomorphy statement that can be proved similarly to the above is as follows.   The $\H$-valued function
\be\label{staval}  F(x_1,x_2,\cdots,x_n)=\phi(x_1)\phi(x_2)\cdots \phi(x_n)|\Omega \ra\ee
(or  the inner product of this function with any other state) is holomorphic if the imaginary part of 
$x_1$ and of $x_{i+1}-x_i$, $i=1,\cdots,n-1$
is future timelike.  (It is continuous up to the boundary of that domain.)  This is proved 
by writing\footnote{We work in signature $-++\cdots +$, so $x\cdot P=-tH+\vec x\cdot \vec P$ where $H$ is the Hamiltonian;
 this operator is negative semidefinite for $t>|\vec x|$, so $|\exp(-\i x\cdot P)|\leq 1$ for $\mathrm{Im}\,x$ future timelike. This ensures  that for such $x$,
 the operator
 $\exp(-\i x\cdot P)$ is defined for all states and holomorphically varying.}
\begin{align}\label{aval}F(x_1,x_2,\cdots,x_n)=  &\bigl[\exp(-\i x_1\cdot P)\phi(0)\exp(\i x_1\cdot P)\bigr]\bigl[\exp(-\i x_2\cdot P)\phi(0)\exp(\i x_2\cdot P)\bigr] \cr &\cdots\bigl[ \exp(-\i x_{n-1}\cdot P)\phi(0)\exp(\i x_{n-1}\cdot P)\bigr]\bigl[\exp(-\i x_n\cdot P)\phi(0)\bigr]|\Omega\ra\end{align}
and using the fact that  $\exp(-\i x_1\cdot P)$ and each $\exp(-\i (x_{j}-x_{j-1})\cdot P)$ is bounded and
varies holomorphically under the stated condition on the
$x$'s.

\subsection{Vectors Of Bounded Energy-Momentum}\label{bounded}

In proving the Reeh-Schlieder theorem, we used the fact that the energy-momentum operators $P^\mu,\,\mu=0,\cdots,D-1$ annihilate 
the vacuum state $|\Omega\rangle$.
This implies, in particular, that for any $D$-vector $c$, $\exp(\i c\cdot P)|\Omega\rangle=|\Omega\rangle$.  
 However\cite{Borchersold}, in the proof
it would be sufficient to know that, for a general $D$-vector $c^\mu$,  $\exp(\i c\cdot P)|\Omega\rangle$ varies holomorphically with the
components $c^0,c^1,\cdots, c^{D-1}$ of $c$.    Then in the above argument, we could not drop the factor $\exp(\i u \st\cdot P)|\Omega\rangle$,
but its presence would not affect the discussion of holomorphy.  

If a state $\Psi$ has the property that $\exp(\i c\cdot P)|\Psi\rangle$ is holomorphic in $c$, we say that the translation group
acts holomorphically on $\Psi$. 
 This is not true for an arbitrary $\Psi$, since if $c$ has a future timelike imaginary
part, $\exp(\i c\cdot P)$ is an unbounded operator and $\exp(\i c\cdot P)|\Psi\rangle$ may not make sense in Hilbert space.\footnote{\label{remember} An unbounded operator on a Hilbert space is defined at most on a dense set of vectors.
Suppose, for example, that in some orthonormal basis
$\psi_n$ of a Hilbert space $\H$, an operator $X$ acts by $X\psi_n=\lambda_n\psi_n$.  For $X$ to be unbounded means
that the $\lambda_n$ are unbounded.  In this case, there is a vector $\Psi=\sum_n c_n \psi_n$ with $\sum_n|c_n|^2<\infty$ (so
$\Psi\in \H$) but $\sum_n |\lambda_n|^2|c_n|^2=\infty$ (so $X\Psi$ does not make sense as a vector in $\H$).}

  A source of many vectors  on which the translation group has a holomorphic action 
is the following.  The $P^\mu$ are a set of $D$ commuting, self-adjoint operators.  This leads to a spectral decomposition of the Hilbert
space $\H$ on which the $P^\mu$ act.  For every closed set $S$ in momentum space, there is a corresponding projection operator $\Pi_S$
onto the subspace $\H_S$ of Hilbert space consisting of states whose energy-momentum is contained in the set $S$.  (We cannot actually
diagonalize the $P^\mu$ in Hilbert space, since states of definite energy-momentum -- other than the vacuum -- are not normalizable.)
If $S$ is compact, then in any Lorentz frame, the energy of a state $\Psi$ that is in the image of $\Pi_S$ is bounded.
This  gives, for any $c$, an upper bound on the norm of $\exp(ic\cdot P)\Psi$ and ensures that the translation group acts holomorphically
on $\Psi$.

If $\Psi$ is any state and $S$ is compact, the projection $\Pi_S\Psi$ to states with energy-momentum in $S$ is a state on which
the translation group acts holomorphically.  Moreover,  $\Pi_S\Psi$ is nonzero for sufficiently large $S$ and in fact converges to $\Psi$
as $S$ becomes large.  So every state can actually be approximated by states that could be used instead of the vacuum in the
Reeh-Schlieder theorem.

As an example of why this is useful, we can consider superselection sectors.  In general, the ``vacuum sector'' $\H_0$, consisting
of states that can be created from the vacuum by a product of local operators, is not the full Hilbert space $\H$ of a quantum field theory.
In part, this is because there may be conserved charges that are not carried by any local operator.  For example, in four spacetime dimensions, a theory with a massless $\UU(1)$ gauge field has conserved
electric and magnetic charges that are not carried by any local operators.\footnote{Below four spacetime dimensions,
it may not be possible to fully characterize superselection sectors by conserved charges.  An example is given by
three-dimensional theories with nonabelian statistics.  (For a treatment of this situation in algebraic quantum
field theory, see \cite{FRS}.) 
  Likewise, soliton sectors in two spacetime dimensions cannot always be
fully characterized by conserved charges.
However, the following remarks about the Reeh-Schlieder theorem do not depend on whether a given superselection sector
can be characterized by conserved charges.}  
  Let $\H'$ be the subspace of Hilbert space characterized by 
particular values of these charges.  Such an $\H'$ is called a superselection sector.  In a nontrivial superselection sector (not containing
the vacuum), there is no state of lowest energy that we could use instead of the vacuum in the Reeh-Schlieder theorem.\footnote{To minimize
the energy of, say, a magnetic monopole, we would want to take it to have zero momentum.  But such a state is not normalizable.}
However, in such a sector, there is no problem to construct states of bounded energy-momentum, and for any such state $\Lambda$,
the analog of the Reeh-Schlieder theorem holds:  whatever can be created by local operators acting on $\Lambda$ can be created
by local operators that act on $\Lambda$ in the small open set $\U_\V$.

What happens to the Reeh-Schlieder theorem if Minkowski spacetime $M_D$ is replaced by another globally hyperbolic spacetime $M$?
In curved spacetime, there is no natural analog of the vacuum state, and there are, of course, also no natural translation 
generators $P^\mu$.   However, it is natural to expect that the Reeh-Schlieder theorem should have an analog for any spacetime $M$
that is globally hyperbolic and real analytic.   An analog of a vector on which spacetime translations
act holomorphically is  a vector whose evolution is holomorphic in the following sense.  
In general, a vector $\Psi_\Sigma$ defined in quantization
on a Cauchy  hypersurface $\Sigma\subset M$ can be evolved forwards or backwards in time to a vector $\Psi_{\Sigma'}$ on
any other such hypersurface $\Sigma'$.  If $M$ is real analytic, it can be ``thickened'' slightly to a complex analytic manifold
$\hat M$, and we can ask whether $\Psi_{\Sigma'}$ evolves holomorphically with $\Sigma'$ if $\Sigma'$ is displaced slightly
away from $M$ in $\hat M$.  If so, we say that $\Psi_\Sigma$ has holomorphic evolution and a reasonable analog
of the Reeh-Schlieder theorem would say  that states $\a\Psi_\Sigma$,
where $\a$ is supported in some given open set, are dense in Hilbert space.  For results in this direction, see \cite{SVW,GW}.  There is
also a version of the Reeh-Schlieder theorem adapted to Anti de Sitter space and holography \cite{morrison}, and
there are attempts to generalize the theorem to curved spacetime without assuming real analyticity
\cite{KS}.

\subsection{An Important Corollary}\label{corollary}

The Reeh-Schlieder theorem has an important and immediate corollary.   Let us assume that the open set $\V\subset \Sigma$
is small enough so that its closure $\bar\V$ is not all of $\Sigma$.  Then the complement of $\bar \V$ in $\Sigma$ is another open set $\V'$,
disjoint from $\V$.  $\V'$ and $\V$ are spacelike separated, and they are contained in  small open sets $\U_\V, \U_{\V'}\subset M_D$ that
are also spacelike separated.  One also may choose to let $\U_\V$ and $\U_{\V'}$ be as large as possible, while remaining at
spacelike separation.  The precise choice of $\U_\V$ and $\U_{\V'}$ is not important in this section.

Let $\a$ be any operator
supported in the spacetime region $\U_\V$, not necessarily constructed from a product of finitely many local operators.
Because the regions $\U_\V$, $\U_{\V'}$ are spacelike separated, $\a$ commutes with local operators in $\U_{\V'}$;
\be\label{reddo} [\phi(x), \a]=0,~~~x\in \U_{\V'}.\ee
 Conversely, an operator
$\a'$ supported in $\U_{\V'}$ satisfies
\be\label{eddo}[\phi(x),\a']=0,~~~x\in\U_\V.\ee

The Reeh-Schlieder theorem applies equally well to $\V$ or to $\V'$, as they are both nonempty open sets in the initial value hypersurface
$\Sigma$.   This has the following consequence.  
Suppose that an operator $\a$ supported in $\U_\V$ annihilates the vacuum state 
\be\label{peddo} \a|\Omega\rangle=0. \ee
Because $\a$ commutes with the local operators $\phi(x_i)$, $x_i\in \U_{\V'}$,  the vanishing of $\a|\Omega\rangle $ implies
that
\be\label{leddo} \a \phi(x_1)\phi(x_2)\cdots \phi(x_n)|\Omega\rangle =0, ~~~x_i\in \U_{\V'}. \ee
But the Reeh-Schlieder theorem tells us that the states $\phi(x_1)\phi(x_2)\cdots \phi(x_n)|\Omega\rangle$, $x_i\in\U_{\V'}$ are
dense, in the vacuum sector $\H_0$ of Hilbert space.  So the vanishing of the left hand side of eqn. (\ref{leddo}) for all $n$ and all
$x_i\in \U_{\V'}$ implies that the operator $\a$  is identically 0, in the vacuum sector.

For an open set $\U$ in spacetime, let us define $\A_\U$ to be the algebra of operators supported in $\U$.  We will call this
a ``local algebra'' of the quantum field theory.  In section \ref{furtherp}, we will be more specific about what we mean by ``all operators.'' For
now we leave this open.
In the present discussion, we have considered two open sets, namely $\U=\U_\V$ and $\U'=\U_{\V'}$, which are thickenings of $\V$ and $\V'$,
respectively, so there are two algebras to consider, namely $\A_\U$ and $\A_{\U'}$.

By way of terminology, a vector $\Psi$ in a Hilbert space $\H_0$ is called a cyclic vector for an algebra such as $\A_\U$ if the states
$\a|\Psi\rangle$, $\a\in \A_\U$ are dense in $\H_0$.     $\Psi$ is said to be separating for $\A_\U$ if the condition $\a|\Psi\rangle=0$,
$\a\in\A_\U$ implies that $\a=0$.  The Reeh-Schlieder theorem says that the vacuum vector $\Omega$ is cyclic for $\A_\U$ and
for $\A_{\U'}$.  As we have just explained,  a state that is cyclic for one of these algebras is separating for the other, so in fact the
vacuum is cyclic and separating for $\A_\U$ and for $\A_{\U'}$.

More generally, the Reeh-Schlieder theorem implies that, in each superselection sector,
any vector on which the translation group acts holomorphically is
cyclic and separating for $\A_\U$ and for $\A_{\U'}$.

As we have seen,  if $\U$ and $\U'$ are a pair of spacelike separated open sets, then many vectors are cyclic and separating for
$\A_\U$ and for $\A_{\U'}$, but it is certainly not
true that every vector has this property.  For a simple counterxample, consider a theory with a complex free fermion $\psi$.  
Then for a smearing function $f$ supported
in $\U$,
$\psi_f=\int \d^4x f(x)\psi(x)$ obeys $\psi_f^2=0$.  It therefore annihilates any vector of the form $\psi_f\chi$.   If one defines
the local algebras to consist of bosonic operators only (as does Haag \cite{Haag}), then one
can pick a pair of smearing functions $f,g$ supported
in $\U$ and set $\O_{f,g}=\psi_f\psi_g$.  Then $\O_{f,g}$ is a bosonic operator supported in $\U$ and obeying
 $\O_{f,g}^2=0$, so $\O_{f,g}$ annihilates any
state $\O_{f,g}\chi$.  So $\psi_f\chi$ or $\O_{f,g}\chi$  is a state that is not separating for $\A_\U$, or cyclic
for $\A_{\U'}$.

The fact that the vacuum is separating for the algebra $\A_\U$ has interesting consequences for the energy density
in quantum field theory \cite{EGJ}.  Of course, the total energy $H$ is positive semidefinite, and annihilates only the vacuum state.  It can be defined as the integral  of the energy density $T_{00}$ over an initial value surface $t=0$.
However, in contrast to classical physics, the energy density $T_{00}(x)$ is not positive-semidefinite in quantum field theory, and the same holds for any smeared operator $T_f=\int_{\U_\V} \d^Dx \,f(x) T_{00}(x)$,
where $f$ is any real smearing function with support in  $\U_\V$.  Poincar\'{e} invariance and the fact that $H\Omega=0$ imply that the
vacuum has vanishing energy density, $\la \Omega|T_{00}(x)|\Omega\ra=0$.     However, the separating property of the vacuum for the algebra $\A_\U$
implies that $T_f|\Omega\ra\not=0$.  Let $\chi$ be some state with $\la \chi|T_f|\Omega\ra\not=0$.  Let $\mathcal W$ be the two-dimensional
subspace of Hilbert space generated by $\Omega$ and $\chi$.   If we write a vector in $\mathcal W$ as a column vector with
$\Omega$ and $\chi$ corresponding to the upper and lower components, then  $T_f$ restricted to $\mathcal W$ takes the form
\be\label{zorof}\bp 0 & b\cr \bar b & c\ep ,\ee
with $b=\la\chi|T_f|\Omega\ra\not=0$.  Such a matrix is not positive semidefinite, implying that $T_f$ has a negative expectation value
in some state $\t\chi\in{\mathcal W}\subset \H$.

\subsection{Discussion} \label{discussion}

\def\MM{{\sf M}}
The Reeh-Schlieder theorem may seem paradoxical at first.  It implies that by acting on the vacuum with an operator $\a$
supported in a small region $\U_\V$, one can create whatever one wants -- possibly a complex body such as the Moon --  in a faraway, spacelike separated region of spacetime.

To understand this better, let $\V^*$ be a distant region in which we want to create the Moon.  Let $\MM $ be an operator supported in region $\U_{\V^*}$ that to good approximation has expectation value 0 in states
that do not contain a moon in region $\V^*$ and 1 in states that do contain one.  Thus
\be\label{turnof}\langle \Omega|\MM |\Omega\rangle \approx 0 ,\ee
but according to the Reeh-Schlieder theorem, there is some operator $\a$ supported in $\U_\V$ such that the state $\a\Omega$,
to very good approximation, contains a moon in region $\V^*$.   Thus $\langle \a\Omega|\MM |\a\Omega\rangle\approx 1$, so
$\langle\Omega|\a^\dagger \MM  \a|\Omega\rangle\approx 1.$    As $\a^\dagger $ is supported in region $\U_\V$ and $\MM $ is supported in the
spacelike separated region $\U_{\V^*}$, these operators commute and thus
\be\label{urnof}\langle\Omega| \MM  \a^\dagger \a|\Omega\rangle \approx 1. \ee

Is there a conflict between (\ref{turnof}) and (\ref{urnof})?   If we could choose the operator $\a$ to be unitary, we would have $\a^\dagger \a=1$,
and then there would indeed be a conflict.   However, the Reeh-Schlieder theorem does not say that there is a {\it unitary} operator
supported in $\U_\V$ that will create the Moon in some distant region; it merely says that there is {\it some} operator supported in 
$\U_\V$ that will do this.  

If one asks about not mathematical operations  in Hilbert space but physical operations that 
are possible in the real world, then the only physical way that one can modify a quantum state is by perturbing the Hamiltonian by which it
evolves, thus bringing about a unitary transformation.  If one is able to couple a given quantum field theory to some auxiliary quantum system,
then one can implement a unitary transformation on the combined system.  It is not possible by such a unitary transformation supported
in $\U_\V$ to make any change in observations in a spacelike separated region $\V^*$.  That is what we learn from the above computation,
which shows that for any operator $\MM $ supported in $\U_{\V^*}$ and any unitary operator $\a$ supported in $\V$,
$\langle \a\Omega|\MM |\a\Omega\rangle =\langle \Omega|\MM |\Omega\rangle$.  
This computation is unaffected if $\a$ acts also on some auxiliary quantum system, as long as $\a$ is unitary and commutes with operators
in $\V^*$.

While it is not possible for a physical operation in one region to influence a measurement in another region, there can be correlations in the vacuum
 between operators in the two regions.  This happens all the time in quantum field theory, even in free field
theory.  We are seeing such correlations in eqn. (\ref{urnof}), which shows that $\langle \Omega|\MM \a^\dagger \a|\Omega\rangle
\not= \langle \Omega|\MM |\Omega\rangle \langle \Omega|\a^\dagger \a|\Omega\rangle$.

The Reeh-Schlieder theorem can be given an intuitive interpretation by considering a finite-dimensional quantum system with a tensor
product Hilbert space $\H=\H_1\otimes \H_2$.   For what follows, the most interesting case is that $\H_1$ 
and $\H_2$ have the same dimension
$n$.  We let $\A_1$ be the algebra of $n\times n$ matrices acting on $\H_1$, and $\A_2$ the algebra of $n\times n$ 
matrices acting on $\H_2$. (In language that we will introduce shortly, these are $*$-algebras and they are each other's commutants.)
 A generic state $\Psi$ of the composite system is entangled.  For any given $\Psi$, it is possible to
choose a basis $\psi_k$, $k=1,\dots,n$ of $\H_1$ and another basis $\psi'_k$, $k=1,\dots,n$ of $\H_2$ such that
\be\label{morx} \Psi =\sum_{k=1}^n c_k \psi_k\otimes \psi_k', \ee
with some coefficients $c_k$.
It is convenient to write $|k\rangle$ and $|k'\rangle$ for $\psi_k$ and $\psi_k'$, so that this formula becomes
\be\label{orx}\Psi=\sum_{k=1}^n c_k |k\rangle \otimes |k\rangle'. \ee
The vector $\Psi$ is cyclic and separating for $\A_1$ and for $\A_2$ if and only if the $c_k$ are all nonzero,
or equivalently if the reduced density matrices on $\H_1$ and on $\H_2$ are invertible.  We will return to this setup in section \ref{fincase}.

The Reeh-Schlieder theorem says that, in quantum field theory, if $\A_\V$ and $\A_{\V'}$ are the algebras 
of operators supported in complementary
regions of spacetime, then similarly the vacuum is a cyclic separating vector for this pair of algebras.\footnote{This remains so if $\V$ is replaced by a smaller
region, and $\V'$ by a correspondingly larger one.  That fact would have no natural analog for a finite-dimensional quantum system,
and shows in a different way from what is explained in the text
the limitations of the analogy between the vacuum of a quantum field theory and a fully entangled state of a finite-dimensional quantum system.}
  This might make one suspect
that the Hilbert space $\H$ should be factored as $\H=\H_\V\otimes \H_{\V'}$, with the vacuum being a fully entangled vector in the
sense that the coefficients analogous to $c_k$ are all nonzero.  This is technically not correct.   If it were correct, then picking $\psi\in
\H_\V$, $\chi\in \H_{\V'}$, we would get a vector $\psi\otimes\chi\in \H$ with no entanglement between observables in $\V$ and
those in $\V'$.  This is not what happens in quantum field theory.  In quantum field theory, the entanglement entropy 
between adjacent
regions has a universal ultraviolet divergence, independent of the states considered. The leading ultraviolet divergence is the same in any state as it is in the vacuum, because every state looks like the vacuum at short
distances.   The universality of this ultraviolet divergence means
that it reflects not a property of any particular state but rather the fact that $\H$ cannot be factored as
$\H_\V\otimes \H_{\V'}$.     

It is also not correct, technically, to write $\H$ as a direct sum or integral of Hilbert spaces $\H_\V^\zeta$
and $\H_{\V'}^\zeta$, where $\zeta$ is some discrete or continuous variable and each $\H_\V^\zeta$, $\H_{\V'}^\zeta$ is supposed
to furnish a representation of $\A_\V$ or $\A_{\V'}$.  If one had $\H=\oplus_\zeta \H_{\V}^\zeta\otimes \H_{\V'}^\zeta$ (where the direct
sum over $\zeta$ might be a continuous integral), then there would be operators -- such as any function of $\zeta$ -- that commute with
both $\A_\V$ and $\A_{\V'}$.  Bounded functions of the $\zeta$'s would be bounded Hilbert space operators, defined on all states. Moreover, because the leading
ultraviolet divergence in the entanglement entropy is proportional to the
area of the boundary between these two regions, these operators would have to be local along the boundary.  There is nothing
like that in quantum field theory.  What we usually call a local operator $\phi(x)$ has to be smeared just to make a densely defined
unbounded operator (let alone a bounded operator, defined on all of Hilbert space), 
and such a smeared operator does not commute with $\A_\V$ and $\A_{\V'}$.

Despite all this, many statements that one could deduce from a naive factorization $\H=\H_\V\otimes \H_{\V'}$ and whose analogs
are true for entangled quantum systems of finite dimension are actually true in quantum field theory.  
Tomita-Takesaki theory, which we introduce in section \ref{relmood}, is an important tool in proving such statements.

\subsection{The Local Algebras}\label{furtherp}

In section \ref{corollary}, we introduced the notion of associating to an open set $\U$ in spacetime a ``local algebra'' $\A_\U$
consisting of ``all operators'' supported in $\U$.

\begin{figure}
 \begin{center}
   \includegraphics[width=5.5in]{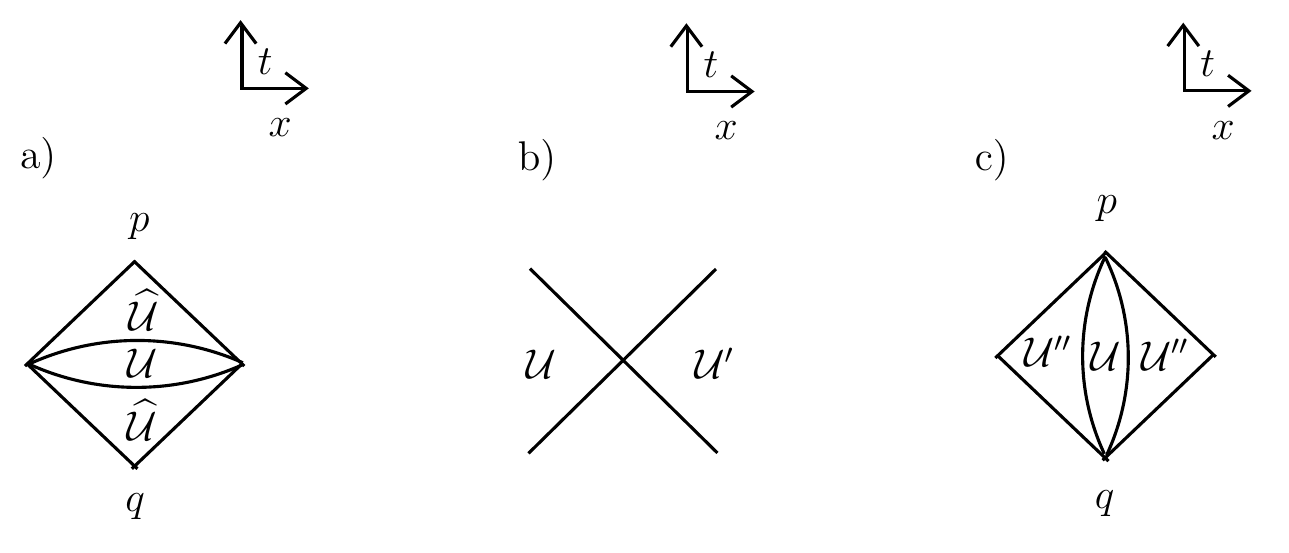}
 \end{center}
\caption{\small (a) An open set $\U$ in Minkowski spacetime, and its domain of dependence $\h\U$ (the union of $\U$ with the
regions labeled as $\h\U$ in the figure), which in this
case is a causal diamond and coincides with the causal completion $\U''$ of $\U$. (b) The two open sets $\U$ and $\U'$
are causal complements; each is the largest open set that is spacelike separated from the other.  (c) A quite different open set  $\U$ whose
causal completion  $\U''$ (the union of $\U$ and the regions labeled $\U''$) is the same causal diamond as in (a).  \label{Fig1.5}}
\end{figure}

But what do we mean by ``all operators''?   The operators that we have considered so far are what one might call simple operators,
namely polynomials in smeared local fields.  However, there are serious drawbacks to considering only simple operators.\footnote{The simple operators
also have important advantages, of course; they are the basis of a standard and powerful machinery of renormalization theory, operator product expansions, and
so on.}   For
one thing,  one would like to be able to claim \cite{HS} that if $\U$ is an open set in spacetime and $\h\U$ is a larger open set that is its
domain of dependence (fig.  \ref{Fig1.5}(a)) then the algebras $\A_\U$ and $\A_{\h\U}$ coincide.  The logic behind this is that
the dynamical time evolution of the theory determines operators in the larger region $\h\U$ in terms of operators in $\U$.   This
is true, but operators supported in regions of $\h\U$ that are to the future or the past of $\U$ are in general exceedingly complex
functions of operators in $\U$.  Thus we can only get a simple relation $\A_\U=\A_{\h\U}$ if we include in $\A_\U$ all operators
that can be made from the simple ones.

What sort of operators can we make from simple ones?    Some elementary operations come to mind.  For example, if $f$
is a real smearing function and $\phi_f=\int \d^Dx\, f\phi$, we can consider the operator $\exp(\i \phi_f)$, which actually is a {\it bounded}
operator made from $\phi_f$.  More generally, if $F$ is any bounded function of a complex variable, we can consider $F(\phi_f)$ (now with a possibly complex-valued
smearing function $f$);
this again is a bounded operator.
Still more generally, if $f_1,\dots,f_n$ are $n$ smearing functions and $F$ is a bounded function of $n$ complex variables,
we can consider $F(\phi_{f_1},\phi_{f_2},\cdots,\phi_{f_n})$.  

The reason to consider bounded operators is that they  are defined on all of Hilbert space, so  they 
can be multiplied without any trouble, and  naturally form an algebra.  Unbounded
operators in general cannot be multiplied, as they are defined on different dense subspaces of Hilbert space.
If we try to define  ``all unbounded functions'' of the $\phi_f$'s and hope to make them into an algebra, we will probably
have a lot of trouble.

We could go on with elementary constructions.  To complete the story, what is really needed is to include limits 
of the operators we already have.   
To decide what sort of limits to allow, let us think for a moment about what is involved in {\it measuring} an operator, such as
the weak Hamiltonian that is involved in beta decay.   What an experiment gives us is a measurement of 
finitely many matrix elements of an operator, each with some experimental error.  If $\a_1,\a_2,\cdots $ is a sequence
of operators  all of whose matrix elements $\la\psi|\a_n|\chi\ra $   converge for large $n$ to the corresponding matrix
elements $\la\psi|\a|\chi\ra$ of some operator $\a$,  
this means that any given experiment will not distinguish $\a_n$ from $\a$ once $n$ is large enough.
In such a situation, it is reasonable physically to say that $\a=\lim_{n\to\infty}\a_n$.   What we have just described (following Haag \cite{Haag}
in this reasoning) is the
mathematical notion of a weak limit of a sequence of operators.
 
 It is reasonable to believe that we should define $\A_\U$ to be closed under such weak limits.\footnote{However, a result of
 von Neumann shows that if we define $\A_\U$ to be closed only under a more restricted type of limit called a strong limit,
 we will actually get the same algebra.  A sequence $\a_1,\a_2,\cdots$ of operators has an operator $\a$ as its strong limit
 if for any Hilbert space state $\chi$, $\lim_{n\to\infty} \,\a_n\chi=\a\chi$.}    One also expects $\A_\U$
 to be closed under a more trivial operation.   
 The set of smeared fields in a given region is closed under hermitian conjugation.  (If $\phi_f=\int \d^Dx \,f(x)\phi(x)$ is a smeared
field supported in a given region, then so is $\phi_f^\dagger=\int \d^Dx\,\bar f(x)\phi(x)$.)   Any reasonable set of operations
that builds new operators from old ones, starting from a set of operators that is closed under hermitian conjugation,
 will give a set of operators that remains closed under hermitian conjugation.
  An algebra acting on a Hilbert space and closed under hermitian
conjugation is called a $*$-algebra.  Thus any reasonable choice of what we would mean by $\A_\U$ will be a $*$-algebra.  

A $*$-algebra of bounded operators on a Hilbert space that is closed under weak limits  (and contains the identity operator)
is called a von Neumann algebra.   Thus
we are led in this way to the notion that the local algebra $\A_\U$ of an open set $\U$ should be a von Neumann algebra.

If $\A$ is a $*$-algebra of bounded operators on a Hilbert space $\H$, then its commutant $\A'$, defined as 
the set of all bounded operators on $\H$  that commute
with $\A$,   is another $*$-algebra.
   $\A'$ is always a von Neumann algebra
even if $\A$ is not.\footnote{\label{later}The nontrivial point is that $\A'$ is closed under weak limits.
 If $\a_1',\a_2',\cdots$ is a sequence of bounded operators that commute with $\A$ and 
has weak limit $\a'$, then for any states $\psi,\chi\in\H$ and any $\a\in \A$, one has $\la\psi|[\a,\a']|\chi\ra=\lim_{n\to\infty}\la\psi|[\a,\a'_n]|\chi\ra
=0$; vanishing of $\la\psi|[\a,\a']|\chi\ra$ for all $\psi,\chi$ means $[\a,\a']=0$ and therefore $\a'\in\A'$, showing that $\A'$ is closed
under weak limits.}  If $\A$ is a von Neumann algebra, then the relation
between $\A$ and $\A'$ is reciprocal: each is the commutant of the other.    This is von Neumann's theorem that if $\A$ is a von Neumann
algebra, then $\A''=(\A')'$ satisfies $\A''=\A$.

Operators at spacelike
separation commute, so one expects that if $\U$ and $\U'$ are spacelike separated, then\footnote{In the presence of fermions,
one has anticommutativity as well as commutativity of operators at spacelike separation.  In the algebraic approach, one can consider
a von Neumann algebra with an automorphism that distinguishes even and odd operators.  For one approach, see \cite{GL}.}
\be\label{motto} [\A_\U,\A_{\U'}]=0,\ee
which is an abbreviated way to say that $[\a,\a']=0$ if $\a\in\A_\U$, $\a'\in\A_{\U'}$. 
Thus one expects that $\A_{\U'}$ is always contained in $\A_\U'$.  

It was proposed by Haag \cite{HG} and by Haag and Schroer \cite{HS} that if $\U$ and $\U'$ are causal complements, meaning that they are
maximal open sets under the condition of being spacelike separated, then the corresponding algebras $\A_\U$ and $\A_{\U'}$
are commutants, meaning that they are maximal under the condition of commuting with each other.  This condition,  sometimes
called Haag duality,  can be written
\be\label{poffo} \A_{\U'}=\A'_\U.  \ee
 This condition is stated in
\cite{Haag} as part of Tentative Postulate 4.2.1.   The rest of the postulate says that if $\U$ is a union of open sets $\U_\alpha$,
then $\A_\U$ is the smallest von Neumann algebra containing the $\A_{\U_\alpha}$, and that if $\U$, $\t\U$ are two open sets
then $\A_{\U\cap\t\U}=\A_\U\cap \A_{\t\U}$.    Haag duality is known to be true in many circumstances; for example, it
was proved by Bisognano and Wichmann \cite{BiWi}  for complementary Rindler regions in Minkowski spacetime (this is explained at the
end of section \ref{pathint}).  Haag duality and the rest of Postulate 4.2.1 are apparently valid in an interesting class of quantum
field theories and for some open sets in a wider class, but it  appears that 
in some theories and for some classes of open sets,  Haag duality and other
parts of Tentative Postulate 4.2.1 can fail
\cite{LRT,DL,SchroerP, HO}.

 We will give an example of the simplification that occurs if two algebras are commutants.
If $\A$ and $\A'$ are commutants, then
 a vector $\Omega\in\H$ is separating for $\A$ if and only if it is cyclic for $\A'$, and vice-versa.   The ``if'' part of
 this statement only depends on $\A$ and $\A'$ commuting and was explained in  section
 \ref{corollary}.   What we gain if $\A$ and $\A'$ are commutants is the ``only if'' statement.  Suppose in fact that a vector 
$\Omega$ is not cyclic for $\A'$.   Then the vectors $\a'|\Omega\rangle$, $\a'\in \A'$
generate a Hilbert space  $\H'$ that is a proper subspace of $\H$.  
 Let $\Pi:\H\to\H$ be the orthogonal projection onto $\H'_\perp$.  Then $\Pi$ is bounded and commutes
with $\A'$, so if the two algebras are commutants, $\Pi\in\A$.   But $\Pi\Omega=0$ (since $1\in\A'$, certainly
$\Omega=1\cdot\Omega$ is of the form $\a'\Omega,$ $\a'\in\A'$, and therefore $\Omega\in\H'$).   Thus if $\Omega$
is not cyclic for $\A'$, then $\Pi\in\A$ annihilates $\Omega$ and $\Omega$ is not separating for $\A$.

We conclude by describing an analogy between algebras and open sets that is developed in \cite{Haag}.  In the analogy,
a $*$-algebra corresponds to an open set in spacetime,  a von Neumann algebra corresponds to a causally complete open set,
and commutants correspond to causal complements.

  Let $\A$
be a $*$-algebra of bounded operators on $\H$ (not necessarily a von Neumann algebra) and $\A'$ its commutant.
Then $\A'$ is a von Neumann algebra as explained in footnote \ref{later}.  In particular, the commutant $\A''=(\A')'$ of $\A'$ is
a von Neumann algebra.  Clearly $\A\subset \A''$ ($\A''$ consists of all bounded operators that commute with $\A'$,
and the definition of $\A'$ ensures that any element of $\A$ commutes with $\A'$).   $\A''$ is called the von Neumann algebra closure of $\A$; it is the smallest
von Neumann algebra containing $\A$.  If $\A$ was a von Neumann algebra to begin with, then $\A=\A''$.  On the other hand
$\A'$ is always a von Neumann algebra so one always has $\A'=\A'''$.  If $\A$ is a von Neumann algebra, $\A$ and $\A'$ are each other's
commutants. 

Now consider open sets.  If $\U$ is an open set, then as above, its causal complement $\U'$ is the union of all open sets that
are spacelike separated from $\U$ (equivalently, it is the largest open set spacelike separated from $\U$).
The causal complement   $\U''=(\U')'$ of $\U'$ always contains $\U$, since $\U$ is an open set spacelike separated from $\U'$.  
 One always has $\U'''=\U'$.   (Indeed, since $\U\subset \U''$,
the condition for a point to be spacelike separated from $\U''$ is stronger than the condition for it to be spacelike separated from
$\U$, so $\U'''= (\U'')'\subset \U'$.  The opposite inclusion $\U'\subset \U'''$ just says that the open set $\U'$ is contained in $(\U')''=\U'''$.)   $\U$ is said to be causally complete if $\U''=\U$.   The result $\U'''=\U'$ means that $\U'$
is always causally complete.     In general, $\U''$ (which also is always causally complete since $\U'=\U'''$ implies $\U''=\U''''$) is the smallest
causally complete set containing $\U$ and is called 
the causal completion of $\U$.  If $\U$ is causally complete, then
 $\U$ and $\U'$ are each
other's causal complements.

If Haag duality holds in some theory 
for all open sets, not necessarily causally complete, then it implies that $\A_\U =\A_{\U''}$ for all $\U$,
a property stated in  \cite{Haag}, (III.1.10).   (Indeed, Haag duality says that $\A_{\U''}=(\A_{\U'})'=(\A_\U)''=\A_\U$, where in the
last step we use the fact that $\A''=\A$ for any von Neumann algebra $\A$.)  The conditions for this to hold
 do not appear to be known,\footnote{As a counterexample if $\U$  is not required to be connected,
in two-dimensional spacetime, let $\U$ be the union of small balls centered at the two points $(t,x)=(\pm 1,0)$.   Then $\U''$
is again a (slightly rounded) causal diamond.  Massless fields are functions only of $x_\pm = x\pm t$.  In $\U''$, one can measure
modes of massless fields in the whole range $-1\leq x_\pm \leq 1$, but in $\U$, one only see values of $x_\pm$ near $\pm 1$.}
    but it does have a surprisingly wide range of validity.  Two illustrative cases are
shown in figs. \ref{Fig1.5}(a) and (c).  In fig. \ref{Fig1.5}(a), $\U''$ is a causal diamond, and coincides with the domain of dependence
$\hat \U$  of $\U$.    Causality would lead
us to expect in this example
 that $\A_\U=\A_{\U''}$ and this was indeed an input to the discussion above.  In fig. \ref{Fig1.5}(c), $\U$
is a thin ``timelike tube'' (with corners at the top and bottom) whose causal completion  $\U''$ is the same causal diamond.     In this case, there is no simple
reason of causality  to expect that $\A_\U=\A_{\U''}$, but this can be proved with a more sophisticated use of the 
ingredients that went into proving the Reeh-Schlieder theorem.  The result is sometimes called the Borchers timelike tube theorem \cite{OtherBorchers, OtherAraki, Wightman,BorchersBook}.

\section{The Modular Operator And Relative Entropy In Quantum Field Theory}\label{relmood}

\subsection{Definition and First Properties}\label{defop}

In some quantum field theory in Minkowski spacetime with Hilbert space $\H$, let  $\A_\U$ be the algebra of
observables in a spacetime region $\U$, and let $\A_\U'$ be its commutant.  (If Haag duality holds, then $\A'_\U$ coincides
with $\A_{\U'}$; but we do not need to assume this.)  If the context is clear, we sometimes write
just $\A$ and $\A'$ for $\A_\U$ and $\A'_\U$.
   Let $\Psi$ be a vector -- such as the vacuum vector -- that is cyclic and separating for both regions.  
   
The Tomita operator for the state $\Psi$ is an antilinear operator $S_\Psi$ that, roughly speaking,
is defined by
\be\label{moxy} S_\Psi \a|\Psi\ra =\a^\dagger|\Psi\ra,\ee
for all $ \a\in \A_\U$.
To understand this definition, note first of all that because $\Psi$ is a separating vector for $\A_\U$,  the state $\a|\Psi\ra$
is nonzero for all nonzero $\a\in\A_\U$.  Therefore, we avoid the inconsistency that would arise in this definition
if some $\a$ would satisfy
$\a|\Psi\ra=0$, $\a^\dagger|\Psi\ra\not=0$.    Second, because the states $\a|\Psi\ra$, $\a\in\A_\U$ are dense in $\H$, eqn. (\ref{moxy})
does define the action of $S_\Psi$ on a dense subspace of $\H$.

The definition of eqn. (\ref{moxy}) will lead to an unbounded operator $S_\Psi$ for the following reason.   In the region $\U$,
given that it is small enough that its causal complement contains another open set $\U'$, it is not possible to make a mode
of definite positive or negative frequency.  But by using modes of very short wavelength, we can construct an operator $\a$
in region $\U$ that is arbitrarily close to being an annihilation operator (one that lowers the energy) while $\a^\dagger$ is equally
 close
to being a creation operator.  So $\a|\Omega\rangle$ can be arbitrarily small while $\a^\dagger|\Omega\ra$ is not small.
Thus $S_\Psi$ is unbounded.

An unbounded operator cannot be defined on all states in Hilbert space (recall footnote \ref{remember}).  But it is important to slightly extend the definition of $S_\Omega$ as follows.  If $\a_n$, $n=1,2,3,\cdots$ is a sequence of elements of $\A_\U$ such
that both limits
\be\label{zonto}x=\lim_{n\to\infty}\a_n|\Psi\ra,~~~~y=\lim_{n\to\infty}\a_n^\dagger|\Psi\ra\ee
exist, then we define\footnote{\label{closeable} For this definition to make sense, it must be that if $\lim_{n\to\infty}\a_n\Psi=0$ then
also $\lim_{n\to\infty}\a_n^\dagger\Psi=0$.   
Suppose that $y=\lim_{n\to\infty}\a_n^\dagger|\Psi\ra$ exists and 
is nonzero.   As it is separating for $\A_\U$, the state $\Psi$ is cyclic for $\A'_\U$.  
So there is $\a'\in\A'_\U$ with nonzero $C=\la \a'\Psi|y\ra=\lim_{n\to\infty}\la \a'\Psi|\a_n^\dagger\Psi\ra$.  Then $\bar C=\lim_{n\to\infty}
\la \a_n^\dagger\Psi|\a'\Psi\ra =\lim_{n\to\infty}\la\a'^\dagger\Psi |\a_n\Psi\ra$ is also nonzero.  This implies that $x=\lim_{n\to\infty}\a_n|\Psi\ra$
is nonzero.
 Mathematically, we have proved that the operator $S_\Psi$ is ``closeable.''  The importance  will become clear in section \ref{theproof}.}
\be\label{ronto}S_\Psi x = y. \ee
Extending the definition of $S_\Psi$ in this way gives what technically is known as a ``closed'' operator, meaning that its
graph is closed; see section \ref{theproof}.

The definition (\ref{moxy}) makes it clear that 
\be\label{plonto}S_\Psi^2=1,\ee
so in particular $S_\Psi$ is invertible.  Another obvious fact is that
\be\label{lonto}S_\Psi|\Psi\ra=|\Psi\ra. \ee

We could of course similarly define the modular operator $S'_\Psi$ for the commuting algebra $\A'_{\U}$.  In fact, these
operators are hermitian adjoints:
\be\label{wonto} S'_\Psi=S_\Psi^\dagger.\ee
The definition of the adjoint of an antilinear operator $W$ is that for any states $\Lambda, \chi$, 
\be\label{donto} \la \Lambda| W\chi\ra = \overline {\la W^\dagger \Lambda|\chi\ra}=\la\chi|W^\dagger\Lambda\ra.\ee
A special case of this which we will use shortly is that if $W$ is antiunitary, meaning that it is antilinear and satisfies
$W^\dagger W=W W^\dagger=1$, then
\be\label{ponto}\la W\Lambda| W\chi\ra=\overline{\la\Lambda|\chi\ra}=\la\chi|\Lambda\ra.\ee

To show that $S'_\Psi=S_\Psi^\dagger$, we have to show that for all states $\Lambda,\chi$, we have
$\la S'_\Psi \Lambda|\chi\ra = \la S_\Psi \chi|\Lambda\ra$.  It is enough to check this for a dense set of states,
so we can take $\chi=\a\Psi$, $\Lambda=\a'\Psi$, with $\a\in\A_\U$, $\a'\in\A_\U'$.  Using the definitions of $S_\Psi$ and $S'_\Psi$ and
of a hermitian adjoint 
and the fact that $\a$ and $\a'$ commute, we get
\begin{align}\label{zano} \la S'_\Psi \a'\Psi|\a\Psi\ra =&\la \a'{}^\dagger\Psi|\a\Psi\ra=\la \Psi|\a'\ \a \Psi\ra =\la\Psi|\a\a'\Psi\ra
 = \la \a^\dagger\Psi|\a'\Psi\ra =\la S_\psi \a\Psi|\a'\Psi\ra \end{align}
 as desired.\footnote{This argument really only shows that $S^\dagger_\Psi$ is an extension of $S'_\Psi$ (meaning that the two
 operators act in the same way on any vector on which $S'_\Psi$ is defined). For the proof that it is not a proper extension (meaning that $S^\dagger_\Psi$ cannot be defined, consistent with $\la S^\dagger_\Psi\chi|\Lambda\ra=\la S_\Psi \Lambda|\chi\ra$,
 on any vector on which $S'_\Psi$ is not defined), see for example Theorem 13.1.3  in \cite{Jones}.}
 
Since it is invertible, $S_\Psi$ has a {\it unique} polar decomposition
\be\label{unipo}S_\Psi = J_\Psi \Delta_\Psi^{1/2}, \ee
where $J_\Psi$ is antiunitary and $\Delta_\Psi^{1/2}$ is hermitian and positive-definite. 
This implies that
\be\label{warplo} \Delta_\Psi = S_\Psi^\dagger S_\Psi. \ee
$\Delta_\Psi$ and $J_\Psi$ are called the modular operator and the modular conjugation.
Since $S_\Psi \Psi=S_\Psi^\dagger\Psi=\Psi$, we can deduce the important result
\be\label{zarplo}\Delta_\Psi|\Psi\ra =|\Psi\ra.\ee
From eqn. (\ref{zarplo}), it follows that for any function $f$,
\be\label{arol}f(\Delta_\Psi)|\Psi\ra =f(1)|\Psi\ra. \ee

In addition, because $S_\Psi^2=1$, we have $J_\Psi \Delta_\Psi^{1/2}J_\Psi \Delta_\Psi^{1/2}=1$
or 
\be\label{rplo} J_\Psi \Delta_\Psi^{1/2}J_\Psi=\Delta_\Psi^{-1/2}. \ee
  Hence 
\be\label{narplo} J_\Psi^2 (J_\Psi^{-1}\Delta_\Psi^{1/2}J_\Psi)   =\Delta_\Psi^{-1/2}=1\cdot \Delta_\Psi^{-1/2}. \ee
Since $J_\Psi^{-1}\Delta_\Psi^{1/2}J_\Psi$ is positive, this
 gives two different polar decompositions of the operator $\Delta_\Psi^{-1/2}$.  By the uniquess of the
polar decomposition, we must have
\be\label{arplo} J_\Psi^2=1 .\ee
Therefore
\be\label{barplor} S'_\Psi = S_\Psi^\dagger = \Delta_\Psi^{1/2}J_\Psi =J_\Psi\Delta_\Psi^{-1/2}. \ee
Comparing this to the polar decomposition $S'_\Psi=J'_\Psi \Delta'_\Psi{}^{1/2}$, we find
\be\label{tarplor} J'_\Psi = J_\Psi,~~~ \Delta'_\Psi = \Delta_\Psi^{-1}. \ee

Finally, because $J_\Psi \Delta_\Psi J_\Psi =\Delta_\Psi^{-1}$, we have $J_\Psi f(\Delta_\Psi) J_\Psi = \bar f(\Delta_\Psi^{-1})$
for any function $f$.  In particular, taking $f(x)=x^{\i s}$ for real $s$, we get
\be\label{arplor} J_\Psi \Delta^{\i s} J_\Psi = \Delta^{\i s},~~~~s\in\R. \ee

The operators that we have introduced have a number of other important properties, which we will explain in section \ref{fte}
after exploring these definitions for finite-dimensional quantum systems.

\subsection{The Relative Modular Operator}\label{relmod}

Now let $\Phi$ be a second state.   The relative Tomita operator\footnote{We should warn the reader that what we call $S_{\Psi|\Phi}$ is often denoted $S_{\Phi|\Psi}$ (or $S_{\Phi/\Psi}$, $S_{\Phi,\Psi}$,
etc.).   The purpose of our convention is to agree with quantum information theory, where it has become standard to define the relative entropy between density matrices $\rho,\sigma$
as $\S(\rho||\sigma)=\Tr\,\rho (\log \rho-\log\sigma)$.    In the relation to information theory, $\Psi$ and $\Phi$ correspond respectively to $\rho$ and $\sigma$, as we will learn in section
\ref{fincase}, so we put $\Psi$ before $\Phi$ just as $\rho$ is conventionally put before $\sigma$ in  $\S(\rho||\sigma)$.  We should note that some of the classic papers used the opposite
ordering for both $S_{\Psi|\Phi}$ and $\S(\rho||\sigma)$. } $S_{\Psi|\Phi}$ for the algebra $\A_\U$ is defined by
\cite{Araki2}
\be\label{zorf} S_{\Psi|\Phi} \a|\Psi\rangle =\a^\dagger|\Phi\ra. \ee
In this definition, we usually assume that
\be\label{orf}\la\Psi|\Psi\ra =\la\Phi|\Phi\ra=1.\ee
The definition of $S_{\Psi|\Phi}$  is completed by taking limits as in eqn. (\ref{zonto}). 

As before, for  $S_{\Psi|\Phi}$ to make sense as a densely defined operator, the state $\Psi$ must be cyclic and separating
for the algebra $\A_\U$.  But $\Phi$ can be any state at all.  
If $\Phi$ is cyclic separating, then we can define
\be\label{oorf}S_{\Phi|\Psi}\a|\Phi\ra=\a^\dagger |\Psi\ra.\ee
In this case $S_{\Phi|\Psi}S_{\Psi|\Phi}=1$ and in particular $S_{\Psi|\Phi}$ is invertible.    A calculation similar to that of eqn. 
(\ref{zano})  shows  that $S_{\Psi|\Phi}$  for one algebra $\A_\U$ is the adjoint of $S_{\Psi|\Phi}$ for the commutant $\A_\U'$.

The relative modular operator is defined by
\be\label{omorf}\Delta_{\Psi|\Phi} =S^\dagger_{\Psi|\Phi}S_{\Psi|\Phi}. \ee
It is positive semidefinite, and is positive definite  if and only if $S_{\Psi|\Phi}$ is invertible.   If $\Phi=\Psi$, $S_{\Psi|\Phi}$
reduces to $S_\Psi$ and $\Delta_{\Psi|\Phi}$ reduces to the usual modular operator:
\be\label{limorf} \Delta_{\Psi|\Psi}=\Delta_\Psi.\ee

The polar decomposition of the relative modular operator is
\be\label{nomorof} S_{\Psi|\Phi}=J_{\Psi|\Phi}\Delta_{\Psi|\Phi}^{1/2}, \ee
where $J_{\Psi|\Phi}$ is the relative modular conjugation.  Here we have to be careful.  If $\Phi$ is not separating, then $S_{\Psi|\Phi}$
has a kernel, which is also a kernel  of $\Delta_{\Psi|\Phi}$ and $\Delta_{\Psi|\Phi}^{1/2}$.  
In such a situation, to make the polar decomposition unique, $J_{\Psi|\Phi}$ is defined to annihilate this kernel.  Also, if $\Phi$ is
not cyclic, then the image of $S_{\Psi|\Phi}$ is not a dense subspace of $\H$.  In general, $J_{\Psi|\Phi}$ is an antiunitary map
from the orthocomplement of the kernel of $S_{\Psi|\Phi}$ to its image.  However, if $\Phi$ is cyclic separating, then $J_{\Psi|\Phi}$
is antiunitary.

Now let us discuss what happens if $\Phi$ is replaced by $\a'\Phi$, where $\a'$ is a {\it unitary} element  of the commuting
algebra $\A_\U'$.   For $\a\in\A_\U$, we  get $S_{\Psi|\a'\Phi}\a\Psi =\a^\dagger \a'\Phi=\a'\a^\dagger\Phi$, since $\a^\dagger$ and $\a'$ commute.  So
$S_{\Psi|\a'\Phi}=\a'S_{\Psi|\Phi}$.  With $\a'$ unitary, this implies
\be\label{nomorf}\Delta_{\Psi|\a'\Phi}=\Delta_{\Psi|\Phi}.\ee

If it is important to specify the region $\U$, we write  $\Delta_{\Psi|\Phi:\U}$ for the relative modular 
operator for the algebra $\A_\U$ and the states $\Psi,\Phi$, and similarly for $S_{\Psi|\Phi;\U}$.

The following gives a useful characterization of the relative modular operator:
\be\label{usefulone} \la\a^\dagger\Psi|\Delta_{\Psi|\Phi}|\b\Psi\ra=\la \a^\dagger\Psi|S^\dagger_{\Psi|\Phi}S_{\Psi|\Phi}|\b\Psi\ra
=\la S_{\Psi|\Phi}\b\Psi|S_{\Psi|\Phi}\a^\dagger\Psi\ra=\la\b^\dagger \Phi|\a\Phi\ra.\ee

\noindent{\bf{Remark}} For future reference, observe that the definition of $S_{\Psi|\Phi}$ and $\Delta_{\Psi|\Phi}$
does not require that $\Psi$ and $\Phi$ are vectors in the same Hilbert space.  Let $\H$ and $\H'$ be two different
Hilbert spaces with an action of the same algebra $\A_\U$.  For example, $\H$ and $\H'$ might be different superselection
sectors in the same quantum field theory.  If $\Psi$ is a cyclic separating vector in $\H$ and $\Phi$ is any vector in $\H'$
then eqn. (\ref{zorf})  makes sense and defines an antilinear operator $S_{\Psi|\Phi}:\H\to \H'$.   Its adjoint is an antilinear
operator $S_{\Psi|\Phi}^\dagger:\H'\to \H$.   The product $S^\dagger_{\Psi|\Phi}S_{\Psi|\Phi}$ 
is a nonnegative self-adjoint operator, the modular
operator $\Delta_{\Psi|\Phi}:\H\to \H$.
When not otherwise noted, we usually assume $\H=\H'$.

\subsection{Relative Entropy In Quantum Field Theory}\label{relen}

A primary application of the relative modular operator in these notes will be to study the relative entropy.    
Relative entropy was defined in  classical information theory by Kullback and Leibler \cite{KL} and in nonrelativistic quantum mechanics 
by Umegaki \cite{Umegaki}; a
definition suitable for quantum
field theory was given by Araki \cite{Araki,Araki2}.  The relative entropy $\S_{\Psi|\Phi}(\U)$ between two states $\Psi$ and $\Phi$,
for measurements in the region $\U$, is
\be\label{onorf}\S_{\Psi|\Phi}(\U)= -\la\Psi|\log \Delta_{\Psi|\Phi}|\Psi\ra. \ee 
(In this section, $\U$ is kept fixed and we write $\Delta_{\Psi|\Phi}$ for $\Delta_{\Psi|\Phi;\U}$.)
In general, $\S_{\Psi|\Phi}(\U)$  is a real number or $+\infty$. For example, $\S_{\Psi|\Phi}(\U)$ may be $+\infty$ if $\Delta_{\Psi|\Phi}$
has a zero eigenvalue, which will occur if $\Phi$ is not separating for $\A_\U$.
How this definition is related to what may be more familiar definitions of relative entropy will be explained in section \ref{fte}.
In this section, we simply discuss the properties of the relative entropy.

An important elementary property is that $\S_{\Psi|\Phi}(\U)$ is always non-negative, and vanishes precisely if $\Phi=\a'\Psi$
where $\a'$ is a unitary element of the commuting algebra $\A_\U'$.  This condition implies that $\la\Phi|\a|\Phi\ra=\la\Psi|\a|\Psi\ra$
for all $\a\in\A_\U$, so it 
means that $\Phi$ and $\Psi$ cannot be distinguished by a measurement in region $\U$.
To see the vanishing if $\Phi=\a'\Psi$, with $\a'\in\A_\U'$, note that in this case, according to (\ref{limorf}) and (\ref{nomorf}),
 $\Delta_{\Psi|\Phi}$ is the ordinary modular operator $\Delta_\Psi$.  So using eqn. (\ref{arol}) with $f(x)=\log x$,
we get $\log \Delta_{\Psi|\Phi}|\Psi\ra=0$ for $\Phi=\a'\Psi$, whence $\S_{\Psi|\Psi}(\U)=0$.

To show that $\S_{\Psi|\Phi}(\U)>0$ if $\Phi$ is not of the form $\a'\Psi$, one uses \cite{Araki} the inequality
for a non-negative real number $\log \lambda \leq \lambda -1$.    This inequality for numbers implies the operator inequality
$\log \Delta_{\Psi|\Phi} \leq \Delta_{\Psi|\Phi} -1$, or $-\log \Delta_{\Psi|\Phi}\geq 1-\Delta_{\Psi|\Phi}$.   So
\be\label{zonorf}\S_{\Psi|\Phi}(\U)\geq  \la\Psi|(1-\Delta_{\Psi|\Phi})|\Psi\ra=\la\Psi|\Psi\ra -\la\Psi|S^\dagger_{\Psi|\Phi}S_{\Psi|\Phi}
|\Psi\ra =\la\Psi|\Psi\ra - \la\Phi|\Phi\ra=0, \ee
since we assume $\la \Psi|\Psi\ra=\la \Phi|\Phi\ra=1$. 

Because the inequality $\log \lambda\leq \lambda -1$ is only saturated at $\lambda=1$, to saturate the inequality (\ref{zonorf})
we need $\Delta_{\Psi|\Phi}$ to equal 1 in acting on $\Psi$, that is we need $\Delta_{\Psi|\Phi}\Psi=\Psi$.  But as we will
show, this implies
 that $\Phi=\a'\Psi$ for some unitary $\a'\in \A_\U'$.   The statement that $\Delta_{\Psi|\Phi}\Psi=\Psi$ implies that for
any state $\chi$, 
\be\label{gocon}\la\chi|\Delta_{\Psi|\Phi}\Psi\ra=\la\chi|\Psi\ra.\ee    In particular, this must be so if $\chi=\a\Psi$
for $\a\in \A_\U$.  We calculate
\be\label{pikl} \la \a\Psi|\Delta_{\Psi|\Phi}\Psi\ra =\la \a\Psi|S^\dagger_{\Psi|\Phi}S_{\Psi|\Phi}\Psi\ra
 = \la \a\Psi|S_{\Psi|\Phi}^\dagger\Phi\ra=\la\Phi|S_{\Psi|\Phi}\a\Psi\ra=\la\Phi|\a^\dagger\Phi\ra =\la \a\Phi|\Phi\ra.\ee
 We used $S_{\Psi|\Phi}\Psi=\Phi$ and the definition of the adjoint of an antilinear operator.
 The condition (\ref{gocon}) then is that  $\la \a\Phi|\Phi\ra =\la \a\Psi|\Psi\ra$ for all
 $\a\in \A_\U$.  Accordingly, for $\a,\b\in\A_\U$,
 \be\label{tongo}\la \a\Phi|\b\Phi\ra=\la \b^\dagger\a\Phi|\Phi\ra =\la \b^\dagger \a\Psi|\Psi\ra =\la \a\Psi|\b\Psi\ra. \ee
 Since states of the form $\a\Psi$ or $\b\Psi$ are dense in $\H$, we can define a densely defined linear operator that takes
 $\a\Psi$ to $\a\Phi$.  Eqn. (\ref{tongo}) says that this operator is unitary (and so, being bounded, it can be naturally extended to
 all of $\H$), and as it clearly commutes with $\A_\U$,
 it is given by multiplication by a unitary element $\a'\in \A_\U'$.   Thus $\a\Phi=\a'\a\Psi$ for all $\a$, and in particular $\Phi=\a'\Psi$,
 as claimed.

Positivity of relative entropy has various applications in quantum field theory, for instance in the interpretation and 
proof \cite{Casini,LongoXu} of the Bekenstein bound on the energy, entropy, and size of a quantum system.  The more subtle property of monotonicity of relative entropy, to which we come next, also has various applications,
for instance in the proof of a semiclassical generalized second law of thermodynamics that includes black hole entropy \cite{Wall}.

\subsection{Monotonicity of Relative Entropy}\label{monrel}

In quantum field theory,  in the definition of the algebra of observables and the associated
modular operators, we can replace the open set $\U$ by a smaller open set $\t \U\subset \U$.     Thus, for given $\Psi$ and $\Phi$, we can define Tomita operators $S_{\Psi|\Phi;\,\U}$ and $S_{\Psi|\Phi;\,\t U}$ and associated
modular operators $\Delta_{\Psi|\Phi;\,\U}$ and $\Delta_{\Psi|\Phi;\,\t\U}$.   Then we have the relative entropy $\S_{\Psi|\Phi}(\U)$ for measurements in $\U$,
\be\label{zoonorf}\S_{\Psi|\Phi}(\U)= -\la\Psi|\log \Delta_{\Psi|\Phi;\,\U}|\Psi\ra \ee 
and the corresponding relative entropy for measurements in $\t\U$,
\be\label{xonorf}\S_{\Psi|\Phi}(\t\U)= -\la\Psi|\log \Delta_{\Psi|\Phi;\,\t\U}|\Psi\ra. \ee 
Monotonicity of relative entropy says that if $\t\U$ is contained in $\U$, then 
\be\label{wonorf}\S_{\Psi|\Phi}(\U) \geq \S_{\Psi|\Phi}(\t\U). \ee

In nonrelativistic quantum mechanics, a version of monotonicity of relative entropy was proved by Lieb and Ruskai \cite{LiebRuskai}, along
with strong subadditivity of quantum entropy, to which it is closely related.  The proof used a lemma of Lieb \cite{Lieb}.  A
 more general form of monotonicity of relative entropy was proved by Uhlmann \cite{Uhlmann}.  In a form that encompasses the statement (\ref{wonorf}) in quantum field
theory, monotonicity of relative entropy was proved by Araki \cite{Araki,Araki2}.
  Petz \cite{Petz}, with later elaboration by Petz and Nielsen \cite{PetzNielsen},
formulated a proof for nonrelativistic quantum mechanics that drew partly on Araki's framework.   Some of these matters will be explained in section \ref{fte},
but for now we just concentrate on understanding eqn. (\ref{wonorf}).

The states $\Psi$ and $\Phi$ will be kept fixed in the rest of this section, so to lighten the notation we usually just write $S_\U$ for $S_{\Psi|\Phi;\,\U}$ and $\Delta_\U$
for $\Delta_{\Psi|\Phi;\,\U}$, and similarly for $\t \U$.  The inequality (\ref{wonorf}) is a direct consequence of an operator inequality
\be\label{opin} \Delta_{\t \U}\geq \Delta_\U. \ee
A self-adjoint operator $P$ is called positive if $\la\chi|P|\chi\ra\geq 0$ for all $\chi$; in that case, one writes $P\geq 0$.
 If $P$ and $Q$ are bounded self-adjoint
operators, one says $P\geq Q$ if $P-Q\geq 0$.    (The reason for assuming here that $P$ and $Q$ are bounded is that it ensures
that $\la\chi|P-Q|\chi\ra=\la\chi|P|\chi\ra -\la\chi|Q|\chi\ra$ is defined for all $\chi$; we explain shortly how to interpret
the statement $P\geq Q$ in general.)
  If $P,Q\geq 0$,
an equivalent statement to $P\geq Q$ is
\be\label{zopin}\frac{1}{s+P}\leq \frac{1}{s+Q}, \ee
for all $s>0$.      
 (If $P$ and $Q$ are strictly positive, one can take $s=0$.)  To show this, consider the family of operators $R(t)=tP+(1-t)Q$,
 $t\in\R$.  Writing $\dot R=\d R/\d t$, we  see that $\dot R=P-Q\geq 0$.
We have
\be\label{wopin}\frac{\d}{\d t}\frac{1}{s+R(t)}=-\frac{1}{s+R(t)} \dot R\frac{1}{s+R(t)}.\ee
The right hand side is $\leq 0$ since it is of the form $-ABA$ with $A$ self-adjoint and $B\geq 0$.
Integrating eqn. (\ref{wopin}) in $t$ from $t=0$ to $t=1$, we learn that  $1/(s+R(1))\leq 1/(s+R(0))$, which is (\ref{zopin}).    We describe this result by saying that $1/(s+P)$ is a decreasing function of $P$, or equivalently that $-1/(s+P)$ is an increasing function of $P$.     The opposite inequality that (\ref{zopin}) implies $P\geq Q$ is proved in the same way, writing
$P=1/T-s,$ with $T=1/(s+P)$. 

So far we have assumed that $P$ and $Q$ are bounded.
If $P$ and $Q$ are  densely defined unbounded operators, but non-negative, then it is reasonable to interpret (\ref{zopin}) as the {\it definition} of what we mean by $P\geq Q$.  In general, $P$ and $Q$ are defined on 
different (dense) subspaces, so it can be hard to interpret the statement that $\la\chi |P|\chi\ra\geq\la\chi|Q|\chi\ra$ for all $\chi$. But $1/(s+P)$ and $1/(s+Q)$ are bounded, and so defined for all $\chi$.  The statement (\ref{zopin}) just means that
\be\label{polm}\left\la\chi\left|\frac{1}{s+P}\right|\chi\right\ra\leq\left\la \chi\left|\frac{1}{s+Q}\right|\chi\right\ra, ~~\forall \chi\in\H.\ee
This is a much stronger and more useful statement than just saying that  $\la\chi |P|\chi\ra\geq\la\chi|Q|\chi\ra$ for all $\chi$ on which
both $P$ and $Q$ are defined.

Using
\be\label{lopin}\log R=\int_0^\infty\d s \left(\frac{1}{s+1}-\frac{1}{s+R}\right),\ee
we see that since $1/(s+R)$ is a decreasing function of $R$, $\log R$ is an increasing function of $R$.   
Thus $P\geq Q$ or its equivalent  $1/(s+P)\leq 1/(s+Q)$ implies
\be\label{ropin}\log P\geq \log Q. \ee
So eqn. (\ref{opin}) implies that
\be\label{ztopin} \log \Delta_{\t\U}\geq \log \Delta_\U. \ee 
The monotonicity statement (\ref{wonorf}) is simply the expectation value of this operator inequality in the state $\Psi$.

 The proof of the crucial inequality (\ref{opin}) is rather short and is explained in section \ref{theproof}.  However, we first explain some background and motivation in section \ref{practice}.
The goal of section \ref{practice} is to ensure that the reader will consider the result obvious before actually getting to the proof.

To conclude this section, we  will explain another monotonicity statement that will be useful later, and then, to help the reader appreciate the subtlety of such
statements, we will explain a superficially
similar version that is false.  For $0<\alpha<1$, we have
\be\label{zox}R^\alpha =\frac{\sin \pi\alpha}{\pi}\int_0^\infty \d s\, s^\alpha \left(\frac{1}{s}-\frac{1}{s+R}\right). \ee
If $R$ depends on a parameter $t$, and  $\dot R=\d R/\d t$, we get
\be\label{wox}\frac{\d}{\d t}R^\alpha=\frac{\sin\pi\alpha}{\pi}\int_0^\infty \d s\, s^\alpha \frac{1}{s+R}\dot R\frac{1}{s+R}. \ee
This is nonnegative if $\dot R\geq 0$, so $R^\alpha$ is an increasing function of $R$ in this range of $\alpha$.   If, however, $\alpha>1$, then $R^\alpha$ is
in general not an increasing function of $R$.  For $\alpha>1$, the representation (\ref{zox}) is not valid.  But if $1<\alpha<2$, we can
write $R^\alpha = R \cdot R^\beta$, with $0<\beta<1$, and then use (\ref{zox}) for $R^\beta$.  So in this range of $\alpha$,
\be\label{zoggo}R^\alpha=\frac{\sin\pi(\alpha-1)}{\pi}\int_0^\infty \d s \, s^{\alpha-1}\left(\frac{R}{s}-1+\frac{s}{s+R}\right),\ee
and hence
\be\label{woggo}\frac{\d}{\d t}R^\alpha =\frac{\sin\pi(\alpha-1)}{\pi}\int_0^\infty \d s\, s^{\alpha-1}\left(\frac{\dot R}{s}-s\frac{1}{s+R}\dot R\frac{1}{s+R}\right).\ee
This is not necessarily non-negative for $\dot R\geq 0$, since the last term is negative-definite and can dominate. For an example with $2\times 2$ matrices, set
$R=\begin{pmatrix} 2& 0 \cr 0& 1\end{pmatrix}$, $\dot R=\begin{pmatrix}1&1\cr 1&1\end{pmatrix}$, and $\chi=\begin{pmatrix}1\cr -1\end{pmatrix}$.  Then
\be\label{oggo} \left\la\chi \left|\frac{\d}{\d t}R^\alpha\right|\chi\right\ra<0. \ee

\subsection{Examples}\label{practice}

The relation between $S_{\U}$ and $S_{\t\U}$ is as follows. They are both defined on a dense set of states by the same formula $S\a\Psi=\a^\dagger\Psi$ (together with limiting cases as described in eqn. (\ref{zonto})).
The only difference is that  the dense subspace on which $S_\U$ is defined is larger than the dense subspace on which $S_{\t\U}$ is defined.  In the case of 
$S_{\t\U}$, $\a$ is an element of the algebra $\A_{\t\U}$, while in the case of $S_\U$, $\a$ is an element of the larger algebra $\A_\U$.

Let $X$ and $Y$ be unbounded operators\footnote{A much more systematic explanation of the 
requisite facts
can be found in \cite{ReedSimon}, chapter VIII, and \cite{Simon}, chapter VII.5. The example with the Dirichlet and Neumann Laplacians
is analyzed in the latter reference.}  on a Hilbert space $\H$ (either both linear or both antilinear).   If $X$ is defined whenever $Y$ is defined and they act in the same way on any vector on which they are both
defined, then $X$ is called an extension of $Y$.  In this situation, as we will see, it is always true that $X^\dagger X\leq Y^\dagger Y$, and therefore that $\log X^\dagger X \leq \log Y^\dagger Y. $  Applied to the case $X=S_\U$,
$Y=S_{\t\U}$, this is the inequality we want.

The following remarks apply for either $\U$ or $\t\U$, so we drop the subscripts from $S$ and $\Delta$.  The operator $\Delta=S^\dagger S$ is associated to the hermitian form $F(\chi,\eta)=\la S\chi|S\eta\ra$, which
is defined on the dense set of vectors $\chi,\eta\in\H$ in the domain of $S$.   This hermitian form is positive-definite in the sense that $F(\chi,\chi)\geq 0$ with equality only if $\chi=0$.  Formally
\be\label{tofo} \la S^\dagger S\eta|\chi\ra=\la S\chi|S\eta\ra.\ee
The way we interpret  this statement is that if, for some $\eta$ in the domain of $S$, the relation
$\la\zeta|\chi\ra=\la S\chi|S\eta\ra$ holds for all $\chi$ on which $S$ is defined, then we define
\be\label{fogo} S^\dagger S\eta=\zeta. \ee
In other words, we define $S^\dagger S$ on every vector on which it can be defined so as to make (\ref{tofo}) true.

If $F$ and $G$ are two hermitian forms on $\H$, we say that $F$ is an extension of $G$ if it is defined whenever $G$ is defined and they agree where they are both defined.  In our problem, we have two
hermitian forms  $W_\U(\chi,\eta)=\la S_\U\chi|\S_\U\eta\ra$ and $W_{\t\U} (\chi,\eta)=\la S_{\t\U}\chi|S_{\t\U}\eta\ra$.  $W_\U$ is an extension of $W_{\t\U}$ because $S_\U$ is an extension of $S_{\t\U}$.
The claim that we will motivate here and prove in section \ref{theproof} is that in this situation, the operators $\Delta_\U=S^\dagger_\U S_\U$ and $\Delta_{\t\U}=S^\dagger_{\t\U}S_{\t\U}$ associated to the two
hermitian forms satisfy $\Delta_{\t\U}\geq \Delta_\U$.  In these statements, it does not matter if $S$ is linear or antilinear or if $S$ maps a Hilbert space $\H$ to itself or to some other Hilbert space $\H'$.

To motivate the claim, we will consider a more familiar example.  Let $M$ be a compact region in $\R^n$ with boundary $N$.  Let $\H$ be the Hilbert space of square-integrable functions on $M$, and
$\H'$ the Hilbert space of square-integrable 1-forms on $M$.   Roughly speaking, we want to consider the exterior derivative $\d$ acting from functions to 1-forms.  But we will consider two different versions of
this operator.  We let $T_0$ be the exterior derivative acting on continuous functions $\phi$ on $M$ such that $\d\phi$ is square-integrable 
and $\phi$ vanishes along the boundary of $M$.  Such functions are dense in $\H$,
so $T_0$ is a densely-defined unbounded operator.    We let $T_1$ be the exterior derivative acting on continuous
functions $\phi$ on $M$ such that $\d\phi$ is square-integrable but with no restriction on $\phi$ along the boundary of $M$.
Clearly $T_1$ is an extension of $T_0$.  The corresponding hermitian form $F_1$ is likewise an extension of the hermitian form $F_0$:
\begin{align} \label{guyter} F_0(\phi,\rho) & = \la T_0\phi|T_0\rho\ra= \int_M\d^n x \,\sum_i \frac{\partial\bar \phi}{\partial x_i} \frac{\partial \rho}{\partial x_i} \\
                                            F_1(\phi,\rho) & = \la T_1\phi|T_1\rho\ra= \int_M\d^n x \,\sum_i \frac{\partial\bar \phi}{\partial x_i} \frac{\partial \rho}{\partial x_i} \end{align}
                         The only difference between $F_0$ and $F_1$ is that in the definition of $F_0$, $\phi$ and $\rho$ are required to vanish along $N=\partial M$, while $F_1$ is defined without this condition.    
                         
Now let us compute the operators $T_0^\dagger T_0$ and $T_1^\dagger T_1$ associated to the quadratic forms $F_0$ and $F_1$.  Since $T_0$ and $T_1$ are both defined by the exterior derivative on some class of functions,
it is natural to expect that $T_0^\dagger T_0$ and $T_1^\dagger T_1$ will both equal, in some sense, the Laplacian
\be\label{monz}\Delta=\d^\dagger\d =-\sum_{i=1}^n\frac{\partial^2}{\partial x_i^2}. \ee
The identity that we need in order to show that $T^\dagger T\phi = \Delta\phi$ for some function $\phi$ (where $T$ may be $T_0$ or $T_1$) is that 
\be\label{onz}\int_M \d^Dx \left(  -\sum_{i=1}^n\frac{\partial^2\bar\phi}{\partial x_i^2}\right)\rho \overset{?}{=} \int_M\d^n x \sum_{i=1}^n\frac{\partial\bar\phi}{\partial x_i}\frac{\partial \rho}{\partial x_i} \ee
for all $\rho$ in the appropriate domain.
When we try to prove this identity by integration by parts, we run into a surface term
\be\label{worz}\int_N\d\mu\,\, \bigl(-\partial_\perp \bar\phi\bigr) \,\,\rho, \ee
where $\d\mu$ is the Riemann measure of $N$ and $\partial_\perp$ is the inward normal derivative along $N$.

If we are trying to define $T^\dagger_0T_0$, then $\rho$ and $\phi$ are constrained to vanish along $N$.   Therefore, the surface term (\ref{worz}) vanishes.  Accordingly, the identity (\ref{onz}) is satisfied for any functions
$\phi$, $\rho$ in the domain of $T_0$, that is, any functions (continuous and with square-integrable exterior derivative) that vanish along $N=\partial M$.  Thus $T^\dagger_0 T_0$ is the Laplacian $\Delta$
acting on functions that are constrained to vanish on the boundary.  This is usually called the Dirichlet Laplacian, and we denote it as 
$\Delta_D$.

If we are trying to define $T^\dagger_1 T_1$, then there is no constraint on $\rho$ along the boundary, and hence to  make the surface term vanish we have to require $\partial_\perp \phi=0$ along $N$.   The Laplacian
acting on such functions is usually called the Neumann Laplacian, and we will denote it as $\Delta_N$.   

Thus the inequality $T_0^\dagger T_0\geq T_1^\dagger T_1$ corresponds in this case to $\Delta_D\geq \Delta_N$.   To make it obvious that one should expect such an inequality, we can interpolate between $F_0$ and $F_1$
in the following way.  For $\lambda\geq 0$, we define the hermitian form
\be\label{ronz} G_\lambda(\phi,\rho) =\int_M \d^n x \,\sum_i \frac{\partial\bar \phi}{\partial x_i} \frac{\partial \rho}{\partial x_i} +\lambda\int_N\d\mu\, \bar\phi \rho, \ee
which is defined for continuous functions $\phi,\rho$, with square-integrable first derivative, and also square-integrable restriction to $N$.
The associated quadratic form  $G_\lambda(\phi,\phi)$ is increasing with $\lambda$ for generic $\phi$ and nondecreasing for all $\phi$.
We therefore expect that the operator  associated with this quadratic form, which we will call $X_\lambda$,
 will be increasing with $\lambda$.  $X_\lambda$ will again be the Laplacian, with some boundary condition, since $G_\lambda$ coincides
with the hermitian forms considered earlier except for a boundary term.    

To identify the boundary condition in $X_\lambda$, we observe that in order to have $X_\lambda\phi=\Delta\phi$ for some function $\phi$, the identity we need is
\be\label{thid} \la \Delta\phi|\rho\ra =G_\lambda(\phi,\rho)=\int_M \d^n x \,\sum_i \frac{\partial\bar \phi}{\partial x_i} \frac{\partial \rho}{\partial x_i} +\lambda\int_N \bar\phi \rho, \ee
for all $\rho$ in the domain of $G_\lambda$.  In trying to prove this identity, we run into a surface term, which now is
\be\label{nid}\int_N\d\mu(-\partial_\perp \bar\phi +\lambda\bar\phi)\rho. \ee
The boundary condition that we need is therefore $-\partial_\perp \phi+\lambda\phi=0$.    The operator $X_\lambda$ is the Laplacian with this boundary condition.  

$X_\lambda$ coincides with the Neumann Laplacian $\Delta_N$ at $\lambda=0$, and with the Dirichlet Laplacian $\Delta_D$ in the limit $\lambda\to\infty$.  Since $X_\lambda$ is increasing with $\lambda$,
this accounts for the inequality $\Delta_D\geq \Delta_N$.  

A more brief way to say some of this is that to go from the Neumann quadratic form to the Dirichlet quadratic form, we impose a constraint on the wavefunction: it should vanish on the boundary.  This naturally increases
the energy, so it leads to our inequality.  

It is useful -- especially with a view to  section \ref{fte} -- to consider a somewhat similar situation in finite dimensions.   Let $X$ be a positive
hermitian matrix acting on $\C^{n+m}=\C^n\times \C^m$. We write
\be\label{blocks}X=\begin{pmatrix} A & B \cr B^\dagger & C\end{pmatrix},\ee
where $A$ and $C$ are blocks of size $n\times n$ and $m\times m$, acting on a column vector  $\Psi=\begin{pmatrix}\psi\cr \chi
\end{pmatrix},$ with $\psi\in \C^n$, $\chi\in \C^m$.  For real $\lambda>0$, let 
\be\label{nocks}X_\lambda =  \begin{pmatrix} A & B \cr B^\dagger & C+\lambda\end{pmatrix}.\ee
Clearly $X_\lambda$ is  increasing with $\lambda$, and in particular,  for $s\geq 0$,
\be\label{pnocks} \frac{1}{s+X}\geq \frac{1}{s+X_\lambda}. \ee
On the other hand, for very large $\lambda$, $1/(s+X_\lambda)$ simplifies, because the upper and lower components decouple:
\be\label{otocks} \frac{1}{s+X_\lambda} \sim \begin{pmatrix}{1}/{(s+A)} & \O(1/\lambda)\cr\O(1/\lambda)
     & 1/\lambda\end{pmatrix},~~~~\lambda>>0.\ee
The inequality (\ref{pnocks}) means that for any $\Psi\in\C^{n+m}$,
\be\label{lnocks} \left\la\Psi\left|\frac{1}{s+X}\right|\Psi\right\ra \geq \left\la\Psi\left|\frac{1}{s+X_\lambda}\right|\Psi\right\ra. \ee
Let us evaluate this for $\Psi=\begin{pmatrix}\psi\cr 0\end{pmatrix}$.  The right hand side, for $\lambda\to\infty$, reduces to
$\la\psi|(s+A)^{-1}|\psi\ra$.  If we define an isometric embedding $U:\C^n\to \C^{n+m}$ by $U(\psi)=\begin{pmatrix}\psi \cr 0\end{pmatrix}$,
then the left hand side is $\la\psi|U^\dagger (s+X)^{-1}U|\psi\ra$.  So for $\psi\in\C^n$,
\be\label{tnocks} \left \la \psi\left|U^\dagger\frac{1}{s+X}U\right|\psi\right\ra \geq\left\la\psi\left|\frac{1}{s+A} \right|\psi\right\ra.\ee
Integrating over $s$ and using (\ref{lopin}), we get
\be\label{znocks}\left \la \psi\left|U^\dagger(\log X) U\right|\psi\right\ra\leq \left\la\psi\left|\log A \right|\psi\right\ra .\ee
Since $A=U^\dagger X U$, this is equivalent to
\be\label{linocks} \left \la \psi\left|U^\dagger(\log X) U\right|\psi\right\ra\leq \left\la\psi\left|\log (U^\dagger X U) \right|\psi\right\ra .\ee

\subsection{The Proof}\label{theproof}

Now we will complete the proof of monotonicity of relative entropy under reducing the size of a region.

Suppose that $T$ is an unbounded, densely defined operator from one Hilbert space $\H$ to a possibly different Hilbert space $\H'$.
It is convenient to set $\h\H=\H\oplus \H'$ and to consider the {\it graph}  $\Gamma$ of $T$, which is the set of all vectors $(x,Tx)\in \h \H$.  
$\Gamma$ is obviously a linear subspace of $\h\H$.  The operator $T$ is said to be {\it closed} if $\Gamma$ is a closed subspace
of $\h\H$, or equivalently if it is a Hilbert subspace.   For $\Gamma$ to be closed means that if a sequence $(x_n,Tx_n)\in\Gamma$
has a limit $(x,y)\in\h\H$, then this limit is actually in $\Gamma$.  In more detail, this amounts to saying that
 if $(x_n,Tx_n)$ is a sequence of elements
of $\Gamma$ such that both limits
\be\label{formo} x=\lim_{n\to\infty}x_n,~~y=\lim_{n\to\infty}Tx_n\ee
exist,
 then $T$ is defined on $x$ and $Tx=y$.   The reason that in defining the Tomita operator $S_\Psi$
 and its relative cousin $S_{\Psi|\Phi}$, we included limit points (\ref{zonto}) was to ensure that these are closed operators.
 
 If $\Gamma$ is a closed subspace of a Hilbert space $\h\H$, then one can define an orthogonal projection $\Pi:\h\H\to \Gamma$.   $\Pi$ is bounded
 (with eigenvalues 0,1) and so is defined on all states.
 Such an orthogonal projection does not exist if $\Gamma$ is a linear subspace of $\h\H$ that is not closed.  
 
 If $\Gamma$ is the graph of $T$, then the orthogonal projector 
 $\Pi$ onto its graph can be written
 explicitly as a $2\times 2$ matrix\footnote{Since $\Pi$ is bounded,  also the operators  $(1+T^\dagger T)^{-1}$, $(1+T^\dagger T)^{-1}T^\dagger$, etc.,
 appearing as matrix elements of the following matrix are  bounded.  In  particular these operators
 are defined on all states.   That is actually part of why introducing $\Pi$ is
 useful in making a rigorous proof.   For example, momentarily when we write  $\eta=(1+T^\dagger T)^{-1}(\psi+T^\dagger\chi)$, this formula makes sense
 because, although $\chi$ may not be in the domain of $T^\dagger$, it is in the domain of $(1+T^\dagger T)^{-1}T^\dagger$.}
  of operators acting on a column vector $\begin{pmatrix}\psi\cr\chi\end{pmatrix}$ with $\psi\in \H$,
 $\chi\in\H'$:
 \be\label{golfo}\Pi=\begin{pmatrix}({1+T^\dagger T})^{-1}&   ({1+T^\dagger T})^{-1} T^\dagger \cr   T ({1+T^\dagger T} )^{-1} &T   ({1+T^\dagger T})^{-1}T^\dagger   \end{pmatrix}.\ee
It is straightforward to verify that $\Pi$ is hermitian and $\Pi^2=\Pi$, so $\Pi$ is an orthogonal projection operator.  It projects onto the
graph of $T$,  since $\Pi\begin{pmatrix}\psi\cr\chi
\end{pmatrix}=\begin{pmatrix}\eta\cr T\eta\end{pmatrix}$ with $\eta=(1+T^\dagger T)^{-1}(\psi+T^\dagger\chi)$.  Clearly, $\begin{pmatrix}\eta\cr T\eta\end{pmatrix}$
is in the graph of $T$, and  every vector in the graph of $T$ is of this form.

We are finally ready for the proof.   Suppose that $T_0,T_1$ are densely defined operators from $\H$ to $\H'$, with graphs $\Gamma_0$ and $\Gamma_1$.  Let
$\Pi_0$ and $\Pi_1$ be the projectors onto the two graphs. 
If $T_1$ is an extension of $T_0$, then $\Gamma_0$ is a subspace of $\Gamma_1$.  This implies that $\Pi_1\geq \Pi_0$,
so $\la\Psi|\Pi_1|\Psi\ra\geq \la\Psi|\Pi_0|\Psi\ra$ for any state $\Psi=\begin{pmatrix}\psi\cr \chi\end{pmatrix}$.   Specializing to the case
$\chi=0$ and using (\ref{golfo}), we get the inequality
\be\label{wolfo} \left\la\psi\left| \frac{1}{1+T_0^\dagger T_0}\right|\psi\right\ra \leq  \left\la\psi\left| \frac{1}{1+T_1^\dagger T_1}\right|\psi\right\ra.\ee
Repeating this analysis with $T_0/\sqrt s$ and $T_1/\sqrt s$ instead of $T_0$ and $T_1$ for some $s>0$, we get
\be\label{wolfon} \left\la\psi\left| \frac{1}{s+T_0^\dagger T_0}\right|\psi\right\ra \leq  \left\la\psi\left| \frac{1}{s+T_1^\dagger T_1}\right|\psi\right\ra.\ee
Thus $T_1^\dagger T_1\leq T_0^\dagger T_0$ and $\log T_1^\dagger T_1\leq \log T_0^\dagger T_0$.  

Taking $S_{\t\U}$ and $S_{\U}$ for $T_0$ and $T_1$, this is what we needed to prove (\ref{opin}) and thus the monotonicity of relative entropy.
There is perhaps just one more detail to clarify.   $S_\U$ and $S_{\t\U}$ are usually defined as antilinear operators from a Hilbert space
$\H$ to itself.  However, an antilinear operator from $\H$ to $\H$ is the same as a linear operator from $\H$ to $\bar \H$, where $\bar \H$
is the complex conjugate\footnote{The complex conjugate $\bar\H$ of a Hilbert space $\H$ is defined as follows.
Vectors in $\bar\H$ are in 1-1 correspondence with vectors in $\H$.  But a complex scalar that acts on $\H$ as multiplication by $\lambda$
acts on $\bar\H$ as multiplication by $\bar\lambda$, and inner products in $\bar \H$ are complex conjugates of those in $\H$.
 $\bar\H$ satisfies all the axioms of a Hilbert space.}   of the Hilbert space $\H$.  So we can regard $S_\U$ and $S_{\t\U}$ as linear
operators $\H\to\H'$, with $\H'=\bar \H$, and then the above analysis applies precisely.

We have followed Borchers \cite{Borchers} in this explanation of why $\Delta_\U$ increases as the region $\U$ is made
smaller.  Borchers uses this inequality not to analyze the relative entropy but for another application. 
The computation involving the projection on the graph is much older \cite{Stone,Halmos}.  

It might be helpful to analyze the graphs $\Gamma_0$ and $\Gamma_1$ in the example considered in section \ref{practice}.   In doing this, for simplicity, we will work in one dimension, so we take $M$ to be the unit interval $[0,1]$ on the $x$-axis.
The operators $T_0$ and $T_1$ reduce to $\d/\d x$, acting on functions that are or are not required to vanish at the endpoints
in the case of $T_0$ or $T_1$, respectively.  The graph $\Gamma_0$ consists of pairs $(f(x), \d f(x)/\d x)$, where $f$ vanishes
at the endpoints, and the graph $\Gamma_1$ consists of pairs $(g(x),\d g(x)/\d x)$ with no such constraint on $g$ at the endpoints.
We claim that $\Gamma_0$ is a proper subspace of $\Gamma_1$.  To show this, we will show that there are pairs
$(g,g')\in\Gamma_1$ that are orthogonal to all $(f,f')\in\Gamma_0$.  The condition of orthogonality is
\be\label{pongo}\int_0^1\d x \,\bar f g +\int_0^1 \d x\, \frac{\d \bar f}{\d x}\frac{\d g}{\d x}=0.\ee
We want to find $g$ such that this is true for all $f$.  The requisite  condition is that 
\be\label{wongo} \left(1-\frac{\d^2}{\d x^2}\right)g=0. \ee
In verifying that (\ref{wongo}) implies (\ref{pongo}) for all $f$, one has to integrate by parts; there is no surface term as $f$ vanishes
at the endpoints.  Eqn. (\ref{wongo}) has a two-dimensional space of solutions $g(x)=Ae^x+Be^{-x}$, so $\Gamma_0$ is of
codimension two in $\Gamma_1$.

Directly explaining the relation between the unbounded operators $T_0$ and $T_1$ is subtle because one has to talk about
two dense but non-closed subspaces of Hilbert space, one of which is larger than the other.
  Passing to the graphs 
brings the essential difference into the open, as it now involves a comparison of the Hilbert spaces $\Gamma_0$ and $\Gamma_1$.

\section{Finite-Dimensional Quantum Systems And Some Lessons}\label{fte} 

In this section, we will explore the modular operators for finite-dimensional quantum systems and draw some lessons.

\subsection{The Modular Operators In The Finite-Dimensional Case}\label{fincase}

In finite dimensions, the interesting case is a tensor product Hilbert space $\H=\H_1\otimes \H_2$ with tensor factors $\H_1$ and $\H_2$.
Such a tensor product describes what is called a bipartite quantum system.
We let $\A$ be the algebra of linear operators acting on $\H_1$ and $\A'$ the algebra of linear operators acting on $\H_2$.  A linear
operator $\a:\H_1\to \H_1$ is taken to act on $\H$ as $\a\otimes 1$, while $\a':\H_2\to\H_2$ similarly acts on $\H$ as
$1\otimes \a'$.  The algebras $\A$ and $\A'$ are each other's commutants, since a linear transformation of $\H$ that commutes
with $\a\otimes 1$ for all $\a$ is of the form $1\otimes \a'$, and vice-versa.    So from section \ref{furtherp}, 
we know that  a vector is cyclic for $\A$ if and only
if it is separating for $\A'$, and vice-versa.    

Any vector $\Psi\in\H$ has an expansion 
\be\label{expvec}\Psi=\sum_{k=1}^n c_k \psi_k\otimes \psi'_k,\ee
where $\psi_k$ are orthogonal unit vectors in $\H_1$
and $\psi'_k$ are orthogonal unit vectors in $\H_2$. Moreover, we can assume the $c_k$ to be all nonzero (or we could omit
some terms from the sum).   We have
\be\label{zombo} (\a\otimes 1)\Psi=\sum_k c_k \a\psi_k\otimes \psi'_k,\ee
so $\a\otimes 1$ annihilates $\Psi$ if and only if $\a$ annihilates all of the $\psi_k$.  If the $\psi_k$ are a complete basis for $\H_1$,
this implies that $\a=0$;  otherwise, there is some nonzero $\a$ that annihilates all of the $\psi_k$.  Thus $\Psi$ is separating for the algebra
$\A$ if and only if the $\psi_k$ are a basis of $\H_1$; likewise it is separating for $\A'$ if and only if the $\psi'_k$ are a basis for $\H_2$.
Since  $\Psi$ is cyclic for one algebra if and only if it is separating for the other, it follows that $\Psi$ is cyclic and separating for $\A$ and for $\A'$ precisely if the $\psi_k$ and the $\psi'_k$ are orthonormal bases for
their respective spaces.  In particular, this is possible precisely if $\H_1$ and $\H_2$ are of equal dimension.  Conversely, if $\H_1$
and $\H_2$ are of the same dimension $n$, then a generic vector $\Psi\in\H_1\otimes \H_2$ has an expansion as in eqn. (\ref{expvec})
with all $c_k$ nonzero, and thus is cyclic and separating for the two algebras.
As a matter of notation, we will write $\psi_k=|k\ra$, $\psi'_k=|k\ra'$.   We also abbreviate $|j\ra\otimes |k\ra'$ as $|j,k\ra$.  Thus
\be\label{nexp}\Psi=\sum_{k=1}^n c_k|k\ra|k\ra'=\sum_{k=1}^n c_k|k,k\ra.\ee

As a check on some of this, we observe that as $\H_1$ and $\H_2$ have dimension $n$, $\H$ has dimension $n^2$.   The algebras
$\A$ and $\A'$ are algebras of $n\times n$ matrices, so they likewise are of dimension $n^2$.  So the linear map
$\A\to\H$ that takes $\a\in \A$ to $(\a\otimes 1)\Psi\in \H$ is surjective if and only if it has trivial kernel.  In other words, $\Psi $ is separating
for $\A$ if and only if it is cyclic.  Both properties are true precisely if the $c_k$ are all nonzero.  

We would like to find the modular operators in this situation.   The definition of $S_\Psi:\H\to \H$ is
\be\label{pilgo}S_\Psi((\a\otimes 1)\Psi) = (\a^\dagger\otimes 1)\Psi. \ee
To work out the consequences of this, pick some $i$ and $j$ in the set $\{1,2,\cdots,n\}$, and let $\a$ be the elementary matrix
that acts on $\H_1$ by
\be\label{nilgo}\a|i\ra=|j\ra, ~~~~\a|k\ra=0~~{\mathrm{if}}~k\not=i.\ee
Its adjoint acts by
\be\label{wilgo}\a^\dagger|j\ra=|i\ra,~~~~\a^\dagger|k\ra=0~~~~{\mathrm{if}}~k\not=j. \ee
So
\be\label{ilgo} (\a\otimes 1)\Psi = c_i|j,i\ra,~~~(\a^\dagger\otimes 1) \Psi=c_j|i,j\ra.\ee
Thus the definition of $S_\Psi$ implies
\be\label{kilgo} S_\Psi(c_i|j,i\ra)=c_j|i,j\ra.   \ee
Recalling that $S_\Psi$ is supposed to be antilinear, this implies
\be\label{filgo}S_\Psi |j,i\ra=\frac{c_j}{\bar c_i}|i,j\ra. \ee
That gives a complete description of $S_\Psi$, since the states $|i,j\ra$ are a basis of $\H$.
The adjoint $S_\Psi^\dagger $ acts by
\be\label{utilgo} S^\dagger_\Psi |i,j\ra=\frac{c_j}{\bar c_i}|j,i\ra.\ee
The modular operator $\Delta_\Psi=S_\Psi^\dagger S_\Psi$ hence acts by
\be\label{wutilgo} \Delta_\Psi|j,i\ra =\frac{|c_j|^2}{|c_i|^2}|j,i\ra.\ee
To get this formula, one must recall that $S^\dagger_\Psi $ is antilinear.

We also want to find the antiunitary operator $J_\Psi$ that appears in the polar
decomposition $S_\Psi=J_\Psi \Delta_\Psi^{1/2}$.  Since
\be\label{putilgo}\Delta_\Psi^{1/2}~|j,i\ra=\sqrt{\frac{|c_j|^2}{|c_i|^2}}~~ |j,i\ra, \ee
we have
\be\label{zutilgo}J_\Psi ~|j,i\ra =\sqrt{\frac{c_jc_i}{\bar c_j \bar c_i}}~~|i,j\ra. \ee

If $\Phi$ is a second state in $\H$, we can work out in a simple way the relative operators $S_{\Psi|\Phi}$ and $\Delta_{\Psi|\Phi}$.
In some orthonormal bases $\phi_\alpha$ of $\H_1$ and $\phi_\alpha'$ of $\H_2$, $\alpha=1,\dots,n$, we have
\be\label{zelb}\Phi=\sum_{\alpha=1}^nd_\alpha \phi_\alpha\otimes \phi'_\alpha, \ee
with some coefficients $d_\alpha$.
We write $|\alpha\ra$ and $|\alpha\ra'$ for $\phi_\alpha$ and $\phi'_\alpha$,
and abbreviate $|\alpha\ra\otimes |\beta\ra'=|\alpha,\beta\ra$, 
and similarly $|\alpha\ra\otimes |i\ra'=|\alpha,i\ra $, $|i\ra\otimes |\alpha\ra' = |i,\alpha\ra$, etc.
The state $\Phi$ is cyclic and separating for both algebras if and only if the $d_\alpha$ are all nonzero; we do not assume this.
We will determine the operator $S_{\Psi|\Phi}$ directly from the definition
\be\label{omko}S_{\Psi|\Phi}((\a\otimes 1)\Psi) =(\a^\dagger\otimes 1)\Phi,~~~\forall \a\in \A.\ee
For some $i,\alpha \in\{1,2,\cdots,n\}$, suppose that $\a\in\A$ acts by
\be\label{micop} \a|i\ra=|\alpha\ra,~~~\a|j\ra=0~~{\mathrm{for}}~j\not=i.\ee
Then
\be\label{nicop} \a^\dagger|\alpha\ra=|i\ra,~~~\a^\dagger|\beta\ra=0~~{\mathrm{for}}~\beta\not=\alpha.\ee
So 
\be\label{licop}(\a\otimes 1)\Psi=c_i |\alpha,i\ra,~~~ (\a^\dagger\otimes 1)\Phi=d_\alpha |i,\alpha\ra.\ee
Accordingly
\be\label{ricop} S_{\Psi|\Phi}|\alpha,i\ra = \frac{d_\alpha}{\bar c_i}|i,\alpha\ra. \ee
The adjoint is characterized by
\be\label{ticop}S^\dagger_{\Psi|\Phi}|i,\alpha\ra =\frac{d_\alpha}{\bar c_i}|\alpha,i\ra. \ee
It follows that
\be\label{wicop}\Delta_{\Psi|\Phi}|\alpha,i\ra = \frac{|d_\alpha|^2}{|c_i|^2}|\alpha,i\ra . \ee

Some of these formulas can be conveniently described in terms of density matrices.   Let us assume that  $\Psi,\Phi$ 
are unit vectors:
\be\label{miffo}\sum_i|c_i|^2 = \sum_\alpha |d_\alpha|^2 =1. \ee
To the state $\Psi\in\H_1\otimes \H_2$,
one associates a density matrix $\rho_{12}=|\Psi\ra \la\Psi|$.   It is a matrix acting on $\H$ by $|\chi\ra \to |\Psi\ra \la\Psi|\chi\ra$;
in other words it is the projection operator onto the subspace generated by $|\Psi\ra$.   In particular, it is positive and has trace 1:
\be\label{wiffo} \Tr_{12}\,\rho_{12}=1.\ee
Here $\Tr_{12}$ represents the trace over $\H=\H_1\otimes \H_2$.  By taking a partial trace over $\H_2$ or $\H_1$, one defines
reduced density matrices $\rho_1=\Tr_2\,\rho_{12}$, $\rho_2=\Tr_1\,\rho_{12}$.  Here $\rho_1$ and $\rho_2$ are positive matrices
acting on $\H_1$ and $\H_2$ respectively.  They have trace 1 since for example $\Tr_1\,\rho_1=\Tr_1\Tr_2\,\rho_{12}=\Tr_{12}\,\rho_{12}=1$.
Likewise, one defines a density matrix $\sigma_{12}=|\Phi\ra\la\Phi|$ associated to $\Phi$ and reduced density matrices 
$\sigma_1=\Tr_2\,\sigma_{12}$, $\sigma_2=\Tr_1\,\sigma_{12}$, all positive and of trace 1.

For the state $\Psi$ defined in eqn. (\ref{expvec}), the corresponding reduced density matrices are
\be\label{tolgo} \rho_1=\sum_i|c_i|^2 |i\ra\la i|,~~~\rho_2=\sum_i|c_i|^2 |i\ra'\la i|'. \ee  Clearly, $\rho_1$ and $\rho_2$
are invertible if and only if the $c_i$ are all nonzero, that is if and only if $\Psi$ is cyclic separating for both algebras.
Similarly the reduced density matrices of $\Phi$ are
\be\label{woolgo} \sigma_1 =\sum_\alpha |d_\alpha|^2|\alpha\ra\la\alpha|,~~~ \sigma_2=\sum_\alpha |d_\alpha|^2|\alpha\ra'\la\alpha|'. \ee

Comparing these formulas to (\ref{wutilgo}) and (\ref{wicop}),
the modular operator $\Delta_\Psi$ and the relative modular operator $\Delta_{\Psi|\Phi}$ can be conveniently written in terms of the
reduced density matrices:
\be\label{nolgo} \Delta_\Psi=\rho_1\otimes \rho_2^{-1}. ~~~ \Delta_{\Psi|\Phi}=\sigma_1\otimes \rho_2^{-1}. \ee

 The density matrix $\rho_2$ is conjugate to $\rho_1$ under
the exchange $|i\ra\leftrightarrow |i\ra'$, and similarly for $\sigma_1$ and $\sigma_2$.

It can be convenient to pick the phases of the states $|i\ra'$ relative to $|i\ra$ to ensure that the $c_i$ are all positive.  If we do this,
the antiunitary operator $J_\Psi$ becomes a simple flip:
\be\label{wolgo}J_\Psi |i,j\ra =|j,i\ra. \ee
The existence of a natural antiunitary operator $J_\Psi$ that flips the two bases in this way suggest that it is natural (once a cyclic
separating state $\Psi$ is given) to identify $\H_2$ as the dual of $\H_1$, by thinking of an element of $\H_1$ in the basis $|i\ra$ as a column vector and an element of $\H_2$
in the basis $|i\ra'$ as a row vector.  Then an element of $\H=\H_1\otimes \H_2$ is regarded as an $n\times n$ matrix, acting on $\H_1$.
The Hilbert space inner product of $\H$ is interpreted in terms of matrices $x,y:\H_1\to\H_1$ as
\be\label{ubb} \la x|y\ra =\Tr_{\H_1}\,x^\dagger y. \ee
The action of $\a\in \A$ on $\H$ becomes
\be\label{nubb} x\to \a x \ee
and the action of $\a'\in\A'$ on $\H$ becomes
\be\label{zubb} x\to x \a'\,^{\mathrm{tr}},\ee
where $b^\mathrm{tr}$ is the transpose of a matrix $b$.   With states reinterpreted in this way as matrices, the state $\Psi$ becomes
\be\label{prim} \Psi = \rho_1^{1/2}. \ee
This follows upon comparing (\ref{expvec}) and (\ref{tolgo}), remembering that we now take the $c_k$ to be positive
and interpret $\psi_k\otimes \psi'_k$ as a matrix $|k\ra\la k|$.  

When states are reinterpreted as matrices, eqn. (\ref{nolgo}) for the action of $\Delta_{\Psi|\Phi}$ on a state $x$ becomes
$\Delta_{\Psi|\Phi}(x)=\sigma_1 x(\rho^\tr _2)^{-1}$.   But once we identify $\H_2$ as the dual of $\H_1$, $\rho_2^\tr=\rho_1$  so
\be\label{zurim} \Delta_{\Psi|\Phi}(x)=\sigma_1 x\rho_1^{-1}. \ee
For future reference, we note that this implies
\be\label{wurim}\Delta_{\Psi|\Phi}^\alpha(x)=\sigma_1^\alpha x\rho_1^{-\alpha},\ee 
leading to a formula that will be useful later:
\be\label{yurim} \la\Psi |\Delta_{\Psi|\Phi}^\alpha|\Psi\ra = \Tr_{\H_1}\, \rho_1^{1/2}\Delta_{\Psi|\Phi}^\alpha(\rho_1^{1/2})
=\Tr_{\H_1}\,\rho_1^{1/2}\sigma_1^{\alpha}\rho_1^{1/2}\rho_1^{-\alpha}=\Tr_{\H_1}\,\sigma_1^\alpha \rho_1^{1-\alpha}. \ee

The identification of $\H_2$ with the dual of $\H_1$ depended on the choice of a cyclic separating vector $\Psi$, so we do not automatically
get an equally simple relation between $\Phi$ and its reduced density matrices $\sigma_1$ and $\sigma_2$.   However, if we are only
interested in $\sigma_1$ and not $\sigma_2$, we can act on $\Phi$ with a {\it unitary} element of $\A'$ without changing $\sigma_1$.
In general, once we identify $\H$ with the space of matrices acting on $\H_1$, $\Phi$ corresponds to such a matrix.  As such it has a polar decomposition $\Phi=P U$, where $P$ is
positive and $U$ is unitary.  In general $P=\sigma_1^{1/2}$.  Acting with a unitary element of $\A'$ to eliminate $U$, one reduces to
$\Phi=\sigma_1^{1/2}$.

\subsection{The Modular Automorphism Group}\label{mod group}

All of the properties of the operators $S_\Psi$, $\Delta_\Psi$, etc., that we deduced in general in sections \ref{defop} and \ref{relmod}
are of course still true in this finite-dimensional setting.

However,  some important additional properties are now more transparent.  Most of these involve what is called the modular
automorphism group.  This is the group of unitary transformations of the form $\Delta_\Psi^{\i s}$, $s\in\R$. We already know
(eqn. (\ref{arplor})) that $\Delta_\Psi^{\i s}$ commutes with $J_\Psi$.    In the finite-dimensional setting, we have the explicit formula
(\ref{nolgo}) for $\Delta_\Psi$.  By virtue of this formula, $\Delta_\Psi^{\i s}=\rho_1^{\i s}\otimes \rho_2^{-\i s}$.   So for any $\a\otimes 1\in \A$, 
\be\label{zotto}\Delta_\Psi^{\i s}(\a\otimes 1)\Delta_\Psi^{-\i s}=\rho_1^{\i s}\a\rho_1^{-\i s}\otimes 1.  \ee
The important fact here is that the right hand side of (\ref{zotto}) is of the form $\b\otimes 1$ for some $\b$, so it is in $\A$.
In other words, conjugation by the modular group maps $\A$ to itself.  It similarly maps $\A'$ to itself. We summarize this as 
\be\label{notto}\Delta_\Psi^{\i s}\A \Delta_\Psi^{-\i s}=\A,~~~~~\Delta_\Psi^{\i s}\A'\Delta_\Psi^{-\i s}=\A'. \ee
On the other hand, conjugation by $J_\Psi$ exchanges the two algebras $\A$ and $\A'$:
\be\label{potto}J_\Psi \A J_\Psi=\A',~~~ J_\Psi \A' J_\Psi = \A. \ee  For example, if we choose the phases of the
states
so that $J_\Psi$ flips basis vectors $|i,j\ra$  as in eqn. (\ref{wolgo}), then $J_\Psi(\a\otimes 1)J_\Psi =1\otimes \a^*$ (where $\a^*$
is the complex conjugate matrix to $\a$) and likewise
$J_\psi(1\otimes \a)J_\Psi =\a^*\otimes 1$.

The group of unitary transformations $\Delta_{\Psi|\Phi}^{\i s}$, $s\in\R$, is called the relative modular group.
In the finite-dimensional setting, eqn. (\ref{nolgo}) leads to 
\be\label{blottor}\Delta_{\Psi|\Phi}^{\i s}(\a\otimes 1)\Delta_{\Psi|\Phi}^{-\i s}=\sigma_1^{\i s}\a\sigma_1^{-\i s}\otimes 1.  \ee
Again, conjugation by the relative modular group maps $\A$ (or $\A'$) to itself.  But now we see the additional important property
that this conjugation depends only on $\Phi$ and not on $\Psi$.  Thus if $\Psi$ and $\Psi'$ are two cyclic separating vectors, we
have 
\be\label{blotto}\Delta_{\Psi|\Phi}^{\i s}(\a\otimes 1)\Delta_{\Psi|\Phi}^{-\i s}=\Delta_{\Psi'|\Phi}^{\i s}(\a\otimes 1)\Delta_{\Psi'|\Phi}^{-\i s}.\ee

The properties just stated are regarded as the main theorems of Tomita-Takesaki theory.  For general infinite-dimensional von Neumann
algebras with cyclic separating vectors, these properties are not so easy to prove.  However, there is a relatively simple proof
\cite{LongoSImpleProof} in the case of an infinite-dimensional algebra $\A$ that is a limit of matrix algebras.  This is believed to be the
case in quantum field theory for the algebra $\A_\U$ associated to an open set $\U$ in spacetime.    The statement means roughly
that one can think of the degrees of freedom in region $\U$ as an infinite collection of qubits.  Taking just $n$ of these qubits,
one gets an algebra $M_n$ of $2^n\times 2^n$ matrices that is an approximation of $\A_\U$.  Adding qubits, one gets an ascending chain
of algebras $M_1\subset M_2\subset \cdots \subset M_n\subset \cdots \subset\A_\U$ with $\A_\U$ as its limit.\footnote{We will
discuss algebras defined in this way in section \ref{algex}.}  It is believed that this
picture is rigorously valid in quantum field theory.
At each finite step
in the chain, one defines an approximation\footnote{This is done as follows.  If $\Psi\in \H$ is a cyclic
separating vector, then for each $n$, $\H_n=M_n\Psi$  is a subspace of $\H$ of dimension $2^{2n}$.
$M_n$ acts on $\H_n$ with cyclic separating vector $\Psi$, so one can define the modular operator $\Delta_\Psi^{\la n\ra}:\H_n\to \H_n$.
One defines $\Delta_\Psi^{(n)}:\H\to\H$ to coincide with $\Delta_\Psi^{\la n\ra}$ on $\H_n$ and to equal 1 on the orthocomplement.}
    $\Delta_\Psi^{(n)}$ to the  modular operator (or similarly  to $J_\Psi$ or $\Delta_{\Psi|\Phi}$).   Each such approximation obeys
    eqns. (\ref{zotto}), and the nature of this statement is such that if it is true at each step, it remains true in the limit.  Of course
    the main point of the proof is to show that $\Delta_\Psi^{(n)}$ does in an appropriate sense converge to $\Delta_\Psi$.
    
    Similarly the statements (\ref{potto}) and (\ref{blotto}) have the property that if true in a sequence of approximations, they remain
    true in any reasonable limit.  So one should expect these statements to hold in quantum field theory.
    
    The infinite-dimensional case becomes essentially different from a finite-dimensional matrix algebra when one considers the
    behavior of $\Delta_\Psi^{\i s }$ (or $\Delta_{\Psi|\Phi}^{\i s}$) when $s$ is no longer real.  For a matrix algebra, there is no problem;
    $\Delta_\Psi^{\i z}=\exp(\i z\log \Delta_\Psi)$ is an entire matrix-valued function of $z$.   In quantum field theory, 
    $\Delta_\Psi$ is unbounded and
    the analytic properties of $\Delta_\Psi^{\i z }\chi$ for a state $\chi$ depend very much on $\chi$.  By taking spectral projections, we can
    find states $\chi$ such that $\Delta_\Psi^{\i z}\chi$ is entire in $z$, just as in section \ref{bounded} we found vectors
    on which $\exp(ic\cdot P)$ acts holomorphically.    At the opposite extreme, we can also find states $\chi$ on 
    which $\Delta_\Psi^{\i z}\chi$ can only be defined if
    $z$ is real.
    
    Frequently, however, we are interested in the action of $\Delta_\Psi$ on a vector $\a\Psi$, $\a\in \A$ (or $\a'\Psi$, $\a'\in\A'$).  
    Here we have some simple holomorphy.   First of all, $\Delta_\Psi^{1/2} \a\Psi$ has finite norm and so makes sense as a Hilbert
    space vector:
    \be\label{prox} |\Delta_\Psi^{1/2}\a\Psi|^2=\la \Delta_\Psi^{1/2}\a\Psi|\Delta_\Psi^{1/2}\a\Psi\ra=\la \a\Psi|\Delta_\Psi|\a\Psi\ra
    =\la \a\Psi|S^\dagger_\Psi S_\Psi|\a\Psi\ra = \bar{\la S \a\Psi|S \a\Psi\ra}=\bar{\la \a^\dagger\Psi|\a^\dagger\Psi\ra}<\infty. \ee
    On the other hand, for $0\leq r\leq 1$,  the inequality $\lambda^r<\lambda+1$ for a positive real number $\lambda$ implies
    $\Delta_\Psi^r<\Delta_\Psi+1$.   So
    \be\label{woox} \la \Delta_\Psi^{r/2}\a\Psi|\Delta_\Psi^{r/2}\a\Psi\ra< \la \Delta_\Psi^{1/2}
    \a\Psi|\Delta_\Psi^{1/2}\a\Psi\ra +\la \a\Psi|\a\Psi\ra <    \infty, ~~0\leq r\leq 1.\ee
    The unitary operator $\Delta_\Psi^{\i s}$, $s\in \R$ does not change the norm of a state  so  $\Delta_\Psi^{\i s}\Delta_\Psi^{r/2}\a\Psi$
    also has finite norm for $s\in\R$,
    $0\leq r\leq 1/2$.   The upshot of this is that $\Delta_\Psi^{\i z} \a\Psi$ is continuous in the strip 
    $0\geq\mathrm{Im}\,z\geq -1/2$ and holomorphic in the interior of the strip.  Replacing $\A$ with $\A'$ has the effect of replacing
    the modular operator $\Delta_\Psi$ with its inverse, as we learned in section (\ref{defop}), so $\Delta_\Psi^{\i z} \a'\Psi$ is
   continuous in the strip $1/2\geq \mathrm{Im}\,z\geq 0$ and holomorphic in the interior of the strip.

    In section \ref{more}, we will find in a basic quantum field theory example that the holomorphy statements that we have just made
    are the best possible: generically, $\Delta^z \a\Psi$ and $\Delta^z \a'\Psi$ cannot be continued outside the strips that we have
    identified.    
    
    Now for $\a,\b\in\A$, let us look at the analytic properties of the function
    \be\label{nipppo} F(z)=\la \Psi| \b \Delta_\Psi^{\i z } \a |\Psi\ra, \ee
    initially defined for real $z$.  If $z=s-\i r$, this is
    \be\label{fippo}\la \b^\dagger\Psi | \Delta_\Psi^{\i s}\Delta_\psi^r \a|\Psi\ra
    =\la \Delta_\Psi ^{r/2} \b^\dagger \Psi| \Delta_\Psi^{\i s} |\Delta_\Psi^{r/2} \a\Psi\ra . \ee
    For $r\leq 1$, the states $\Delta_\Psi^{r/2}\a\Psi$ and $\Delta_\Psi^{r/2}\b^\dagger\Psi$ are normalizable,
    as we have already discussed.  So the function $F(z)$ is continuous in the strip $0\geq \mathrm{Im}\,z\geq -1$
    and holomorphic in the interior of the strip.  On the upper boundary of the strip, we have 
    \be\label{wippo} F(s)=\la \Psi|\b\Delta_\Psi^{\i s} \a|\Psi\ra. \ee
    Let us determine the boundary values on the lower boundary of the strip.  We have
    \begin{align}\label{ippo}F(-\i+s)=&\la\Psi|\b \Delta_\Psi^{1+\i s} \a|\Psi\ra
    =\la \Delta_\Psi^{1/2} \b^\dagger\Psi|\Delta_{\Psi}^{\i s}|\Delta_\Psi^{1/2}\a\Psi\rangle
    =\la J_\Psi S_\Psi \b^\dagger\Psi|\Delta_{\Psi}^{\i s}|J_\Psi S_\Psi \a\Psi\ra\cr =
    &\la J_\Psi \b\Psi| \Delta_{\Psi}^{\i s}|J_\Psi \a^\dagger\Psi\ra=\la J_\Psi \b\Psi| J_\Psi \Delta_{\Psi}^{\i s}\a^\dagger\Psi\ra
    =\la \Delta_\Psi^{\i s}\a^\dagger\Psi|\b\Psi\ra=\la\Psi|\a \Delta_\Psi^{-\i s } \b|\Psi\ra.\end{align}
    We used the fact that $J_\Psi$ is  antiunitary and commutes with $\Delta_\Psi^{\i s}$.      
     
     To understand what these statements mean for a finite-dimensional quantum system with $\H=\H_1\otimes\H_2$ and $\A$ acting on the
     first factor, consider again the density matrix $\rho_{12}=|\Psi\ra\la\Psi|$ and the reduced density matrix $\rho_1=\Tr_2\,\rho_{12}$.  
     The ``modular Hamiltonian'' $H$ is defined by $\rho_1=\exp(-H)$.   In the definition of $F(z)$, $\Delta_\Psi^{\i z}\a \Psi$
     can be replaced by $\Delta_\Psi^{\i z} \a \Delta_\Psi^{-\i z} \Psi$ since $\Delta_\Psi\Psi=\Psi$.  As in eqn. (\ref{zotto}),  $\Delta_\Psi^{\i z} \a \Delta_\Psi^{-\i z} \Psi=\rho_1^{\i z}\a\rho_1^{-\i z} \Psi
     =e^{-\i z H}\a e^{\i z H}\Psi.$  Moreover,  for any $\O$ that acts on $\H_1$, $\la\Psi|\O|\Psi\ra=\Tr_{\H_1} \rho_1\O=
     \Tr_{\H_1}e^{-H} \O$.    
     Hence 
     \be\label{zormo}F(z)=\Tr_{\H_1} e^{-H} \b e^{-\i z H} \a e^{\i z H}. \ee
     From this it is clear that the values for $z=s$ and $z=-\i+s$  are
     \be\label{tufo} F(s)= \Tr_{\H_1} \, e^{-H} \b e^{-\i sH} \a e^{\i sH},~~~~F(-\i+s)= \Tr_{\H_2}\, e^{-H}  e^{-\i s H} \a e^{\i sH} \b. \ee
    In the usual physical interpretation, $s$ represents real time, $\a(s)=e^{-\i s H} a e^{\i s H}$ is a Heisenberg operator at time $-s$, 
     and these functions are real time two-point functions in
    a thermal ensemble with Hamiltonian $H$ (and inverse temperature 1), with different
     operator orderings.  The fact that the different operator orderings  can be obtained from
     each other by analytic continuation is important, for example, in the derivation of a general bound on quantum chaos
     \cite{ChaosBound}, and in  many other applications.
     
     For a finite-dimensional quantum system, $F(z)$ is an entire function.   Let us, however, relax the assumption of finite-dimensionality,
     while still assuming a factorization $\H=\H_1\otimes \H_2$ of the Hilbert space.  The definition $\rho_1=e^{-H}$ implies that
     $H$ is nonnegative, but in the infinite-dimensional case, $H$ is inevitably unbounded above, given that $\Tr\,\rho_1=1$.
     For the trace in eqn. (\ref{zormo}) to be well-behaved, given that $H$ is unbounded above, both $\i z$ and $1-\i z$ must have nonnegative real part.
     This leads to the strip $0\geq \mathrm{Im}\,z \geq -1$, which we identified earlier without assuming the factorization $\H=\H_1\otimes\H_2$.

     Assuming the factorization $\H=\H_1\otimes \H_2$,  one would actually predict further holomorphy of correlation functions.
     For example, generalizing eqn. (\ref{zormo}), a three-point function
     \be\label{threp} F(z_1,z_2) =\Tr_{\H_1} e^{-H} {\c} \,e^{-\i  z_1H} \b \,e^{-\i (z_2-z_1)H} \a\, e^{\i z_2 H} .\ee
         should be holomorphic for $\mathrm{Im}\,z_1,\mathrm{Im}\,(z_2-z_1),\,-1-\mathrm{Im}\,z_2<0$.   Such statements can actually be
         proved without assuming a factorization of the Hilbert space.  See section 3 of \cite{ArakiFive} and also Appendix \ref{holoproof} below.

    All  statements we have made about holomorphy still apply if $\Delta_\Psi$ is replaced by the relative modular operator $\Delta_{\Psi|\Phi}$.

\subsection{Monotonicity of Relative Entropy In The Finite-Dimensional Case}\label{finitemon}

Using results of section \ref{fincase}, we can compare Araki's definition of relative entropy, which we used in discussing
quantum field theory, to the standard definition in nonrelativistic quantum mechanics.

We recall that Araki's definition for the relative entropy between two states $\Psi,$ $\Phi$, for measurements in a spacetime
region $\U$, is
\be\label{molno} \S_{\Psi|\Phi;\U}=-\la\Psi|\log \Delta_{\Psi|\Phi;\U}|\Psi\ra. \ee
Here $\Psi$ is a cyclic separating vector for a pair of commuting algebras $\A_\U$, $\A'_{\U}$.

In nonrelativistic quantum mechanics, we do not in general associate algebras with spacetime regions.
But we do have the notion of a vector $\Psi$ that is cyclic separating for a commuting pair of algebras $\A$, $\A'$.  Given a second
vector $\Phi$ we have the relative modular operator $\Delta_{\Psi|\Phi}$.    Given this, we could imitate  in nonrelativistic quantum
mechanics Araki's definition, which in terms of the density matrix $\rho_{12}=|\Psi\ra\la\Psi|$ is
\be\label{olno}\S_{\Psi|\Phi}=-\la\Psi|\log \Delta_{\Psi|\Phi}|\Psi\ra =-\Tr_{12} \rho_{12}\log \Delta_{\Psi|\Phi} .\ee
From eqn. (\ref{nolgo}), $\Delta_{\Psi|\Phi}=\sigma_1\otimes \rho_2^{-1}$, so 
$\log \Delta_{\Psi|\Phi}=\log \sigma_1\otimes 1-1\otimes \log \rho_2$.  The relative entropy is then
\be\label{bolno} \S_{\Psi|\Phi} =-\Tr_{12}\,\rho_{12}\left(\log \sigma_1\otimes 1-1\otimes \log \rho_2\right). \ee
Here $\Tr\,\rho_{12}(\log\sigma_1\otimes 1)=\Tr_1 \rho_1\log \sigma_1$, as one learns by first taking the trace over $\H_2$.
Likewise $\Tr\,\rho_{12} (1\otimes \log \rho_2)=\Tr_2 \,\rho_2\log \rho_2$.  But $\rho_1$ and $\rho_2$ are conjugate as explained at the end of section \ref{fincase}, so $\Tr_2\,\rho_2\log \rho_2
=\Tr_1\,\rho_1\log \rho_1$.   Finally then 
\be\label{nippo} \S_{\Psi|\Phi}=\Tr \,\rho_1(\log \rho_1-\log \sigma_1). \ee

We have arrived at the usual definition of the relative entropy in nonrelativistic quantum mechanics.  (Of course, that was
Araki's motivation.)  The usual approach runs in reverse from what we have said.
One starts with a Hilbert space $\H_1$ and two
density matrices $\rho_1$ and $\sigma_1$.  The relative entropy between them is defined as
\be\label{zippo} \S(\rho_1||\sigma_1)= \Tr \,\rho_1(\log \rho_1-\log \sigma_1). \ee
After introducing a second Hilbert space $\H_2$, $\rho_1$ and $\sigma_1$ can be ``purified'' by deriving them as the reduced
density matrices of pure states $\Psi,\Phi\in \H_1\otimes\H_2$.  The above formulas make clear that $\S(\rho_1||\sigma_1)$ is the same as
$\S_{\Psi|\Phi}$.     

Now let us discuss properties of the relative entropy.
Using the definition (\ref{olno}), the proof of positivity of relative entropy that was
described in section \ref{relen} carries over immediately to nonrelativistic quantum mechanics.

There is also an analog in nonrelativistic quantum mechanics of the more subtle property of monotonicity of relative entropy.   We will recall the
statement and then explain  how it can be understood
in a way similar to what we explained for quantum field theory in section \ref{relmood}.   In fact, though we explained the idea in
section \ref{relmood} in the context of quantum field theory, Araki's point of view was general enough to encompass nonrelativistic
quantum mechanics.  In our explanation below, we will follow Petz \cite{Petz}, later elaborated by Petz and Nielsen \cite{PetzNielsen},
who developed an approach  based in part on Araki's framework.  

To formulate the problem of monotonicity of relative entropy, the first step is to take what we have been calling $\H_1$
to be the Hilbert space of a bipartite system $AB$.
If $\H_A$ and $\H_B$ are the Hilbert spaces of systems $A$ and $B$, then the Hilbert space of the combined system $AB$
is $\H_A\otimes \H_B$.  In what follows, we will call this $\H_{AB}$ rather than $\H_1$.     If we are given density matrices $\rho_{AB}$ and $\sigma_{AB}$ on $\H_{AB}$, then we can
define the reduced density matrices $\rho_A=\Tr_B\,\rho_{AB}$ and $\sigma_A=\Tr_B\,\sigma_{AB}$ on $\H_A$, and the relative
entropies   $\S(\rho_{AB}||\sigma_{AB})$ and $\S(\rho_A||\sigma_A)$.   Monotonicity of relative entropy is the statement\footnote{This
is the version of monotonicity of relative entropy proved by Lieb and Ruskai \cite{LiebRuskai}.
A more general version of Uhlmann  \cite{Uhlmann}   involves an arbitrary quantum channel.  It can be reduced
to what is stated here by considering the Stinespring dilation of the channel.}
\be\label{nbcn} \S(\rho_{AB}||\sigma_{AB})\geq \S(\rho_A||\sigma_A). \ee
We want to explain how this inequality can be understood in a way similar to what we said in the quantum field theory case in section
\ref{relmood}.   In proving this inequality, we will assume that $\rho_{AB}$ (and therefore $\rho_A$) is invertible.   The general case
can be reached from this case by a limit.

In quantum field theory, the starting point was to study two open sets $\U,$ $\t \U$ with $\t\U\subset \U$.  We associated
to them algebras $\A_\U$, $\A_{\t\U}$.    For the bipartite system $AB$, we can introduce two algebras that will play a somewhat
similar role.  These algebras will be simply the algebras of matrices acting on $\H_{AB}$ and $\H_A$, respectively.  We write 
$\A_{AB}$ and $\A_A$ for these algebras.  

In the quantum field theory case, the smaller algebra 
$\A_{\t\U}$ is naturally a subalgebra of $\A_\U$.  The closest analog of this in nonrelativistic quantum mechanics is that
there is a natural embedding $\varphi:\A_A\to \A_{AB}$ by $\a\to \varphi(\a)= \a\otimes 1$.   

By passing from $\H_{AB}$ to a doubled Hilbert space $\H_{AB}\otimes \H'_{AB}$, we can ``purify'' $\rho_{AB}$ and $\sigma_{AB}$,
in the sense of deriving them as reduced density matrices on $\H_{AB}$ associated to pure states\footnote{The reader may wish to consult \cite{PetzNielsen}, where
Petz and Nielsen
make the specific choice $\Psi_{AB}=\rho_{AB}^{1/2}$, $\Psi_A=\rho_A^{1/2}$, etc., as in eqn. (\ref{prim}) above. This  leads to short and explicit
formulas.  The approach below aims to draw out the analogy with the quantum field theory case.  See also  \cite{NT,GR} for somewhat similar explanations.} 
$\Psi_{AB},\Phi_{AB}\in \H_{AB}\otimes \H'_{AB}$.
Since we assume $\rho_{AB}$ to be invertible, $\Psi_{AB}$ is cyclic separating.  Likewise, $\rho_A$ and $\sigma_A$ are reduced
density matrices associated to pure states $\Psi_A$, $\Phi_A$ in a doubled Hilbert space $\H_A\otimes \H'_A$, and $\Psi_A$ is
cyclic separating.

In quantum field theory, the two algebras $\A_\U$ and $\A_{\t\U}$ naturally act on the same Hilbert space $\H$ with the
same cyclic separating vector $\Psi$. In nonrelativistic
quantum mechanics, it is more natural for the smaller algebra $\A_A$ to act on the smaller Hilbert space $\H_A\otimes \H'_A$,
while the larger algebra $\A_{AB}$ acts on $\H_{AB}\otimes \H'_{AB}$.    The best we can do in nonrelativistic quantum mechanics
to imitate the idea that $\A_\U$ and $\A_{\t\U}$ act on the same space is to find a suitable isometric  embedding 
\be\label{uemb}U:\H_A\otimes \H'_A\to \H_{AB}\otimes \H'_{AB}. \ee
The embedding that will enable us to imitate what we had in quantum field theory is
\be\label{zemb} U(\a\Psi_{A})=(\a\otimes 1)\Psi_{AB}. \ee
Since $\Psi_A$ is cyclic separating, this formula does define a unique linear transformation  
$U:\H_A\otimes \H'_A\to \H_{AB}\otimes \H'_{AB}$,
and since $\Psi_{AB}$ is separating, this linear transformation is an embedding.
To show that it is an isometry, which means that $\la\eta|\chi\ra=\la U\eta|U\chi\ra$ for all $\eta,\chi\in \H_A\otimes \H'_A$,
we observe that as $\Psi_A$ is cyclic, we can take $\eta=\a\Psi_A$, $\chi=\b\Psi_A$.  We need then
$\la \a\Psi_A|\b\Psi_A\ra = \la (\a\otimes 1)\Psi_{AB}|(\b\otimes 1)\Psi_{AB}\ra$.  Indeed
\be\label{remb}\la (\a\otimes 1)\Psi_{AB}|(\b\otimes 1)\Psi_{AB}\ra=\la\Psi_{AB}|(\a^\dagger \b\otimes 1) \Psi_{AB}\ra
=\Tr_{\rho_{AB}} \a^\dagger \b\otimes 1=\Tr_{\rho_A} \a^\dagger \b=\la\Psi_A|\a^\dagger \b|\Psi_A\ra = \la
\a\Psi_A|\b\Psi_A\ra. \ee 
Finally, the isometric embedding that we have defined commutes with the action of $\A_A$ in the sense that for
any $\chi\in \H_A\otimes \H'_A$, we have $U(\a\chi)=\varphi(\a)U(\chi)$.  Indeed, if $\chi=\b\Psi_A$, we have 
\be\label{mirz} U(\a\chi)=U(\a\b\Psi_A)=(\a\b\otimes 1)\Psi_{AB}=(\a\otimes 1)(\b\otimes 1)\Psi_{AB}=\varphi(\a)U(\chi). \ee

This shows that, if we identify $\a$ with $\varphi(\a)$, we can regard $\A_A$ as a subalgebra of $\A_{AB}$ and the action of 
$\A_A$ on $\H_A\otimes \H'_A$ is unitarily equivalent to its action on a subspace of $\H_{AB}\otimes \H_{AB}'$.
  We are almost ready to imitate the proof of section \ref{relmood}, but
we still have to compare the relative modular operators.

We  have a relative modular operator $\Delta_{\Psi_{AB}|\Phi_{AB}}$ for the algebra $\A_{AB}$ acting on 
$\H_{AB}\otimes \H'_{AB}$, and a corresponding relative modular operator $\Delta_{\Psi_A|\Phi_A}$ for the algebra $\A_A$
acting on $\H_A\otimes \H'_A$.  To lighten the notation, we will write just $\Delta_{AB}$ and $\Delta_A$ instead of
$\Delta_{\Psi_{AB}|\Phi_{AB}}$ and  $\Delta_{\Psi_A|\Phi_A}$. 

The last fact that we need for the proof of monotonicity of relative entropy is that our isometric embedding $U:\H_A\otimes \H'_A
\to \H_{AB}\otimes \H'_{AB}$ intertwines the relative modular operators, in the sense that
\be\label{omigo} U^\dagger \Delta_{AB} U =\Delta_A. \ee
Here $U^\dagger:\H_{AB}\otimes \H'_{AB}\to \H_A\otimes \H'_A$ is the adjoint of $U:\H_A\otimes\H'_A\to \H_{AB}\otimes \H'_{AB}$.
It is possible to work out an explicit formula for $U^\dagger$, but we will not need it.  
To prove eqn. (\ref{omigo}), 
it is enough to verify that the left and right hand sides  have the same matrix elements between arbitrary states $\a^\dagger\Psi$ and $\b\Psi$.
This is actually a rather direct consequence of eqn. (\ref{usefulone}).    For the matrix element of $\Delta_A$,
we have
\be\label{plomigo}\la \a^\dagger\Psi_A|\Delta_A|\b\Psi_A\ra = \la\b^\dagger\Phi_A|\a\Phi_A\ra=\la\Phi_A|\b\a|\Phi_A\ra=\Tr_A\sigma_A \b\a.\ee
The corresponding matrix element of $U^\dagger \Delta_{AB}U$ is
\begin{align}\label{pilomigo} \la\a^\dagger \Psi_A|U^\dagger\Delta_{AB}U|\b\Psi_A\ra=&\la U(\a^\dagger\Psi_A)|\Delta_{AB}|U(\b\Psi_A)\ra
=\la (\a^\dagger\otimes 1) \Psi_{AB}|\Delta_{AB}|(\b\otimes 1)\Psi_{AB}\ra\cr=&\la (\b^\dagger\otimes 1)\Phi_{AB}|(\a\otimes 1)\Phi_{AB}\ra
=\la \Phi_{AB}|(\b\a\otimes 1)|\Phi_{AB}\ra\cr =&\Tr_{AB}\sigma_{AB}(\b\a\otimes 1)=\Tr_A \sigma_A\b\a.\end{align}

Eqn. (\ref{linocks}) (which was proved for an arbitrary isometric embedding), 
 when combined with eqn. (\ref{omigo}), gives us an inequality
\be\label{purf} U^\dagger (\log \Delta_{AB} )U \leq \log \Delta_A. \ee

Now we are finally ready to compare the relative entropies
\begin{align}\label{urf}  \S(\rho_A||\sigma_A) &=-\la \Psi_A|\log\Delta_A|\Psi_A\ra \cr 
                                  \S(\rho_{AB}||\sigma_{AB}\ra&=-\la \Psi_{AB}|\log\Delta_{AB}|\Psi_{AB}\ra. \end{align}
Using eqn. (\ref{purf}),
we have
\begin{align}\label{happy}\S(\rho_A||\sigma_A)=&-\la\Psi_A|\log \Delta_A|\Psi_A\ra \leq -\la \Psi_A|U^\dagger (\log \Delta_{AB} )U|\Psi_A\ra
\cr =&-\la U\Psi_A|\log\Delta_{AB}|U\Psi_A\ra = -\la\Psi_{AB}|\log\Delta_{AB}|\Psi_{AB}\ra= \S(\rho_{AB}||\sigma_{AB}). \end{align}
This completes the proof.

Was it obvious that this proof would work, or did it depend on checking tricky details?   Hopefully, we have succeeded in convincing
the reader that this explanation -- which largely follows \cite{Petz} and \cite{PetzNielsen} -- is the natural analog
of what was explained for quantum field theory in section \ref{relmood}.  Philosophically, it might seem obvious that quantum field
theory is not simpler than nonrelativistic quantum mechanics, so that an analogous proof in nonrelativistic quantum mechanics
must work somehow.

The only property of the logarithm that we used was that $\log X$ is an increasing function of a positive operator $X$.  Many other
functions have the same property; an example, as shown in section \ref{monrel}, is the function $X^\alpha$, $0\leq \alpha\leq 1$.   Replacing
$-\log \Delta_{AB}$ in eqn. (\ref{happy}) with $\Delta_{AB}^\alpha$ (and reversing the direction of the inequality because of the sign),
we get
\be\label{weg} \la\Psi_A|\Delta_A^\alpha|\Psi_A\ra \geq \la\Psi_{AB}|\Delta_{AB}^\alpha|\Psi_{AB}\ra.\ee 
Evaluating this with the help of eqn. (\ref{yurim}), we learn that\footnote{For recent applications of this inequality, see \cite{BGMO}.
Those authors consider also the case of $\alpha<0$, which can be analyzed by replacing eqn. (\ref{zox}) with
$R^\alpha\sim \int_0^\infty \d s \,s^\alpha/(s+R)$ (in a certain range of $\alpha$) and more generally $R^\alpha\sim \int_0^\infty \d s\, s^{n+\alpha}/(s+R)^{n+1}$ for
any nonnegative integer $n$.}
\be\label{nego}\Tr_A\, \sigma_A^\alpha \rho_A^{1-\alpha}\geq \Tr_{AB}\,\sigma_{AB}^\alpha\rho_{AB}^{1-\alpha},~~~~0\leq \alpha\leq 1. \ee 
This inequality is saturated at $\alpha=0$, since $\Tr_A\,\rho_A=\Tr_{AB}\,\rho_{AB}=1$.   Expanding around $\alpha=0$, the leading
term in the inequality gives back the monotonicity of relative entropy.    Similarly, the only property of the states $\Psi_A$ and 
$\Psi_{AB}$ that was used was that $U\Psi_A=\Psi_{AB}$.  One can derive further inequalities by replacing $\Psi_A$ and $\Psi_{AB}$
by $\a\Psi_A$ and $U(\a\Psi_A)=(\a\otimes 1)\Psi_{AB}$.   These inequalities (in a formulation
originally in terms of convexity rather than monotonicity) 
go back to  Wigner, Yanase, and Dyson \cite{WY} and  Lieb \cite{Lieb}, with later work by Araki \cite{Araki}
and Petz \cite{Petz}, among others.  

We conclude this section by briefly explaining how positivity and monotonicity of relative entropy are related to other important
concepts in quantum information theory.  The von Neumann entropy $\S(\rho)$ of a density matrix $\rho$ is defined as
\be\label{inz}\S(\rho)=-\Tr \,\rho\log\rho. \ee
Consider a bipartite system $AB$ with Hilbert space $\H_{AB}=\H_A\otimes \H_B$, 
density matrix $\rho_{AB}$ and reduced density matrices $\rho_A=\Tr_B\,\rho_{AB}$,
$\rho_B=\Tr_A\,\rho_{AB}$.  One sets $\S_{AB}=\S(\rho_{AB})$, $\S_A=\S(\rho_A)$, etc.  The mutual information $I(A;B)$ between
subsystems $A$ and $B$ is defined as
\be\label{linz} I(A;B)=\S_A+\S_B-\S_{AB}. \ee
Subadditivity of quantum entropy is the statement that $I(A;B)\geq 0$ for all $\rho_{AB}$.   To prove this, define the product
density matrix $\sigma_{AB}=\rho_A\otimes \rho_B$ for system $AB$.  The relative entropy between $\rho_{AB}$ and $\sigma_{AB}$
is
\be\label{minz} \S(\rho_{AB}||\sigma_{AB})=\Tr_{AB}\,\rho_{AB}\left(\log\rho_{AB}-\log\sigma_{AB}\right).\ee
Since $\log \sigma_{AB}=\log \rho_A\otimes 1+1\otimes \log
\rho_B$, this is
\be\label{plinz}\S(\rho_{AB}||\sigma_{AB})= \Tr_{AB}\rho_{AB}\left(\log \rho_{AB}-\log \rho_A\otimes 1-1\otimes \log
 \rho_B\right)=-\S_{AB}+\S_A+\S_B=I(A;B).\ee
Thus, subadditivity of quantum entropy follows from positivity of relative entropy.   For strong subadditivity of quantum entropy
\cite{LiebRuskai}, one considers a tripartite system $ABC$ with Hilbert space $\H_A\otimes \H_B\otimes \H_C$ and density 
matrix $\rho_{ABC}$.  One can define various reduced density matrices, such as $\rho_{AB}=\Tr_C\rho_{ABC}$, with corresponding
entropy $\S_{AB}$, and likewise for other subsystems.  
Strong subadditivity of quantum entropy is the statement that mutual information is monotonic in the sense
that
\be\label{turinz}I(A;B)\leq I(A;BC). \ee
Expanding this out using the definition of the mutual information, an equivalent statement is
\be\label{princo}\S_B+\S_{ABC}\leq \S_{AB}+\S_{BC}. \ee
To deduce strong subadditivity from the monotonicity of relative entropy, we compare the two tripartite density matrices $\rho_{ABC}$ and
$\sigma_{ABC}=\rho_A\otimes \rho_{BC}$.    As we have just seen, the relative entropy between them is
\be\label{drinco}\S(\rho_{ABC}||\sigma_{ABC})=I(A;BC). \ee
On the other hand, taking a partial trace over system $C$, the reduced density matrices for the $AB$ subsystem are $\rho_{AB}$ 
and $\sigma_{AB}=\rho_A\otimes \rho_B$. The relative entropy between them is
\be\label{crinco}\S(\rho_{AB}||\sigma_{AB})=I(A;B). \ee
Monotonicity of relative entropy tells us that taking the trace over subsystem $C$ can only make the relative entropy smaller,
so
\be\label{srinco} \S(\rho_{AB}||\sigma_{AB})\leq \S(\rho_{ABC}||\sigma_{ABC}). \ee 
Putting the last three statements together, we arrive at strong subadditivity.

Ä
\section{A Fundamental Example}\label{more}

\subsection{Overview}\label{preover}

A certain simple decomposition of Minkowski spacetime provides an important (and well-known) illustration of
some of these ideas.

We factorize $D$-dimensional Minkowski spacetime $M_D$ as the product of a two-dimensional Lorentz signature spacetime $\R^{1,1}$ with
coordinates $t,x$ and a $D-2$-dimensional
Euclidean space $\R^{D-2}$ with coordinates $\vec y=(y_1,\dots,y_{D-2})$.  Thus the metric is
\be\label{mozzo} \d s^2=-\d t^2+\d x^2+\d\vec y\cdot \d\vec y. \ee

\begin{figure}
 \begin{center}
   \includegraphics[width=2.5in]{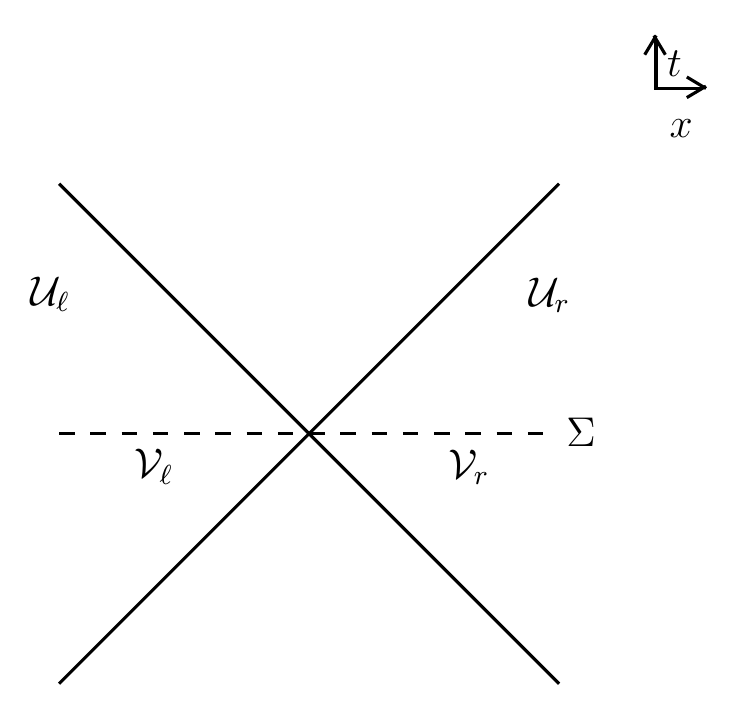}
 \end{center}
\caption{\small  The right wedge $\U_r$ and the left wedge $\U_\ell$ in Minkowski spacetime.  
They are the domains of dependence of the right half and left half of the initial value surface $t=0$, which are labeled as $\V_r$ and $\V_\ell$.
\label{Fig2}}
\end{figure}

In this spacetime, we let $\Sigma$ be the initial value surface $t=0$ (fig. \ref{Fig2}).  
We let $\V_r$ be the open right half-space in $\Sigma$, defined
by $x>0$.  The complement of its closure, which we will call $\V_\ell$, is the left half-space $x<0$.  The domain of dependence of
$\V_r$ is what we will call the right wedge $\U_r$, defined by $x>|t|$.   And the domain of dependence of
$\V_\ell$ is what we will call the left wedge $\U_\ell$, defined by $x<-|t|$.    These wedge-like regions are also often called Rindler spaces \cite{Rindler}.
 Finally, we denote as $\A_r$ and $\A_\ell$ the algebras of observables in $\U_r$ and $\U_\ell$,
respectively.    They commute and we will learn that they are each other's commutants.

Let $\Omega$ be the vacuum state of a quantum field theory on $M_D$.
The goal of this section will be to determine the modular operators $J_\Psi$ and $\Delta_\Psi$ for observations in region $\U_r$.
This problem was first analyzed and solved
 by Bisognano and Wichmann \cite{BiWi}.  Their approach involved the analytic behavior of correlation
functions and will be sketched in section \ref{bw}.    But first, in section \ref{pathint},
we explain a direct path integral approach.  This path integral approach is important in Unruh's thermal interpretation of accelerated motion in Minkowski spacetime
\cite{Unruh}, which we will explain in section \ref{unr}.  It is also closely related to analogous path integral derivations of the thermal
nature of black hole physics \cite{Hawking}, \cite{GibbonsHawking} and of correlation functions in de Sitter spacetime \cite{MoreGibbonsHawking,FHN}.
As this approach is relatively well-known,  we will be brief.

 The $\CPT$ symmetry of quantum field theory will enter in what follows, so we pause to discuss it.
$\CPT$ acts as $-1$ on all  space and time coordinates.
The basic reason that $\CPT$ is an unavoidable 
 symmetry of quantum field theory in $3+1$ dimensions is that in Euclidean signature,\footnote{The
rigorous proof of $\CPT$ invariance can be conveniently found in \cite{StW}.   It depends on the holomorphy statement
of eqn. (\ref{staval}).   Holomorphy is built in for free when one starts in Euclidean
signature, so if one assumes that a quantum field theory can be obtained by analytic continuation from Euclidean signature,
then one can see $\CPT$ without a careful discussion of conditions of holomorphy.} the transformation
that acts as $-1$ on all four 
coordinates is in the connected component of the rotation group. (If we factor $\R^4$ as $\R^2\times \R^2$, then a simultaneous
$\pi$ rotation on each copy of $\R^2$ acts as $-1$ on all four coordinates.)   Therefore,  in Euclidean
signature this operation is inevitably a symmetry of any rotation-invariant theory.  After continuation back to Lorentz signature, this symmetry becomes $\CPT$.  

The statement that a transformation of Euclidean space that acts as $-1$ on all coordinates is in the connected component of the rotation
group is true in and only in even spacetime dimension.  For odd $D$, that operation has determinant $-1$ and is not in the connected
component of the rotation group.   Accordingly, for odd $D$, there is no $\CPT$ symmetry in general.  A better formulation that
is uniformly valid in any dimension is to replace parity -- a sign change of all spatial coordinates -- with a reflection of just one spatial
coordinate.  We will call this operation $\RR$.  Regardless of the spacetime dimension, a
 simultaneous sign change of both the time $t$ and one spatial coordinate
$x$ is  
in the identity component of the rotation group in Euclidean signature, as it is a $\pi$ rotation of the $xt$ plane.  Thus, the universal
symmetry of quantum field theory in any dimension is $\CRT$ rather than $\CPT$.  In $3+1$ dimensions, $\CPT$ is the product of $\CRT$
times a $\pi$ rotation of two spatial coordinates, so the two are essentially equivalent.

Because $\CPT$ or $\CRT$ is antiunitary, it reverses the signs of conserved charges.  
Historically, $\sf P$ and $\sf T$ were defined to be good approximate symmetries of ordinary matter
(until the 1950's, they were assumed to be exact symmetries).  Since ordinary matter is made of leptons and baryons without antileptons
and antibaryons, $\sf P$ and $\sf T$ were defined to commute with baryon number and lepton number.  With this choice, 
the universal discrete symmetry does not coincide with $\sf {PT}$ or $\sf{RT}$ and  deserves to be called $\CPT$ or $\CRT$, to express the fact that it reverses conserved
charges.\footnote{Both $\RR$ and what is usually called $\sf{CT}$ come from the same operation in Euclidean signature (reflection of 
one spatial coordinate), continued back to Lorentz signature in different ways.  So purely from a relativistic point of view, it would be natural to exchange the names $\sf T$ and $\sf{CT}$ and
refer to the universal  discrete symmetry as $\sf{PT}$ or $\sf{RT}$, rather than $\sf{CPT}$ or $\sf{CRT}$.  However, this would
involve too much conflict with standard terminology.}

\subsection{Path Integral Approach}\label{pathint}

\begin{figure}
 \begin{center}
   \includegraphics[width=5.5in]{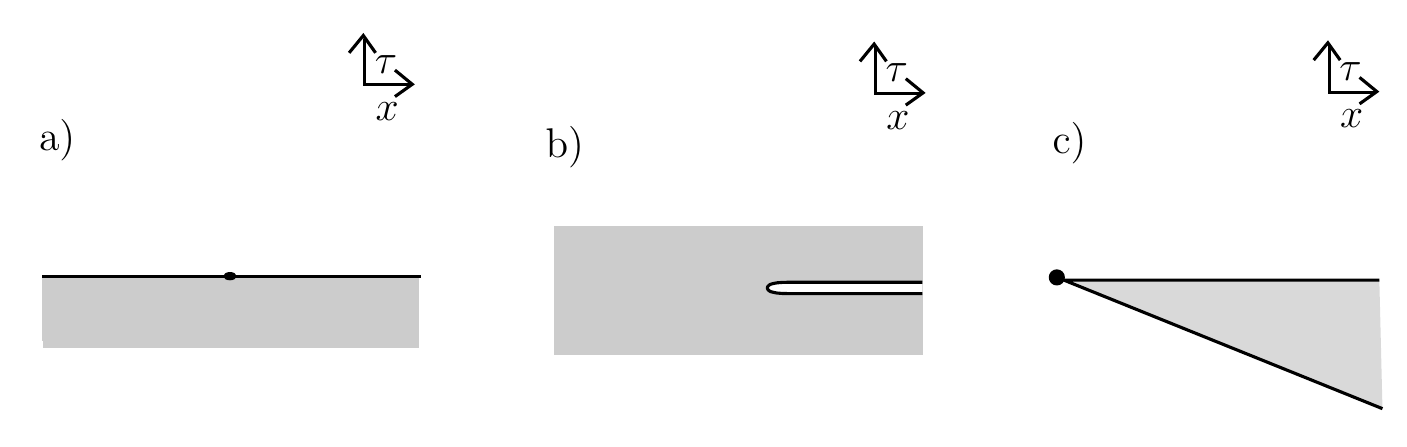}
 \end{center}
\caption{\small (a) The path integral on the half-space $\tau<0$ as a function of boundary values of the fields
gives a way to compute the vacuum wavefunction $\Omega$.  (b) To compute the reduced
 density matrix of the vacuum for the right half of the surface $\tau=0$ by a Euclidean path integral, we use the path
integral on the lower half-space $\tau<0$ to compute a vacuum bra $\la \Omega|$, and the path integral on the upper half-space
$\tau>0$ to compute a vacuum ket $|\Omega\ra$.  Then we glue together the left halves of the boundaries of the $\tau<0$ and $\tau>0$
half-spaces, identifying the field variables on those boundaries in the bra and the ket.  The net effect -- a path integral on the upper
half-space and the lower half-space together with an integral over field variables on half of the $\tau=0$ hypersurface -- produces
a path integral on the space depicted here.  It can be obtained from Euclidean space $\R^D$ by making a ``cut'' along
the half-hyperplane $\tau=0$, $x\geq 0$.  (c) Sketched here is a Euclidean wedge of opening angle $\theta$.
 \label{Fig3}}
\end{figure}

We continue to Euclidean signature, setting $t=-\i \tau$.   Euclidean path integrals are an effective way to compute
the vacuum state $\Omega$ of a quantum field theory.    Thus, the path integral on, say, the half-space $\tau\leq 0$,
as a function of boundary values on the hyperplane $\tau=0$, gives a way to compute $\Omega$ (fig. \ref{Fig3}(a)).

Suppose it were true that the Hilbert space $\H$ of a quantum field theory has a factorization $\H=\H_\ell\otimes \H_r$, where $\H_\ell$
and $\H_r$ are Hilbert spaces of degrees of freedom located at $x<0$ and $x>0$ respectively, and thus acted on by the algebras
$\A_\ell$ and $\A_r$.    In this case, starting with the pure state density matrix $|\Omega\ra\la\Omega|$ and
 taking a partial trace on the degrees of freedom in $\H_\ell$, we could define a reduced density matrix
$\rho_r$ on $\H_r$.    Technically, it is not quite true that $\H$ has the suggested factorization, but assuming that it does
will lead to a correct and illuminating determination of the operators $\Delta_\Omega$ and $J_\Omega$ for the vacuum state.  

To formally construct the density matrix $\rho_r$ for the right half-space, we simply reason as follows. 
Very roughly, think of the vacuum wavefunction
$\Omega$ as a function $\Omega(\phi_\ell,\phi_r)$ that depends on field variables $\phi_\ell$ in the left half-space and $\phi_r$
in the right half-space. (We schematically write $\phi_\ell$ or $\phi_r$ for all the field variables at $x<0$ or $x>0$.)
The density matrix $|\Omega\ra\la\Omega|$ is as usual a function $|\Omega(\phi_\ell',\phi_r')\ra\la\Omega(\phi_\ell,\phi_r)|$ that
depends on two sets of field variables.  A  partial trace over $\H_\ell$ to get the density matrix $\rho_r$
 is carried out by setting  $\phi_\ell'=\phi_\ell$  and integrating over $\phi_\ell$:
\be\label{rodo}\rho_r(\phi_r',\phi_r)=\int D\phi_\ell  |\Omega(\phi_\ell,\phi_r')\ra\la\Omega(\phi_\ell,\phi_r)|. \ee  

This has a simple path integral interpretation.  The bra $\la\Omega(\phi_\ell,\phi_r)|$ can be computed, as already noted
above, by a path integral on the lower half-space $\tau\leq 0$, and similarly the ket $|\Omega(\phi_\ell',\phi_r')\ra$ can
be computed by a path integral on the upper half-space.  To set $\phi_\ell=\phi_\ell'$, we glue together the portion $x<0$
of the boundaries of the upper and lower half-spaces.   This gluing
gives  the spacetime $W_{2\pi}$ that is sketched in fig. \ref{Fig3}(b).  $W_{2\pi}$ is a copy of Euclidean space
except that it has been ``cut'' along  the half-hyperplane $t=0$, $x>0$. (The reason for the notation $W_{2\pi}$ will be clear in a moment.)
In eqn. (\ref{rodo}), the path integral over the lower half-space to get $\la\Omega|$, the path integral
over the upper half-space to get $|\Omega\ra$, and the final integral over $\phi_\ell$  to take a partial trace
all combine together to make  a path integral over
$W_{2\pi}$.   In this path integral, boundary values $\phi_r$ and $\phi_r'$ are specified just below and above the cut.

To identify the modular operator $\Delta_\Psi$, we would like to give a Hamiltonian interpretation to the path integral in $W_{2\pi}$.
For this, we first consider a path integral on a Euclidean wedge $W_\theta$ of opening angle $\theta$ (fig. \ref{Fig3}(c)).  This
path integral can be viewed as computing an operator.  A matrix element of this operator between initial and final states is computed
by specifying an initial state at the lower boundary of the wedge and a final state at the upper boundary.  The wedge operator
is a Euclidean rotation of the $\tau x$ plane by an angle $\theta$.   Thus, the rotation acts by
\be\label{zett}R_\theta \bp \tau\cr x\ep = \bp \cos\theta & \sin\theta \cr -\sin\theta & \cos\theta\ep\bp \tau\cr x\ep. \ee
To identify in familiar terms the operator that acts in this way in Euclidean signature, let us express the formula in terms of
real time $t=-\i\tau$:
\be\label{mett}R_\theta \bp t\cr x\ep = \bp \cos\theta &-\i \sin\theta \cr -\i\sin\theta & \cos\theta\ep\bp t\cr x\ep=
 \bp \cosh(\i\theta) & -\sinh(\i\theta) \cr- \sinh(\i\theta) & \cosh(\i\theta)\ep\bp t\cr x\ep. \ee

 Looking at the right hand side, we see a Lorentz boost of the $tx$ plane by an imaginary boost parameter $-\i\theta$.  The generator
 of such a Lorentz boost can be written as an integral over the initial value surface $t=0$:
 \be\label{yett} K=\int_{t=0}\d x \,\d\vec y \,\, x\,T_{00}. \ee
 It has been defined to map the right wedge forward in time, and the left wedge backward in time. 
  Formally we can write
 \be\label{plett} K=K_r-K_\ell,\ee
 where $K_r$ and $K_\ell$ are partial Lorentz boost generators
 \begin{align}\label{zone} K_r & =\int_{t=0,x\geq 0}\d x\,\d\vec y \,\,xT_{00}\cr 
                                         K_\ell & =-\int_{t=0,x\geq 0}\d x\,\d\vec y \,\,xT_{00}\cr \end{align}
 The minus sign is included so that $K_\ell$ boosts the left wedge forward in time, just as $K_r$ does to the right wedge.\footnote{Rather
 as there is not a rigorous factorization $\H=\H_\ell\otimes \H_r$,  the operators $K_\ell$ and $K_r$ are not really well-defined as
 Hilbert space operators, though of course the difference $K=K_r-K_\ell$ is a well-defined Hilbert space operator.  $K_\ell$ and
 $K_r$ have well-defined matrix elements $\la\Psi | K_\ell|\chi\ra$ and $\la\Psi|K_r|\chi\ra$ between suitable Hilbert space
 states $\chi$ and $\Psi$, but if one tries to compute the norm of the state $K_\ell|\chi\ra$ or $K_r|\chi\ra$, one will find
 a universal ultraviolet divergence, near $x=0$, independent of the choice of $\chi$.  This is related to the fact that the factorization
 $\H=\H_\ell\otimes \H_r$ is not really correct.}                                         
                                         
 The operator $K$ is self-adjoint, and the unitary operator that implements a Lorentz boost  by a real boost parameter $\eta$
 is $\exp(-\i\eta K)$.  Setting $\eta = -\i\theta$, we learn that, in real time language, the path integral on the wedge $W_\theta$
 constructs the operator $\exp(-\theta K_r)$.  
The path integral on the wedge propagates the degrees of freedom on the right half-space only, so the operator in the exponent
is $K_r$, not $K$. 
  To get the density matrix $\rho_r$ of the right wedge, we set $\theta=2\pi$:
 \be\label{poko} \rho_r=\exp(-2\pi K_r). \ee
 
 A precisely similar analysis shows that the density matrix of the left wedge is
 \be\label{noko}\rho_\ell=\exp(-2\pi K_\ell). \ee
 
 We want to combine these results to determine the modular operator $\Delta_\Omega$ for the vacuum state $\Omega$,
 for the algebra $\A_r$ of observables in the right wedge.  Factoring the Hilbert space as $\H=\H_\ell\otimes \H_r$
 and using eqn. (\ref{nolgo}) (where we identify $\H_r$ and $\H_\ell$ with $\H_1$ and $\H_2$), the modular operator is
 \be\label{polgox} \Delta_\Omega=\rho_r\otimes \rho_\ell^{-1}=\exp(-2\pi K_r)\exp(2\pi K_\ell)=\exp(-2\pi K).  \ee
 In the last step, we use the fact that formally
 the operators $K_r$ and $K_\ell$ commute, since they act respectively on $\H_r$ and $\H_\ell$.
 
 Now let us consider a state $\a|\Omega\ra$ obtained by acting on the vacuum with an operator $\a\in\A_r$, supported on the
 right wedge.    For simplicity, we will assume that a well-defined operator $\a$ can be defined by smearing a local operator $\phi$
 in space with no corresponding smearing in time.    This is so if the dimension of $\phi$, measured in the ultraviolet, is less than
 $(D-1)/2$.   It is not true that the operator product algebra of a quantum field theory is always generated by operators of such
 relatively low dimension, so in general the following discussion has to be modified to allow a very slight smearing in time, but we will
 omit this.
  
  \begin{figure}
 \begin{center}
   \includegraphics[width=5.5in]{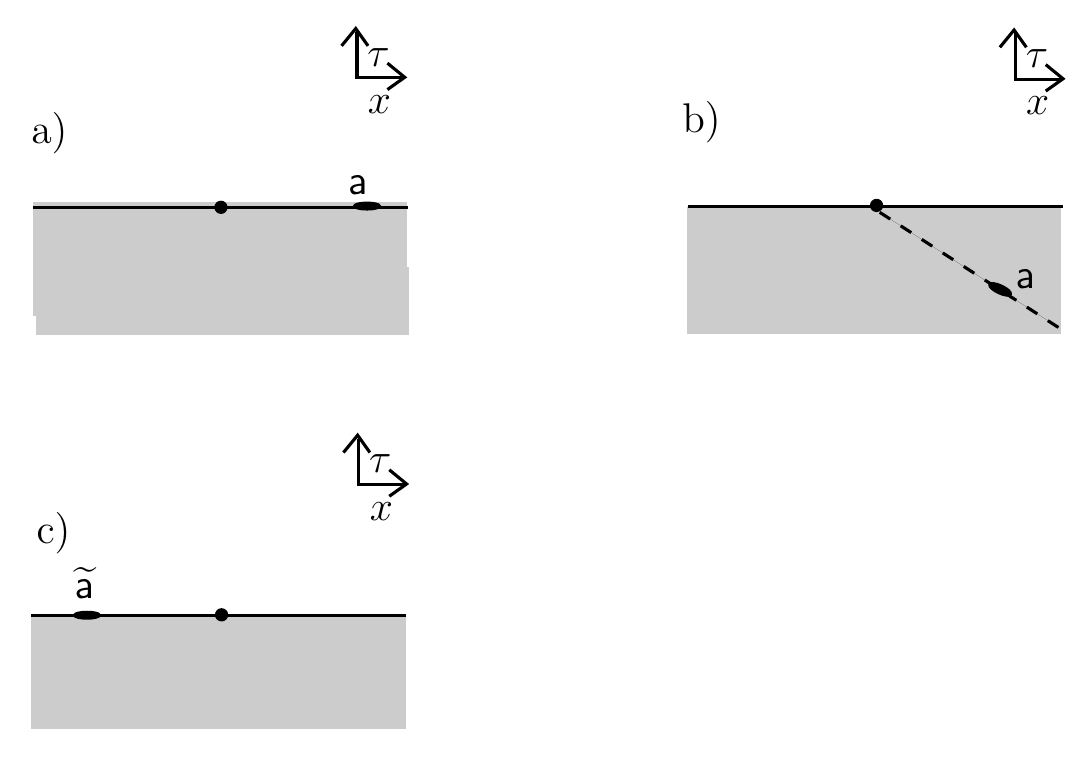}
 \end{center}
\caption{\small   (a) The state $\a|\Omega\ra$ can be obtained by a path integral in the lower half plane, with $\a$ inserted
 on the right half of the boundary.  (b)   Acting with $\exp(2\pi \alpha K_\ell) \exp(-2\pi \alpha K_r) \a|\Omega\ra$
 adds a wedge of opening angle $2\pi\alpha$ to the right boundary and removes one from the left boundary.
 If we rotate the picture so that the boundary is again horizontal, it looks like this; the operator $\a$ is now inserted
 on a ray that is at an angle $2\pi \alpha$ from the horizontal.  (c) By the time we get to $\alpha=1/2$, $\a$ is
 inserted on the left boundary of the lower half plane.  We cannot extend this process farther.\label{Fig4}}
\end{figure}
  
 Under our hypothesis, the state $\a|\Omega\ra$ can be computed by a path integral on the lower half-space, with an insertion of
 the operator $\a$ on the right half of the boundary (fig. \ref{Fig4}(a)).   Now let us consider the state
 \be\label{constate} \Delta_\Omega^\alpha \a|\Omega\ra = \exp(-2\pi \alpha K) \a|\Omega\ra =\exp(2\pi \alpha K_\ell) \exp(-2\pi \alpha K_r) \a|\Omega\ra. \ee
 The operator $\exp(-2\pi \alpha K_r)$ is implemented by gluing on a wedge of opening angle $2\pi \alpha$ to the right half of the boundary
 in fig. \ref{Fig4}(a), while the operator $\exp(2\pi \alpha K_\ell)$ removes such a wedge from the left.
   If we add one wedge and remove the other,
 and also rotate the picture so that the boundary is still horizontal, we arrive at fig. \ref{Fig4}(b).  There is still a path integral on the
 lower half-plane, but now the operator $\a$ is inserted at an angle $-2\pi \alpha$ relative to where it was before.   We can continue
 in this way until we get to $\alpha=1/2$.  This case is depicted in fig. \ref{Fig4}(c).    What at $\alpha=0$ was an operator insertion $\a$
 on the right boundary at $x>0$ has now turned into the insertion of some other operator $\t \a$ on the left boundary at $x<0$.
 As $\t \a$ is inserted on the left boundary, it is an element of the algebra $\A'$.   Thus for $\a\in \A_r$, 
 \be\label{novoc}\Delta_\Omega^{1/2} \a|\Omega\ra = \t \a|\Omega\ra, \ee
 for some $\t \a\in \A_\ell$.  A similar statement holds, of course, with $\A_\ell$ and $\A_r$ exchanged.
 
 We have learned that $\Delta^\alpha \a|\Omega\ra$ is a well-defined Hilbert space state for $0\leq \alpha\leq 1/2$.  But we cannot go farther.
 The operator $\Delta^\alpha$ has removed a wedge of angle $2\pi \alpha$ from the left side of the picture.  By the time we have reached $\alpha=1/2$,
 there is no wedge left to remove on that side and we have to stop.  On the other
 hand, there is no problem in acting on any Hilbert space state with the  unitary operator $\Delta^{\i s}$. 
 So a more general conclusion is that, as was claimed in section \ref{mod group}, $\Delta_\Omega^{\i z}\a|\Omega\ra$ is holomorphic
 in the strip $0>\mathrm{Im}\, z>-1/2$ (and continuous on the boundary of the strip) but not beyond.
 
 Our final goal in this discussion is to determine and exploit
  the  modular conjugation $J_\Omega$.   We will use the fact that $S_\Omega=J_\Omega
 \Delta^{1/2}$ is supposed to satisfy
 \be\label{rovoc} S_\Omega \a|\Omega\ra =\a^\dagger|\Omega\ra,~~\a\in\A_r.\ee
 For simplicity, let us assume that the operator algebra of our theory is generated by a hermitian scalar field $\phi$.   To determine
 what $J_\Omega$ must be, it suffices to consider the case that $\a$ is equal to either $\phi$ or $\dot\phi=\d \phi/\d t$, inserted
 on the right wedge at the initial value surface $t=0$.   Since $\phi$ and $\dot\phi$ are both hermitian, we want
 \be\label{provoc}S_\Omega\phi(0,x,\vec y)|\Omega\ra = \phi(0,x,\vec y)|\Omega\ra, ~~~S_\Omega \dot\phi(0,x,\vec y)|\Omega\ra
 =\dot\phi(0,x,\vec y)|\Omega\ra. \ee    (One could introduce a smearing function in these statements, but this would not change
 what follows.)
 Instead, from eqn. (\ref{novoc}), we have
 \begin{align}\label{lovoc} \Delta_\Omega^{1/2} \phi(0,x,\vec y)|\Omega\ra&=\phi(0,-x,\vec y)|\Omega\ra\cr \Delta_\Omega^{1/2} 
 \dot\phi(0,x,\vec y) |\Omega\ra&=-\dot\phi (0,x,\vec y)|\Omega\ra.\end{align}
 The reason for the minus sign in the second line is that acting with $\Delta_\Omega^{1/2}$ turns a future-pointing time derivative acting
 on $\phi$ in fig. \ref{Fig4}(a) into a past-pointing time derivative in fig. \ref{Fig4}(c), so it reverses the sign of $\d\phi/\d t$.
 Comparing eqns. (\ref{rovoc}) and (\ref{lovoc}), we see that we want
 \be\label{inco} J_\Omega \phi(0,x,\vec y) J_\Omega = \phi(0,-x,\vec y),~~~ J_\Omega\dot\phi(0,x,\vec y)J_\Omega = -\dot\phi(0,-x,\vec y). \ee
 In other words, $J_\Omega$ is supposed to be an antiunitary operator that maps $ x\to -x$, $t\to -t$, $\vec y \to \vec y$.
 
 The antiunitary operator that acts in this way on any hermitian scalar field (with an analogous action on fields of other types)
 is the operator $\CRT$ that was discussed in section \ref{preover}.  Thus
 \be\label{purinco}J_\Omega=\CRT.\ee
 
 Perhaps we should just pause a moment to explain more explicitly why this operator is traditionally called $\CRT$ rather than $\sf{RT}$.
 Consider a theory with two hermitian scalar fields $\phi_1$ and $\phi_2$ rotated by an $SO(2)$ symmetry with generator
 \be\label{zurinco} Q=\int_{t=0}\d x\,\d\vec y\left(\phi_1\dot \phi_2-\dot\phi_1\phi_2\right). \ee
 This charge is odd under $J_\Omega$, since $\phi_1$ and $\phi_2$ are even while $\dot\phi_1$ and $\dot\phi_2$ are odd.
 So $J_\Omega$ reverses the sign of $Q$, and similarly of any other hermitian conserved charge.  Since $\sf R$ and $\sf T$ are
 traditionally defined to commute with Lorentz-invariant conserved charges while $J_\Omega$ reverses their sign, $J_\Omega$
 corresponds to what is traditionally called $\CRT$ rather than $\sf{RT}$.    $\CRT$ is a universal symmetry of
 relativistic quantum field theory, while there is no universal symmetry corresponding to $\sf{RT}$.
 
 In this example, we can explicitly verify the deeper properties of the modular automorphisms $\Delta_\Omega^{\i s}$ and $J_\Omega$
 that were described in section \ref{mod group}.  $\Delta_\Omega^{\i s}$ implements a Lorentz boost with a real boost parameter
 $2\pi s$, so it is an automorphism of the algebras $\A_\ell$ and $\A_r$ of the two wedges.  And $J_\Omega=\CRT$ exchanges the two
 wedges so it exchanges the two algebras.
 
 In general, in Tomita-Takesaki theory, the modular conjugation $J_\Omega$ exchanges an algebra $\A$ with its commutant $\A'$.
 So in the present context, the fact that $J_\Omega$ exchanges $\A_\ell$ and $\A_r$ tells us that these algebras are commutants:
 \be\label{mytz}\A_\ell'=\A_r,~~~\A_r'=\A_\ell. \ee
 This is how Bisognano and Wichmann \cite{BiWi} proved Haag duality for complementary Rindler spaces.

\subsection{The Approach Of Bisognano and Wichmann}\label{bw}

The path integral derivation of the last section is extremely illuminating, and it gives the right result though it is
not altogether rigorous. (The flaws all involve an imprecise treatment of the boundary between the two regions at $x=0$.)
Here, following the presentation by Borchers \cite{Borchers}, we very briefly sketch the original approach of
Bisognano and Wichmann \cite{BiWi}.  The main difference is that instead of a Euclidean path integral and a claimed factorization
$\H=\H_\ell\otimes\H_r$, one uses holomorphy.

Since $J_\Omega=\CRT$ certainly acts as in eqn. (\ref{inco}), to determine $\Delta_\Omega$ and $S_\Omega$,
we have to justify the claim that for $\a\in\A_r$,
\be\label{zornoff}\exp(-2\pi K) \a|\Omega\ra =\t \a|\Omega\ra, \ee
where $\t \a$ is obtained from $\a$ by $t,x,\vec y\to -t,-x,\vec y$.
In checking this, we can take $\a$ to be  a product of field operators
\be\label{wornoff} \a=\phi(t_1,x_1,\vec y_1)\phi(t_2,x_2,\vec y_2)\cdots \phi(t_n,x_n,\vec y_n) \ee
inserted in the right wedge $\U_r$ at points $p_i=(t_i,x_i,\vec y_i)$, $i=1,2,\cdots ,n$.   Moreover, we can take the points $p_i$ to be spacelike separated from each other; as the field operators $\phi(t_i,x_i,\vec y_i)$ thereby commute,
we can order them so that $x_j\geq  x_i$ for $j>i$.  Even more specifically, we can restrict to
\be\label{yrf} x_j-x_i>|t_j-t_i|,~~j>i. \ee It suffices to consider operators $\a$ of this form roughly because
 states $\a|\Omega\ra$ with $\a$ of this type are dense\footnote{One can see this by reviewing the proof
 of the Reeh-Schlieder theorem from section \ref{proof}.  The proof would go through perfectly well if one begins
 by assuming only that the functions $\varphi(x_1,x_2,\dots,x_n)=\langle\chi|\phi(x_1)\phi(x_2)\cdots \phi(x_n)|\Omega\rangle$ (eqn. \ref{goodness})
 vanish under the hypothesis (\ref{yrf}); one can still prove in the same way that these functions vanish identically for all
 $x_1,x_2,\dots,x_n$.} in $\H$, so in
particular they are dense among all states $\a|\Omega\ra$, $\a\in \A_r$.   For a precise statement, see Lemma 3.1.7 in \cite{Borchers}.

For real $s$, the Lorentz boost operator $\exp(-2\pi \i s K)$ is unitary and its action on a state $\a|\Omega\ra$ is straightforward to determine.
The normal coordinates $\vec y$ play no role in what follows so we omit them to simplify the notation. 
A Lorentz boost $\exp(-2\pi\i s K)$ maps $\x =\bp t\cr x\ep$ to 
\be\label{bornoff}\x'(s)=\bp t'(s)\cr x'(s)\cr\ep = \bp \cosh(2\pi s) & \sinh (2\pi s)\cr \sinh(2\pi s) & \cosh(2\pi  s)\ep \bp t\cr x\ep .\ee
The corresponding transformation of operators in the Heisenberg picture is\
 \be\label{gerf}\phi(\x(\eta)) = \exp(2\pi \i\eta K)\phi(\x)\exp(-2\pi \i\eta K).\ee
So for real $\eta$, remembering that $K\Omega=0$,
\be\label{wortoff} \exp(2\pi \i \eta K)\phi(\x_1)\phi(\x_2)\cdots \phi(\x_n)|\Omega\ra =\phi(\x'_1(\eta))\phi(\x'_2(\eta))\cdots \phi(\x'_n(\eta))|\Omega\ra. \ee
We would like to analytically continue this formula in $\eta$.  If it can be continued to $\eta=\i/2$, then, since 
$\x'(i/2)=-\x$, eqn. (\ref{wortoff}) will give the desired result (\ref{zornoff}).

In section \ref{proof}, we learned that the $\H$-valued function 
\be\label{stav} F(\x'_1,\x'_2,\cdots,\x'_n)=\phi(\x'_1)\phi(\x'_2)\cdots \phi(\x'_n)|\Omega \ra\ee
is holomorphic in $\x'_1,\dots,\x'_n$ in a certain domain. To be precise,
if $\x'_i=\u_i+\i \v_i$ with real $\u_i, \,\v_i$, then $F(\x'_1,\x'_2,\cdots,
\x'_n)$ is holomorphic in the domain in which $\v_1$ and $\v_{i+1}-\v_i$ are future timelike.    

We claim that if the points $\x_1,\x_2,\cdots,\x_n$ are chosen as in eqn. (\ref{yrf}), then for $1/2>\mathrm{Im}\,\eta>0$, the points $\x'_1(\eta),\x'_2(\eta),\cdots,\x'_n(\eta)$ are in the domain of
holomorphy that was just described.  Since this statement is manifestly invariant
under real Lorentz boosts, it suffices to verify it for imaginary $\eta$, say $\eta=\i b$, $0<b<1/2$.  Let $\x$ be either $\x_1$ or one of the
  differences 
$\x_{i+1}-\x_i$.   Our assumptions imply in each case that $\x$ is in the right wedge $x>|t|$. We have to show
that  the imaginary part of $\x'(\eta)$, defined in eqn. (\ref{bornoff}) (with $s$ replaced by $\eta=\i b$),
is future timelike for the claimed range of $b$.   We compute
\be\label{zoffo} \bp t'(\eta)\cr x'(\eta)\ep=\bp t\cos 2\pi b +\i x\sin 2\pi b\cr x \cos 2\pi b +\i t\sin 2\pi b\, \ep.\ee
Since $x>|t|$, the imaginary part is future timelike for $0<b
<1/2$, which ensures that $\sin 2\pi b>0$. The $\H$-valued function
on the right hand side of eqn. (\ref{wortoff}) is thus holomorphic for $1/2>\mathrm{Im}\,\eta>0$, and continuous up to the boundary at $\mathrm
{Im}\,\eta=1/2$. (It cannot be continued
holomorphically beyond that.)   This is precisely enough to justify setting $\eta=\i/2$ in eqn. (\ref{wortoff}), and thus to complete the proof.

\subsection{An Accelerating Observer}\label{unr}

\begin{figure}
 \begin{center}
   \includegraphics[width=2.9in]{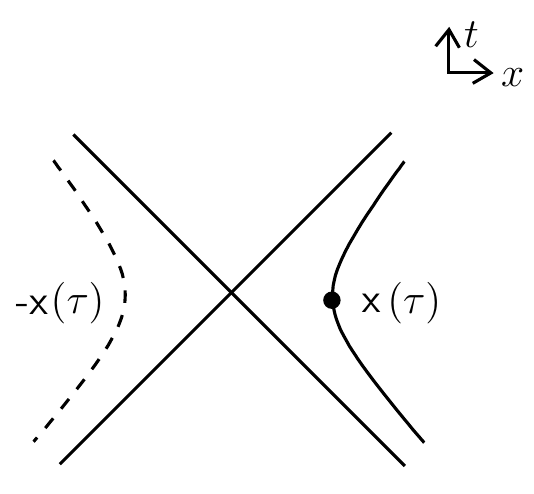}
 \end{center}
\caption{\small An accelerating trajectory ${\sf x}(\tau)$  in the right quadrant of the  $xt$ plane. The point $\tau=0$ is marked.
  Shown in dotted lines, on the left, is
the mirror trajectory $-{\sf x}(\tau)$, which can be obtained from the first by a shift in imaginary time.   The two trajectories are spacelike separated.    \label{Fig8}}
\end{figure}

The problem we have been discussing is closely related to Unruh's  question \cite{Unruh} of what is seen by an  observer undergoing constant acceleration in Minkowski
spacetime, say 
in the $xt$ plane.  The worldline of the observer (fig. \ref{Fig8}) is
\be\label{ombo}\begin{pmatrix} t(\tau)\cr x(\tau)\end{pmatrix} = R\begin{pmatrix}\sinh (\tau/R)\cr \cosh (\tau/R)\end{pmatrix}, \ee
where $\tau$ is the observer's proper time; the proper acceleration is $a=1/R$.   As before, we abbreviate $\begin{pmatrix} t(\tau)\cr x(\tau)\end{pmatrix}$ as $\x(\tau)$.

We suppose that the observer probes the vacuum $\Omega$ of Minkowski spacetime by measuring a local operator $\O$ and its adjoint $\O^\dagger$ along this worldline.
For simplicity, we consider only the two-point functions $\O\cdot \O^\dagger$, but we will consider both operator orderings.  
Thus,  we suppose that the observer has access to $\la\Omega|\O(\x(\tau_1))\O^\dagger(\x(\tau_2))|\Omega\ra$ and $\la\Omega|\O^\dagger(\x(\tau_2))
\O(\x(\tau_1))|\Omega\ra$.   Lorentz invariance implies that these functions depend only on $\tau=\tau_1-\tau_2$, so there is no essential loss to set $\tau_2=0$ and
to consider the two functions:
\begin{align}\label{zza} F(\tau) & = \la\Omega| \O(\x(\tau))\O^\dagger(\x(0))|\Omega\ra \cr
                                      G(\tau)& = \la\Omega|\O^\dagger(\x(0))\O(\x(\tau))|\Omega \ra .\end{align}
 
 Unruh's basic insight was that these correlation functions have thermal properties.  The basic property of real time two-point
 functions in a thermal ensemble, as we already explained in   eqns. (\ref{zormo}) and (\ref{tufo}), is that there is a holomorphic function on a strip in the complex
 plane whose boundary values on the two boundaries of the strip are $F(\tau)$ and $G(\tau)$.   In general, the width of the strip is $2\pi\beta$, where $\beta$ is
 the inverse temperature; in the derivation of eqns. (\ref{zormo}) and (\ref{tufo}), we took $\beta=1$ so the width of the strip was $2\pi$.  We will give two
 derivations of Unruh's result, first starting in real time and deducing the holomorphic properties of the correlation functions, and second starting in Euclidean
 signature and analytically continuing back to real time.

To understand the analytic properties of the real time correlation functions, we first analytically continue the observer's trajectory.
We set $\tau/R = s+\i \theta$ with real $s,\theta$ and compute that
\be\label{wombo} \x(\tau)=R\begin{pmatrix} \sinh s \cos \theta +\i \cosh s \sin\theta  \cr \cosh s \cos\theta+\i \sinh s\sin \theta\end{pmatrix}. \ee                                       
Thus 
\be\label{rombo}\mathrm{Im}\,\x(\tau)=R\sin\theta\begin{pmatrix} \cosh s  \cr  \sinh s\end{pmatrix}. \ee      
$F(\tau)$ is holomorphic when $\mathrm{Im}\,\x(\tau)$ is future timelike and $G(\tau)$ is holomorphic when $\mathrm{Im}\,\x(\tau)$ is past timelike.
So $F(\tau)$ is holomorphic in the strip $0<\theta <\pi$ and continuous on the boundaries of that strip;
we describe this more briefly by saying that $F(\tau)$ is holomorphic in the strip $0\leq \theta\leq \pi$.  Similarly   $G(\tau)$ is holomorphic in the strip $\pi\leq \theta \leq
2\pi$
(or equivalently but less conveniently $-\pi\leq \theta\leq 0$).  

In terms of $\tau$, $F(\tau)$ is holomorphic for $0\leq \mathrm{Im}\,\tau\leq \pi R$.   At $\mathrm{Im}\,\tau=0$, $F(\tau)$ is simply  the original correlation function
$ \la\Omega| \O(\x(\tau))\O^\dagger(\x(0))|\Omega\ra $ on the observer's worldline.    On the other boundary of the strip at $\mathrm{Im}\,\tau =\pi R$, $\x(\tau)$ is again real:
\be\label{omob}\x(\tau+\i\pi R) = -\x(\tau)= - R\begin{pmatrix}\sinh (\tau/R)\cr \cosh (\tau/R)\end{pmatrix}. \ee
So  the boundary values at $\tau=R(s+\i\pi)$ are
\be\label{tromob} F( R(s+\i \pi)) = \la\Omega| \O(-\x(Rs))\O^\dagger(\x(0))|\Omega\ra. \ee

Similarly, $G(\tau)$ at $\mathrm{Im}\,\tau=2\pi R$ is simply the original correlation function  $ \la\Omega|\O^\dagger(\x(0))\O(\x(\tau))|\Omega \ra $ on the observer's worldline.
But at $\mathrm{Im}\,\tau=\pi R$, we get, similarly to (\ref{tromob}),
\be\label{plomob} G(R(s+\i\pi)) = \la\Omega|\O^\dagger(\x(0)) \O(-\x(Rs))|\Omega\ra.   \ee

Crucially, the operators $\O(-\x(Rs))$ and $\O^\dagger(\x(0))$ commute, 
since for all real $s$, $-\x(Rs)$ is spacelike separated from $\x(0)$ (see fig. \ref{Fig8}).  
So the correlation
functions in (\ref{tromob}) and (\ref{plomob}) are equal.  

Thus, we have one function $F(\tau)$ that is holomorphic for $\pi R\geq \mathrm{Im}\,\tau\geq 0$ and another function $G(\tau)$ that is holomorphic
for $2\pi R\geq \mathrm{Im}\,\tau\geq \pi R$; moreover at $\mathrm{Im}\,\tau=\pi R$, these two functions are equal.   It follows that we can define a single
function $H(\tau)$ on the combined strip $2\pi R \geq \mathrm{Im}\,\tau \geq 0$ by
\be\label{sinfn} H(\tau)=\begin{cases} F(\tau) & {\mathrm {if}}~ \pi R\geq \mathrm{Im}\,\tau \geq 0 \cr
 G(\tau) & {\mathrm{ if}}~ 2\pi R\geq \mathrm{Im}\,\tau \geq \pi R.\end{cases}\ee
 This function is holomorphic in the combined strip and continuous on its boundaries.   (For the proof of holomorphy on the line
 $\mathrm{Im}\,\tau =\pi R$ where the two functions were glued together, see fig. \ref{Fig6} in Appendix \ref{holoproof}.)   The boundary values at the top and bottom
 boundaries of the strip are the two correlation functions that we started with, with the two possible operator orderings.  
 
We have arrived  at
   the usual analytic behavior of a real time two-point correlation function in a thermal ensemble: two-point functions with different operator
  ordering are opposite boundary values of a single function that is holomorphic in a strip.  
We  have found a strip of width $2\pi R$, so the effective temperature is $1/2\pi R$.
 
A derivation that  begins with the Euclidean correlation functions might be more transparent.  Let $t_E=\i t$ be the Euclidean time.  
A Euclidean version of eqn. (\ref{ombo}) is
\be\label{zlombo} t_E= R\sin\theta,~~~x = R\cos\theta.\ee
This is the thermal circle that is related to the observations of the accelerated observer.
Let $\x_E=\begin{pmatrix} t_E\cr x \end{pmatrix}$.
In Euclidean space, one considers the correlation function  $\la \O(\x_E(\theta))\O^\dagger (\x_E(0))\ra$.   {\it A priori}, a Euclidean correlation function has
no operator interpretation.  To introduce an operator interpretation, one picks a direction as Euclidean time and introduces a transfer matrix that propagates
operators in that direction.   Then Euclidean correlation functions acquire an operator interpretation, with the operators being ordered in the direction of increasing
Euclidean time.  For example, if $t_E$ is chosen as the Euclidean time direction, then a general Euclidean two-point function is interpreted in the transfer
matrix formalism as 
\be\label{niombo}\la \O(t_E,x) \O^\dagger(t_E',x')\ra =\begin{cases} \la\Omega|\O(t_E,x)\O^\dagger(t_E',x')|\Omega\ra ~ {\mathrm{if}}~ t_E\geq t'_{E}\cr 
 \la\Omega|\O^\dagger(t_E',x')\O(t_E,x)|\Omega\ra ~ {\mathrm{if}}~ t_E'\geq t_{E}.\end{cases}\ee
 As before, this is consistent because if $t_E=t_E'$, the operator ordering does not matter.  
Given this, the operator ordering in the operator interpretation of the Euclidean correlation function $\la \O(\x_E(\theta))\O^\dagger (\x_E(0))\ra$ depends
on the sign of $t_E=R\sin\theta$, as in the above derivation.  
When we analytically continue $\la \O(\x_E(\theta))\O^\dagger (\x_E(0))\ra$ from a function of $\theta$ to a function of $\tau=R(s+\i\theta)$, 
we get the two operator orderings depending on the sign of $\sin\theta$, as above.  This distinction remains in the limit $\theta\to 0^\pm$, where
we recover the real time correlation functions with different operator orderings.

\section{Algebras With a Universal Divergence In The Entanglement Entropy}\label{algex}

\subsection{The Problem}\label{firstpre}

Let $\U$ be an open set in Minkowski spacetime.  It has a local algebra $\A=\A_\U$  with commutant $\A'$ (which, if Haag duality holds,
is  $\A_{\U'}$ for some other open set $\U'$).  
  As in section \ref{furtherp}, we understand
$\A$ and $\A'$ to be von Neumann algebras of bounded operators (closed under hermitian conjugation and weak limits, and containing
the identity operator).  
They act on the Hilbert space $\H$ of the theory in question with
the vacuum state $\Omega$ as a cyclic separating vector.

For a finite-dimensional quantum system, the existence of such a cyclic separating vector would imply a factorization $\H=\H_1\otimes\H_2$,
with $\A$ acting on one factor and $\A'$ on the other.
Such a factorization cannot exist in quantum field theory, for it would imply the existence of tensor product states $\psi\otimes\chi$ with
no entanglement between $\U$ and $\U'$.   Instead, in quantum field theory, there is a universal ultraviolet divergence in the entanglement
entropy.

The essence of the matter is that in quantum field theory, the leading divergence in the entanglement entropy is not a property of the states
but of the algebras $\A$ and $\A'$.  These algebras are not the familiar Type I von Neumann algebras which can act irreducibly in
a Hilbert space.  Instead they are more exotic algebras with the property that the structure of the algebra has the  leading
divergence in the entanglement entropy built in.  In this section, we explain barely enough about von Neumann algebras
to indicate how that comes about.

\subsection{Algebras of Type I}\label{type1}

A Type I von Neumann algebra  $\A$ can  act {\it irreducibly} by bounded operators on a Hilbert space $\K$.
We will only be interested  here in algebras that have trivial centers (consisting only of complex scalars).\footnote{A von Neumann algebra with trivial
center is called a factor.  Factors exhibit the main subtleties of von Neumann algebras, and von Neumann algebras that are not factors
are built from factors in a relatively simple way.  So it is natural to concentrate on factors here.}  Under
this restriction, $\A$ will actually consist of all bounded operators on $\K$.   We also will only consider Hilbert spaces of at most countably
infinite dimension.

If $\K$ has finite dimension $d$, then all operators on $\K$ are bounded.
We say that the algebra  of operators on $\K$ is of type 
 $\I_d$.   If $\K$ is infinite-dimensional, we call the algebra  of bounded operators on $\K$
 an algebra of type $\I_\infty$.  A von Neumann algebra (with trivial center) acting irreducibly on a Hilbert space 
 is always of one of these two types.

A ``trace'' on a von Neumann algebra is a linear function $\a\to \Tr\,\a$ that satisfies $\Tr\,\a\b=\Tr\,\b\a$ and $\Tr\,\a^\dagger \a>0$ for
$\a\not=0$.    Obviously, an algebra of Type $\I_d$ has a trace.  For Type $\I_\infty$, we can define a trace that has the
right properties except that it cannot be
defined on the whole algebra as it may diverge; for instance,
 the trace of the identity operator on an infinite-dimensional Hilbert space is $+\infty$.

In constructing more exotic algebras, we are interested in algebras that can be constructed as limits of matrix algebras. (Such
algebras are called hyperfinite.)
Such constructions were introduced and developed by von Neumann \cite{vonN}, Powers \cite{Powers}, and Araki and Woods \cite{ArakiWoods}.

\subsection{Algebras of Type II}\label{type2}

The first nontrivial example is the hyperfinite Type $\II_1$ factor of Murray and von Neumann.  It can be constructed
as follows from a countably infinite set of maximally entangled qubit pairs.

Let $V$ be a vector space consisting of $2\times 2$ complex matrices, with Hilbert space structure defined by
$\la v,w\ra=\Tr\, v^\dagger w.$
Let $M_2$ and $M_2'$ be two copies of $\I_2$,
the algebra of $2\times 2$ complex matrices.    We let $M_2$ and $M_2'$ act on $V$ on the left and right,
respectively.  Thus $a\in M_2$ acts on $v\in V$ by $v\to av$, and $a'\in M_2'$ acts on $v$ by $v\to v a'{}^\tr$ where $\tr $ is
the transpose.  Obviously, $M_2$ and $M_2'$ are commutants.

We can view $V$ as a tensor product $W\otimes W'$, where $W$ is a space of two-component column vectors acted on by $M_2$
and $W'$ is a space of two-component row vectors acted on by $M_2'$.
Thus $V$ is a bipartite quantum system.  Let $I_2$ be the $2\times 2$ identity matrix. A normalized
maximally entangled  vector in $V$ is given by $I_2'=\frac{1}{\sqrt 2}I_2$.

Now consider a countably infinite set of copies of this construction; thus, for $k\geq 1$, let $V^\kk$ be a space of $2\times 2$
matrices acted on on the left by $M_2^\kk$ and on the right by $M_2'{}^\kk$.

Roughly speaking, we want to consider the infinite tensor product $V^\oone\otimes V^\ttwo \otimes \cdots\otimes V^\kk\otimes \cdots $.
However, taken literally, this infinite tensor product is a vector space of uncountable dimension.  To get a Hilbert space of
countably infinite dimension, we instead proceed as follows.  To start with, we define a space $\H_0$ that consists of tensor
products
\be\label{dofo}v_1\otimes v_2\otimes \cdots \otimes v_k\otimes \cdots \in V^\oone\otimes V^\ttwo \otimes \cdots \otimes V^\kk\otimes \cdots \ee
such that all but finitely many of the $v_k$ are equal to $I'_2$.  This gives a countably  infinite-dimensional
vector space, but not yet a Hilbert space.  To make a Hilbert space, we first define an inner product on $\H_0$.  This is done as follows.
If $v=v_1\otimes v_2\otimes \cdots$ and $w=w_1\otimes w_2\otimes \cdots$ are elements of $\H_0$, then there is some $n$
such that $v_k$ and $w_k$ both equal $I_2'$ for $k>n$.  We truncate $v$ and $w$ at $v_{\dn}=v_1\otimes v_2\otimes \cdots \otimes v_n$,
$w_\dn=w_1\otimes w_2\otimes \cdots \otimes w_n$, and define
\be\label{york}\la v,w\ra= \Tr\,v_\dn^\dagger w_\dn. \ee
This does not depend on where the truncation was made.
  Having defined a hermitian inner product on $\H_0$,
we complete it to get a Hilbert space $\H$, which is called a restricted tensor product of the $V^\kk$.  For
$v_1\otimes v_2\otimes \cdots \otimes v_n\otimes \cdots$ to be a vector in the restricted tensor product, the $v_n$ must
tend rapidly to $I_2'$ for $n\to\infty$.

We do something similar with the algebras.  Roughly speaking, we want to define an algebra $\A$ as an infinite tensor product
$M_2^\oone\otimes M_2^\ttwo \otimes \cdots\otimes M_2^\nn\otimes\cdots$.    However, a general element
${\a}=a_1\otimes a_2\otimes\cdots\otimes a_n\otimes\cdots  $ cannot act on the restricted tensor product $\H$.  (Acting on
$v_1\otimes v_2\otimes \cdots \otimes v_n\otimes \cdots$, it would not preserve the condition that the $v_n$ go rapidly to $I_2'$
for $n\to\infty$.)  To get around this, we first define an algebra $\A_0$ that consists of elements 
${\a}= a_1\otimes a_2\otimes\cdots\otimes a_n\otimes\cdots  $ such that all but finitely many of the $a_i$ are equal to $I_2$.  This
algebra acts on $\H$, and it obeys all the conditions of a von Neumann algebra except that it is not closed.  To make it closed
we add limits.  We say that a sequence ${\a}_{(k)}\in \A_0$ converges if 
$\lim_{n\to\infty} {\a}_{(n)}\chi$ exists for all $\chi\in \H$; if so, we define an operator $\a:\H\to \H$ by
${\a}\chi=\lim_{n\to\infty}{\a}_{(n)}\chi$, and we define $\A$ to include all such limits.  This definition ensures that
for $\a\in \A$, $\chi\in\H$, $\a\chi$ is a continuous function of $\a$.   Note that the definition of $\A$ depends
on a knowledge of the Hilbert space that it is supposed to act on, which entered the question of which sequences 
${\a}_{(n)}$ converge.
This will be important in section \ref{type3}.

The commutant of $\A$ is an isomorphic algebra $\A'$ that is defined in just the same way, as a subalgebra
of $M_2'{}^\oone\otimes M_2'{}^\ttwo\otimes \cdots \otimes M_2'{}^\nn\otimes \cdots$.

The vector
\be\label{torno}\Psi= I_2'\otimes I_2'\otimes \cdots \otimes I_2'\otimes \cdots \in \H\ee
is cyclic separating for $\A$ and for $\A'$.   (To show that ${\a}\Psi\not=0$ for any nonzero $\a\in\A$,
we approximate $\a$ by a linear combination of tensor products $a_1\otimes a_2\otimes\cdots \otimes a_n\otimes \cdots$, where in each term $a_n=I_2$ for sufficiently large $n$,
and observe that a nonzero element of this kind certainly does not annihilate $\Psi$.)

\def\sTr{{\mathrm{Tr}}}
A natural linear function on the algebra $\A$ is defined by $F(\a)=\la\Psi|\a|\Psi\ra$.    Since $\Psi$ is separating for $\A$,
any nonzero $\a\in\A$ satisfies $\a\Psi\not=0$ and 
 hence  $F(\a^\dagger \a)>0$.   We claim that the function $F$ has the defining property of a trace: $F(\a\b)=F(\b\a)$. Indeed,
if ${\a}=a_1\otimes a_2\otimes \cdots \otimes a_n\otimes \cdots $, ${\b}=b_1\otimes b_2\otimes \cdots \otimes b_n\otimes \cdots$
with $a_n,b_n=I_2 $ for $n>k$, then 
\be\label{burop}F(\a\b)=\Tr_{M_2^\oone\otimes M_2^\ttwo \otimes \cdots\otimes M_2^\kk} a_1b_1\otimes a_2b_2
\otimes \cdots \otimes a_kb_k=F(\b\a). \ee
Since elements $\a,\b$ of the form just considered are dense in $\A$, the general result $F(\a\b)=F(\b\a)$ follows by taking limits, given
the way that $\A$ was defined.   Since the function $F(\a)$ has the properties of a trace,
we denote it as $\Tr\,\a$.

We recall that in the case of a Type $\I_\infty$ algebra, one can define a trace on a subalgebra but the trace of the identity element
is infinite.  By contrast, a  hyperfinite Type $\II_1$ algebra has a trace that is defined on the whole algebra, and which we have normalized
so that $\Tr\,1=1$.

Obviously, the entanglement entropy in the state $\Psi$ is infinite,
since each factor of $I_2'$ represents a perfectly entangled qubit pair shared between $\A$ and $\A'$.
Replacing $\Psi$ by another vector in $\H$ will  only change the entanglement entropy by a finite or at least less divergent amount,
because of the way the restricted tensor product was defined.   So the leading divergence in the entanglement entropy is universal,
as in quantum field theory.

  Another fundamental fact -- more or less equivalent to the universal divergence in the entanglement entropy -- is that
the Type $\II_1$ algebra  $\A$ has no irreducible representation.  

$\A$ acts on the Hilbert space $\H$ that we have constructed, but  this
action is far from irreducible, as it commutes with the action of $\A'$ on the same space.
We can make a smaller representation of $\A$ by projecting 
$\H$ onto an invariant subspace.  Set $J_2=\bp 1&0\cr 0&0\ep$
and consider the following element of $\A'$:
\be\label{mork} \Pi_k '=J_2\otimes J_2\otimes \cdots \otimes J_2\otimes I_2\otimes I_2\otimes \cdots \ee
with precisely $k$ factors of $J_2$ and the rest $I_2$.  This 
is a projection operator with\footnote{More generally, for every real $x$ with $0\leq x\leq 1$, $\A'$ has
a projection operator $\Pi'_x$ with $\sTr\,\Pi'_x=x$.  Projecting on the image 
of $\Pi'_x$ (acting on $\H$ on the right) gives a representation of $\A$ whose
``dimension'' in the sense of Murray and von Neumann is $x$.} $\sTr\,\Pi_k' =2^{-k}$.  The subspace $\H \Pi'$ of $\H$
(that is, the set of all elements of $\H$ of the form $\chi \Pi'$ for some $\chi\in\H$)
is a representation of $\A$ that, in a sense that was made precise by 
Murray and von Neumann, is smaller by a factor of $2^k$.
We can keep going and never get to an irreducible representation.  Concretely, $\Pi'_k$ projects onto vectors
$v_1\otimes v_2\otimes \cdots \otimes v_n\otimes \cdots 
 \in \H$ such that $v_1,v_2,\cdots, v_k$ are of the form $\bp s & 0\cr t & 0\ep$.  To get an 
irreducible representation of $\A$, we would have to impose such a condition on  $v_n$ for all $n$, 
but an infinite tensor product of vectors of this type is
not in $\H$.

The Type $\II_1$ algebra that we have considered has some properties in common with local algebras 
in quantum field theory -- they share a
 universal divergence in the entanglement entropy and the absence of an irreducible
representation.  But local algebras in quantum field theory do not possess a trace.

\subsection{Algebras of Type III}\label{type3}

More general algebras can be constructed by proceeding similarly, but with reduced entanglement.

For $0<\lambda<1$, define a matrix 
\be\label{morno} K_{2,\lambda}=\frac{1}{(1+\lambda)^{1/2}}\bp 1&0\cr 0&\lambda^{1/2} \ep.\ee
This matrix describes a pair of qubits with nonzero but also nonmaximal entanglement.  (We sometimes include the 
case $\lambda=1$; note  that $K_{2,1}$ is the matrix $I'_2$ of section \ref{type1}.)

In the construction of a Hilbert space $\H$ in section \ref{type2}, replace $I_2'$ everywhere by $K_{2,\lambda}$.
Thus, consider the space $\H_0$ spanned by vectors $v_1\otimes v_2\otimes \cdots \otimes v_n\otimes \cdots\in
V^\oone\otimes V^\ttwo \otimes \cdots \otimes V^\nn\otimes \cdots$ such that all but finitely many of the
$v_n$ are equal to $K_{2,\lambda}$.  Define $\H_\lambda$ to be the Hilbert space closure of $\H_0$.
Similarly, to define an algebra, start with the same $\A_0$ that we used in section \ref{type2}, and take
its closure in the space of bounded operators acting on $\H_\lambda$.  This gives a von Neumann algebra
$\A_\lambda$.  $\A_\lambda$ differs from the algebra $\A$ constructed in section \ref{type2} because the Hilbert
space $\H_\lambda$ differs from the Hilbert space $\H$ of that section.  In other words, the condition for a sequence
of operators $\a_n\in\A_0$ to converge depends on which vectors the $\a_n$ are supposed to act on, so it
depends on the choice of the matrix $K_{2,\lambda}$.

Again, the commutant $\A'_\lambda$ is defined similarly and is isomorphic to $\A_\lambda$.
The vector $\Psi=K_{2,\lambda}\otimes K_{2,\lambda}\otimes \cdots \otimes K_{2,\lambda}\otimes \cdots$
is cyclic and separating for $\A_\lambda$ and for $\A'_\lambda$.   The corresponding function 
$F(\a)=\la\Psi|\a|\Psi\ra$ does not satisfy $F(\a\b)=F(\b\a)$, and indeed the algebra $\A_\lambda$
does not admit a trace.   

The entanglement entropy betwen $\A_\lambda$ and $\A'_\lambda$ in the state $\Psi$ is divergent,
because $\Psi$ describes an infinite collection of qubit pairs each with the same entanglement.  As
in section \ref{type2},  this divergence is universal; any state in $\H_\lambda$ has the same leading divergence
in the entanglement entropy.

As in section \ref{type2}, the action of $\A_\lambda$ on $\H_\lambda$ is far from irreducible; it can be decomposed
as finely as one wishes 
using projection
operators in $\A'_\lambda$. 
In this case, however, though we will not prove it, the invariant subspaces in which $\H_\lambda$ can be decomposed
are isomorphic, as representations of $\A_\lambda$, to $\H_\lambda$ itself:
a hyperfinite von Neumann algebra of Type III  has only one nontrivial
 representation, up to isomorphism.   All statements in the last three paragraphs also apply to the additional  Type III algebras  that we come to momentarily.

 Powers 
\cite{Powers} proved that $\A_\lambda$ and $\A_{\t\lambda}$ for $\lambda\not=\t\lambda$ are nonisomorphic.
Araki and Woods \cite{ArakiWoods} considered a generalization of this construction involving a sequence
$\lambda_1,\lambda_2,\cdots$, $0<\lambda_i\leq 1$.     Now one considers vectors
$v_1\otimes v_2\otimes \cdots \otimes v_n\otimes \cdots\in
V^\oone\otimes V^\ttwo \otimes \cdots \otimes V^\nn\otimes \cdots$  such that $v_n=K_{2,\lambda_n}$ for
all but finitely many $n$.  Such vectors make a vector space $\H_{0,\vec\lambda}$ whose Hilbert space closure
 gives a Hilbert space $\H_{\vec\lambda}$.   To construct an algebra $\A_{\vec \lambda}$, we start
with the same algebra $\A_0$ as before, and take its closure in the space of bounded operators on
$\H_{\vec\lambda}$.   The commutant $\A'_{\vec\lambda}$ is constructed similarly, and
\be\label{murfy}\Psi_{\vec\lambda}=K_{2,\lambda_1}\otimes K_{2,\lambda_2}\otimes \cdots \otimes K_{2,\lambda_n}\otimes \cdots \ee
is a cyclic and separating vector for this pair of algebras. (The expectation $\la\Psi|\a|\Psi\ra$ is not a trace
unless the $\lambda_i$ are all 1.)

Araki and Woods \cite{ArakiWoods}
showed that if the sequence $\lambda_1,\lambda_2,\cdots $ converges to some $\lambda$
satisfying $0<\lambda<1$, then this construction gives the same Type $\III_\lambda$ algebra as before.  If the
sequence converges to 0, one gets an algebra of type $\I_\infty$ if the convergence is fast enough. If it is not 
fast enough,
one gets a new algebra that is defined to be of Type $\III_0$. 

However, if the sequence $\lambda_1,\lambda_2,\cdots$ does not converge and has at least two limit points in the
interval $0<\lambda<1$, which are generic in a sense that will be described in section \ref{backqft},
 then the algebra $\A_{\vec \lambda}$ is a new algebra that is defined to be of Type $\III_1$.
 
\subsection{Back to Quantum Field Theory}\label{backqft}

Local algebras $\A_\U$ in quantum field theory are of\footnote{This was first shown for free fields by Araki in \cite{ArakiFree}, before the  finer
classification of Type III algebras was known.  See also Longo \cite{LongoKingston} and Fredenhagen \cite{Fredenhagen}.} Type $\III$, since they do not 
 have a trace -- even one defined only on part of the algebra.  In fact, they are believed to be of  Type $\III_1$.   We will give a somewhat heuristic explanation of this statement, by using the spectrum of the modular operator to distinguish
the different algebras.

Because of the way the algebras were constructed from an infinite tensor product of $2\times 2$ matrix algebras, we can understand
the modular operator by looking first at the $2\times 2$ case.  Let us go back to the case of a  single product $M_2\times M_2'$ acting
on a Hilbert space $V$ of $2\times 2$ matrices, with the cyclic separating vector $K_{2,\lambda}$.  We factorize $V=W\otimes W'$
in terms of column and row matrices.   The reduced density matrices for the two factors are
\be\label{zeox}\rho_1=\rho_2=\frac{1}{1+\lambda}\bp 1&0\cr 0&\lambda\ep. \ee
According to section \ref{fincase}, $\Delta_\Psi$ acts on a $2\times 2 $ matrix $x\in V$ by $x\to \rho_1 x(\rho_2^\tr)^{-1}$.
We see that,  in this case, its eigenvalues are $1,\lambda$, and $\lambda^{-1}$.  

Now let us consider the Type $\III_\lambda$ algebra $\A_\lambda$ that was constructed in section \ref{type3}.  
It has the cyclic separating vector
\be\label{tuffy}\Psi=K_{2,\lambda}\otimes K_{2,\lambda}\otimes \cdots \otimes K_{2,\lambda}\otimes \cdots \ee
constructed as an infinite tensor product of copies of
 $K_{2,\lambda}$.  In this case, $\Delta_\Psi$ is an infinite tensor product of the answer that we just found in the $2\times 2$ case.
The eigenvalues of $\Delta_\Psi$ are all integer powers of $\lambda$, each occurring infinitely often.   The accumulation points
of the eigenvalues\footnote{Mathematically, the
``spectrum'' of an unbounded operator is defined to include accumulation points of its eigenvalues, along with the eigenvalues
themselves and a possible continuous spectrum. The accumulation points and the possible continuous spectrum are important in the
following remarks.}  are the powers of $\lambda$ and $0$ (which is an accumulation point as it is the large $n$ limit of $\lambda^n$).
More generally, the vector $\Psi_{\vec\lambda}=K_{2,\lambda_1}\otimes K_{2,\lambda_2}\otimes \cdots \otimes 
K_{2,\lambda_n}\otimes \cdots $ is cyclic separating for $\A_\lambda$ if the $\lambda_k$ approach $\lambda$ sufficiently fast.
The operator $\Delta_{\Psi_{\vec\lambda}}$ now has a more complicated set of eigenvalues, but 0 and the integer powers of $\lambda$
are  still accumulation points.  Still more generally, in the case of a Type $\III_\lambda$ algebra, for any cyclic
separating vector $\Psi$, not necessarily of the form $\Psi_{\vec\lambda}$,
the integer powers of $\lambda$ and 0 are accumulation points of the eigenvalues.    
Roughly this is because any cyclic separating vector can be very well approximated
by only changing the original one in eqn. (\ref{tuffy}) in finitely many factors.

For Type $\III_0$, the $\lambda_k$ are approaching 0 and the only unavoidable
accumulation points of the eigenvalues of $\Delta_{\Psi_{\vec\lambda}}$
are 0 and 1.   These values continue to be accumulation points 
if ${\Psi_{\vec\lambda}}$ is replaced by any cyclic separating vector of a Type $\III_0$ algebra.

Now let us consider a Type $\III_1$ algebra.   Suppose that in eqn. (\ref{murfy}), the $\lambda_k$ take the two values $\lambda$ 
and $\t\lambda$, each infinitely many times.  Then the eigenvalues of $\Delta_{\Psi_{\vec\lambda}}$ consist of the numbers 
$\lambda^n\t\lambda^m$, $n,m\in\Z$, each value occurring infinitely many times.   If $\lambda$ and $\t\lambda$ are generic, then
every nonnegative real number can be approximated arbitrarily well\footnote{The case that this is not true is that there is
some $\lambda'$ with $\lambda=\lambda'^n$, $\t\lambda =\lambda'^m$, $n,m\in\Z$.  Then the spectrum of $\Delta_{\Psi_{\vec\lambda}}$ consists of integer powers of $\lambda'$, and the algebra is of type $\III_{\lambda'}$.} 
as $\lambda^n \t\lambda^m$, with integers $n,m$.   So in this
case all nonnegative real numbers are accumulation points of the eigenvalues.   This is the hallmark of a Type $\III_1$ algebra: for
any cyclic separating vector $\Psi$, the spectrum of $\Delta_\Psi$ (including accumulation points of eigenvalues) comprises
the full  semi-infinite interval $[0,\infty)$.

Now let us return to quantum field theory and consider the case that $\U$ is a wedge region, as analyzed in section \ref{more}.
The modular operator for the vacuum state $\Omega$ is $\Delta_\Omega=\exp(-2\pi K)$, where $K$ is the Lorentz boost operator.
$K$ has a continuous spectrum consisting of all real numbers, so $\Delta_\Omega$ has
a continuous spectrum consisting of all positive numbers.  In particular, all points in $[0,\infty)$ are in that spectrum.  Now suppose we replace $\Omega$ by some other cyclic separating vector $\Psi$.  At short distances, any
state is indistinguishable from the vacuum.  So we would expect that acting on excitations of very short wavelength, $\Delta_\Psi$
can be approximated by $\Delta_\Omega$ and therefore has all points in $[0,\infty)$ in its spectrum. 
See \cite{Fredenhagen} and section V.6 of \cite{Haag} for more precise statements.  Thus the algebra $\A_\U$ is of Type $\III_1$.

What about other open sets $\U\subset M$?
For an important class of examples, 
let $\Sigma$ be an initial value surface, and let $\V\subset \Sigma$ be an open subset whose closure $\bar \V$ has a nonempty
boundary.
 Let $\U_\V\subset M$ be the domain of dependence of $\V$.  Its closure
$\bar\U_\V$ has a ``corner'' along  the boundary  of $\bar\V$. Let $\Delta_\Omega(\U_\V)$ be the modular operator of the state
$\Omega$ for the algebra $\A_{\U_\V}$.   For very high energy excitations localized near the corner, $\U_\V$ looks
like the wedge region $\U$.  So one would expect that for such high energy excitations, $\Delta_\Omega(\U_\V)$ looks like the Lorentz
boost generators and has all positive real numbers in its spectrum.  Again, changing the state will not matter.
So again in this case, the algebra $\A_{\U_\V}$ is of Type $\III_1$.  

According to the Borchers timelike tube theorem, which was already mentioned at the end of section \ref{furtherp},
 for many open sets $\U$ that are not
of the  form $\U_\V$, $\A_\U$ actually coincides with some $\A_{\U_\V}$ where $\U\subset \U_\V$.  So then $\A_\U$ is 
again of Type $\III_1$.

\section{Factorized States}\label{indo}

\subsection{A Question}\label{aq}

\def\alg{{\mathrm{alg}}}

Let $\U$ and $\U'$ be complementary open sets with local algebras $\A_\U$, $\A_{\U'}$. (We recall that
complementary open sets are each other's causal complements and there is no ``gap'' between them.)
If one had a factorization of the Hilbert space $\H=\H_1\otimes \H_2$ 
with each algebra  acting on one of the two factors, then one could specify independently the physics in $\U$ and in $\U'$.
For any $\Psi\in\H_1$, $\chi\in\H_2$, the tensor product state $\Psi\otimes \chi$ would look like $\Psi$
for observations in $\U$ and like $\chi$ for observations in $\U'$.

In fact, there is no such  factorization and 
it is not possible to independently specify the
state in $\U$ and in $\U'$.

\begin{figure}
 \begin{center}
   \includegraphics[width=2.5in]{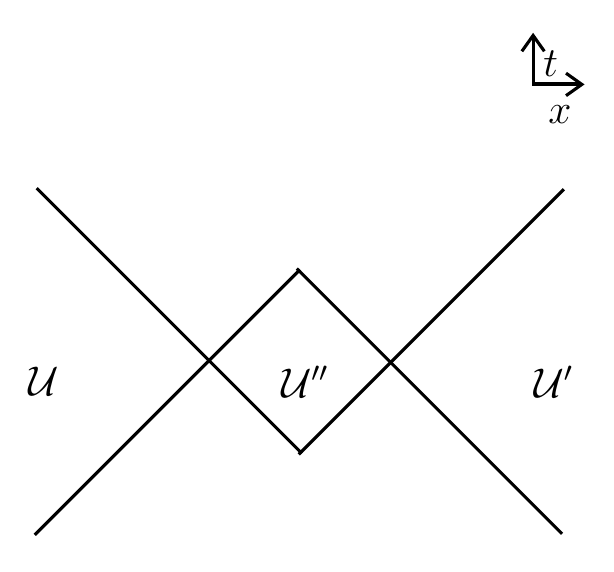}
   \end{center}
\caption{\small  Two spacelike separated open sets $\U$ and $\U'$ in Minkowski spacetime, with a gap between them.
\label{Fig5}}
\end{figure}

Suppose, however, that there is a ``gap''  between $\U$ and $\U'$, leaving room for another
open set $\U''$ that is spacelike separated from both of them (fig. \ref{Fig5}).   Then, given states $\Psi,\chi\in\H$,
 the question of finding a state
looking like $\Psi$ in $\U$ and like $\chi$ in $\U'$
 is not affected by ultraviolet divergences.  But there is still a possible obstruction, which
arises if there is some nontrivial operator $\x$ (not a multiple of the identity) that is in both $\A_\U$ and $\A_{\U'}$.
Such an operator is central in both $\A_\U$ and $\A_{\U'}$ (since these algebras commute with each other).
In Minkowski spacetime, it is reasonable based on what we know from canonical quantization
to expect that $\A_\U$ and $\A_{\U'}$ have trivial center and trivial
intersection, but in general,
in more complicated spacetimes, this might fail \cite{SchroerP,HO}.  If there is some $\x\in \A_\U\cap \A_{\U'}$ with
$\la\Psi|\x|\Psi\ra\not=\la\chi|\x|\chi\ra$, then obviously, since $\x$ can be measured in either $\U$ or $\U'$,
 there can be no state that looks like $\Psi$ in $\U$ and 
like $\chi$ in $\U'$.

In proceeding, we will assume that there is a gap between $\U$ and $\U'$ and that the intersection of the 
two algebras is trivial.    We will impose a further restriction on the boundedness of $\U$ and/or $\U'$ that
is discussed below.  Given this,
it actually is possible,\footnote{This question and similar ones are related to what is called the split property in algebraic
quantum field theory and have been analyzed with increasing detail
in   \cite{Roos}, \cite{Buchholz}, \cite{DL}.}
 for any $\Psi,\chi\in\H$, to find a state that is indistinguishable
from $\Psi$ for measurements in $\U$, and indistinguishable from $\chi$ for measurements in $\U'$.   

We will make use of the gap between $\U$ and $\U'$ in two ways.  First, it ensures that the union of the two open sets, 
$\h\U=\U\cup \U'$, 
is ``small'' enough  so that
the Reeh-Schlieder theorem applies and the vacuum state $\Omega$ is cyclic and separating for the local
algebra $\A_{\h\U}$. (There is another open set $\U''$ that is spacelike separated from $\h\U$, and this is enough to
invoke the theorem.)   

Second, we want to use the gap as an ingredient in  ensuring that there are no subtleties in building observables in $\h\U$
from observables in $\U$ and in $\U'$, in the sense that the algebra $\A_{\h\U}$ is just a tensor product:
\be\label{zoldo} \A_{\h\U}=\A_\U\otimes \A_{\U'}. \ee
However, this point is not straightforward, for several reasons.

First of all, we have to explain what is meant by the tensor product $\A_\U\otimes \A_{\U'}$ of von Neumann algebras.
The {\it algebraic} tensor product $\A_\U\otimes_\alg \A_{\U'}$ is defined in the familiar way; elements are finite linear combinations
$\sum_{i=1}^s \a_i\otimes \a'_i$, with $\a_i\in \A_\U$, $\a'_i\in \A_{\U'}$.  Such finite linear combinations are added and multiplied
in the familiar way.

However, to get a von Neumann algebra, we have to take a completion of $\A_\U\otimes_\alg \A_{\U'}$.  As usual, what we get
when we take a completion depends on what Hilbert space the algebra is acting on.  We have seen several examples of this in section
\ref{algex}.  The completion we want is one in which $\A_\U$ and $\A_{\U'}$ act completely independently.\footnote{It is here
that we assume that the intersection of the two algebras is trivial.  If they have a nontrivial element $\x$ in common,
it is not possible for them to act independently.}
 For this, we introduce
a Hilbert space $\h\H=\H\otimes \H'$ consisting of two copies of the Hilbert space of our quantum field theory, and we consider
the action of $\A_\U\otimes_\alg \A_{\U'}$ on $\h\H$ with $\A_\U$ acting on the first factor and $\A_{\U'}$ acting on the second. 
The von Neumann algebra completion of $\A_\U\otimes_\alg \A_{\U'}$ acting on $\h\H$ is the von Neumann algebra tensor
product $\A_\U\otimes \A_{\U'}$.

This explains what eqn. (\ref{zoldo}) would mean, but it is not true without some further condition on $\U$ and $\U'$.   The gap
between them avoids ultraviolet issues that would obstruct the factorization in eqn. (\ref{zoldo}), but there are still infrared issues.

 Before
explaining this, we consider a simpler question that will actually also be relevant in section \ref{discussing}.
If a given quantum field theory has more than one vacuum state,\footnote{This can happen because of a spontaneously broken
symmetry, but there are other possible reasons.  For instance, vacuum degeneracy not associated to any symmetry can arise
at a first order phase transition, and supersymmetric models often have multiple vacua.}   does the algebra 
$\A_\U$ for an open set $\U$ depend on the choice of vacuum?   If $\U$ is a {\it bounded} open set, with compact closure, 
one expects on physical grounds that the answer will be ``no.'' But in the case of a noncompact region, in general $\A_\U$ does depend on
the vacuum.

To understand this, first pick a  smooth real smearing function $f$ supported in region $\U$ such that
\be\label{orfo}\int_\U \d^Dx \,|f|^2<\infty\ee
but 
\be\label{worfo}\int_U \d^Dx\,f=\infty. \ee  Such an $f$ is, of course, not compactly supported.  
Now pick a local field $\phi$ and consider the question of whether there exists an operator corresponding to
\be\label{corfo}\phi_f=\int_U\d^Dx\, f(x)\phi(x). \ee
A ``yes'' answer means that there is a dense set of Hilbert space states $\Psi$ such that $|\phi_f\Psi|^2<\infty$. If so, then
bounded functions of $\phi_f$ such as $\exp(\i\phi_f)$ would be included in the algebra $\A_\U$.   Actually, 
since we assume (as part of what we mean by saying that $\phi$ is a local field) that $\phi_f$ is a Hilbert space operator
if $f$ is compactly supported, the only concern in the noncompact case is  a possible infrared divergence in computing $|\phi_f\Psi|^2$.
Since any state looks like the vacuum near infinity, such an infrared divergence will not depend on the choice of $\Psi$
and the condition for $\phi_f$ to be a good operator is just that $|\phi_f\Omega|^2<\infty$.
When we compute  $|\phi_f\Omega|^2=\la\Omega|\phi_f\phi_f|\Omega\ra$, we will run into connected and disconnected two-point functions of $\phi$.  Let us
assume for simplicity that our theory has a mass gap.  Then the connected correlation function is short-range and condition
(\ref{orfo}) is sufficient to ensure that there is no infrared divergence in the connected part of the correlation function.  However,
eqn. (\ref{worfo}) means that the disconnected part of the correlation function will make a divergent contribution to $|\phi_f\Omega|^2$
unless $\la\Omega|\phi|\Omega\ra=0$, that is, unless the disconnected part of the correlation function is 0.   The condition
that $\la\Omega|\phi|\Omega\ra=0$ certainly depends on the vacuum, and therefore, the question of which $\phi$ we can use in
constructing $\phi_f$ depends on the vacuum.    Thus, for an unbounded open set $\U$, $\A_\U$ depends on the vacuum.

Somewhat similarly, while keeping fixed the vacuum at infinity, one can ask whether $\A_\U$, for noncompact $\U$,
 depends on the choice of a superselection
sector.  The general answer to this question is not clear to the author.

Now let us go back to the case of $\A_{\h\U}$ with $\h\U=\U\cup\U'$.   For completely general regions $\U$ and $\U'$, there can
be a subtlety analogous to what we encountered in comparing different vacua.  For example,\footnote{This example is discussed
in \cite{Buchholz} and attributed to Araki.}  suppose that $\U$ and $\U'$ are noncompact
and are asymptotically parallel in the sense that there is some fixed vector $b$ such that, at least near infinity, the translation
$x\to x+b$ maps $\U$ to $\U'$.  Then we can pick local fields $\phi_i$ and $\phi'_i$, $i=1,\cdots ,s$  and with $f$ as above,
 we can attempt to define the operator
\be\label{pilbo} X_f=\sum_{i=1}^s\int_\U \d^Dx \,f(x)\phi_i(x)\phi'_i(x+b), \ee
whose support is in $\h\U=\U\cup\U'$.
Assuming again a mass gap, the condition for   $X_f$  to be well-defined is that the relevant vacuum expectation value must vanish.
In the present case, the operator whose vacuum expectation value must vanish
 is $X=\sum_i \phi_i(x)\phi'_i(x+b)$.    The condition for this to vanish in the vacuum
 depends on whether $\A_\U$ and $\A_{\U'}$ (and hence $\phi_i$ and $\phi'_i$) act on the same Hilbert space $\H$ or
on the two factors of $\h\H=\H\otimes \H'$.  When the two algebras act on the same copy of $\H$,
connected two-point functions contribute in the evaluation of $\la\Omega|X|\Omega\ra =\la\Omega|\sum_i \phi_i(x)\phi'_i(x+b) |\Omega\ra$.
There are no such connected contributions if the two algebras act on two different copies of the Hilbert space.
The operators $X_f$ that are well-defined are different in the two cases, and  thus this gives an example of $\U$ and $\U'$ for which the relation (\ref{zoldo}) that we
want is not true.

A sufficient condition that avoids all such questions  is to consider bounded open sets only.
 Indeed, to avoid such issues,  and because of a belief that physics
is fundamentally local in character,
Haag in \cite{Haag} bases the theory on the $\A_\U$
for bounded open sets $\U$.   However, for the specific question under discussion here, we can avoid infrared issues in connected
correlation functions if just $\U$ or $\U'$ is bounded.  Then the well-definedness of an operator such as $X_f$ is the same whether
the two algebras act on the same copy or two different copies of $\H$.  We make this assumption going forward.  For applications
discussed in section \ref{discussing} that involve just one open set $\U$, we assume that $\U$ is bounded.

Now let us suppose that $\U$ and $\U'$ have been chosen to ensure the factorization (\ref{zoldo}).  
Since the Reeh-Schlieder theorem applies to $\h\U$, the algebra $\A_{\h\U}$ acts on the Hilbert space $\H$ of
our quantum field theory with  the vacuum vector $\Omega$ as a cyclic separating vector.   But
eqn. (\ref{zoldo}) means by definition that precisely the same algebra can
act on  $\h\H=\H\otimes\H'$  with $\A_\U$ acting on the first copy and $\A_{\U'}$ acting
on the second.   In $\h\H$, the vector $\Phi=\Omega\otimes \Omega$ is cyclic and separating.

However, whenever the same von Neumann algebra $\A_{\h\U}$ acts on two different Hilbert spaces $\H$ and
$\h\H$, in each case with a cyclic separating vector, there is always a map between the two Hilbert spaces
that maps one action to the other.  (It does not generically map one cyclic separating vector to the other.)
Applied to our problem, this will enable us to find in $\H$ a state that looks like $\Psi$ for observations in $\U$
and like $\chi$ for observations in $\U'$.

We explain the statement about von Neumann algebras in section \ref{statement}.   The application to our
question, and a few other applications, are discussed in section \ref{discussing}.

\subsection{Mapping One Representation To Another}\label{statement}

We assume that the von Neumann algebra $\A$ acts on two Hilbert spaces $\H$ and $\h\H$ with cyclic separating vectors
$\Psi\in\H$ and $\Phi\in\h\H$.  As remarked at the end of section \ref{relmod}, the relative modular operators
$S_{\Psi|\Phi}:\H\to\h\H$ and $\Delta_{\Psi|\Phi}:\H\to\H$ are defined in this generality.    

We will find an isometric or unitary embedding $T:\h\H\to \H$ that commutes with the action of $\A$.
     Using the finite-dimensional formulas
of section \ref{fincase}, one can guess what the map should be.  We define a linear map $T:\h\H\to \H$ by
\be\label{incoc} T(\a|\Phi\ra)=\a\Delta_{\Psi|\Phi}^{1/2}|\Psi\ra.\ee
To begin with $T$ is only defined on the dense set of vectors $\a|\Phi\ra,$ $\a\in \A$.  But once we show
that $T$ is an isometry, this means in particular that it is bounded and  it will automatically extend to all of $\h\H$ as an isometry.

 For $T$ to be an isometry means that for all $\a,\b\in\A$,
 \be\label{zudz}\la \b\Phi|\a\Phi\ra {=} \la \b\Delta_{\Psi|\Phi}^{1/2}\Psi|\a\Delta_{\Psi|\Phi}^{1/2}\Psi\ra. \ee
 The interested reader can show, using formulas of section \ref{fincase}, that this statement is true if the Hilbert space factorizes as
 $\H=\H_1\otimes \H_2$ with each algebra $\A$ and $\A'$ acting on one factor.  Very often, statements that are easy to check
 if one assumes a factorization can be demonstrated in general using Tomita-Takesaski theory.   What follows is fairly illustrative
 of many such arguments.

 The right hand side of eqn. (\ref{zudz}) is
 \be\label{mudz}\la\Psi|\Delta_{\Psi|\Phi}^{1/2}\b^\dagger \a \Delta_{\Psi|\Phi}^{1/2}|\Psi\ra
.\ee
 We want to show that this equals the left hand side of eqn. (\ref{zudz}), but first let us consider
 \be\label{rudz} F(s)=\la\Psi|\Delta_{\Psi|\Phi}^{\i s} \b^\dagger \a \Delta_{\Psi|\Phi}^{1-\i s} |\Psi\ra=\la\Psi|\Delta_{\Psi|\Phi}^{\i s} \b^\dagger \a \Delta_{\Psi|\Phi}^{-\i s} S^\dagger_{\Psi|\Phi}
 S_{\Psi|\Phi}|\Psi\ra\ee
 for real $s$.  

The antiunitarity  of $S_{\Psi|\Phi}$ gives
\be\label{pludz} F(s)= \la S_{\Psi|\Phi}\Psi|S_{\Psi|\Phi} \Delta_{\Psi|\Phi}^{\i s} 
\a^\dagger \b  \Delta_{\Psi|\Phi}^{-\i s}\Psi\ra. \ee
 Now we have to remember that conjugation by $\Delta_{\Psi|\Phi}^{\i s}$ is an automorphism of $\A$, so in 
 particular $ \Delta_{\Psi|\Phi}^{\i s} \a^\dagger \b \Delta_{\Psi|\Phi}^{-\i s}\in\A$.   Moreover, for any $\x\in\A$,
 $S_{\Psi|\Phi}\x\Psi=\x^\dagger\Phi$.  So
 \be\label{orudz}F(s) =\la \Phi| \Delta_{\Psi|\Phi}^{\i s} \b^\dagger \a \Delta_{\Psi|\Phi}^{-\i s}|\Phi\ra. \ee
 Now we remember from section \ref{mod group} 
 that the automorphism $\x\to  \Delta_{\Psi|\Phi}^{\i s} \x \Delta_{\Psi|\Phi}^{-\i s}$ of
 $\A$ depends only on $\Phi$ and not on $\Psi$.  So in evaluating this last formula for $F(s)$, we can set $\Psi=\Phi$,
 whence $\Delta_{\Psi|\Phi}$ reduces to the ordinary modular operator $\Delta_\Phi:\h\H\to\h\H$.  Thus
 \be\label{pludoz} F(s)= \la\Phi|\Delta_{\Phi}^{\i s}\b^\dagger \a\Delta_{\Phi}^{-\i s}|\Phi\ra.\ee
 But $\Delta_\Phi|\Phi\ra=|\Phi\ra$, so $\Delta_{\Phi}^{-\i s}|\Phi\ra=|\Phi\ra.$   Thus finally for real $s$
 \be\label{porudz}F(s) =\la \Phi|\b^\dagger \a|\Phi\ra=\la \b\Phi|\a\Phi\ra.  \ee
 In particular, $F(s)$ is independent of $s$ for real $s$.
 
Suppose we know {\it a priori} that $F(s)$ is holomorphic in the strip $0> \mathrm{Im}\,s>-1/2$
and continuous up to the boundary of the strip.  Then $F(s)$ has to be constant 
even if $s$ is not real, so in this case eqn. (\ref{porudz}) remains valid if we set $s=-\i/2$.  
A look back at the definition (\ref{rudz}) of $F(s)$ shows that eqn. (\ref{porudz}) at $s=-\i/2$ is what we want.  
This formula says precisely that  (\ref{mudz}) equals the left hand side of  (\ref{zudz}).

The desired holomorphy goes beyond what was proved in section \ref{mod group} and is explained in Appendix
\ref{holoproof}.

The result that we have found is useful even if the two Hilbert spaces $\H$ and $\hat\H$ are the same.  There 
are many states that are equivalent to $\Phi$ for measurements by operators in $\A$; any state $\a'\Phi$, where
$\a'\in\A'$ is unitary, has this property.  But in that case $\Delta_{\Psi|\a'\Phi}=\Delta_{\Psi|\Phi}$ (eqn. (\ref{nomorf})) so
$\Delta_{\Psi|\a'\Phi}^{1/2}\Psi=\Delta_{\Psi|\Phi}^{1/2}\Psi$.   Thus once $\Psi$ is chosen, in every equivalence class of
vectors that are equivalent to some $\Phi$ for measurements in $\A$, there is a canonical representative $\Delta_{\Psi|\Phi}^{1/2}\Psi$.
These representatives make up the canonical cone \cite{ArakiCone}, which has many nice properties.

\subsection{Applications}\label{discussing}

Our first application of the result of the last section is to a case discussed in section \ref{aq}.   Thus, $\H$
is the Hilbert space of a quantum field theory, and $\h\H=\H\otimes \H'$ is the tensor product of two copies of $\H$. For
open sets $\U$, $\U'$, at least one of which is bounded, with a gap between them, the same algebra
 $\A_{\h\U}=\A_\U\otimes \A_{\U'}$ can act on $\H$ and also on  $\h\H=\H\otimes \H'$, with in the latter case
 $\A_\U$ acting on the first factor and $\A_{\U'}$ acting on the second. 
For cyclic separating vectors, we take $\Psi\in\H$ to be the vacuum vector $\Omega$, and $\Phi\in\h\H$ to 
be $\Omega\otimes \Omega$.  

The construction of the last section gave an isometric embedding $T:\h\H\to \H$ that commutes
with the action of $\A_{\h\U}$. Because of the way we chose the action of $\A_\U$ and $\A_{\U'}$ on $\h\H$, the vector
$\Psi\otimes\chi\in\h\H$ looks like $\Psi$ for measurements in $\U$ and like
$\chi$ for measurements in $\U'$.   So $T(\Psi\otimes \chi)$ is a vector in $\H$ that has the same property.

This sort of reasoning has other applications.  For example, let $\H_1$ and $\H_2$ be two different superselection 
sectors in the same quantum field theory.  Let $\U$ be a bounded open set; then the same algebra $\A_{\U}$ acts on both
$\H_1$ and $\H_2$.  Both $\H_1$ and $\H_2$ contain cyclic separating vectors for $\A_\U$, by the slight
extension of the Reeh-Schlieder theorem that was described in section \ref{bounded}.   So we can find an
isometric embedding $T:\H_1\to \H_2$ that commutes with $\A_\U$.  If $\Psi$ is a vector in $\H_1$, then
$T\Psi$ is a vector in $\H_2$ that cannot be distinguished from $\Psi$ by measurements in the region $\U$.
As explained in  \cite{HaagKastler}, there is an intuitive reason for this.    For example, superselection sectors
that are defined by the total magnetic charge cannot be distinguished by measurements in region $\U$,
because by such measurements one cannot tell how many magnetic monopoles there are in distant regions.

Similarly, consider a quantum field theory with more than one vacuum state.  Let $\H_1$ and $\H_2$ be the Hilbert spaces
based on these two vacua.  For bounded $\U$, the same algebra $\A_\U$ will act in $\H_1$ and in $\H_2$. 
The same argument as before tells us that measurements in region $\U$ cannot determine which vacuum state we are in.  The intuitive reason is that in the Hilbert space built on one vacuum, there
can be a state that looks like some other vacuum over a very large region of spacetime.

For a final application, let us consider the following question.\footnote{See section V.2.2 of \cite{Haag}, where much more
precise results are stated than we will explain here.}  Suppose that $\rho$ is a density  matrix on $\H$.
Is there a pure state $\chi\in\H$ that is indistinguishable from $\rho$ for measurements in region $\U$?    If the Hilbert space
factored as $\H=\H_1\otimes \H_2$ with $\A_\U$ acting on the first factor, we would answer this question as follows.
For measurements in $\U$, we can replace $\rho$ with the reduced density matrix $\rho_1=\Tr_{\H_2}\, \rho$ on $\H_1$.
Then, picking a purification $\chi$ of $\rho_1$ in $\H_1\otimes \H_2$, $\chi$ would be indistinguishable from $\rho$
for measurements in $\U$.

To answer the question without such a factorization, we can use something called the Gelfand-Neimark-Segal (GNS) construction.  Consider the function on $\A_\U$ defined
by $F(\a)=\Tr_\H\,\rho\a$; this function  is called a faithful normal state on the algebra $\A_\U$.  Given this function,
the GNS construction produces
a Hilbert space $\K$ with action of  $\A_\U$ and a cyclic separating vector $\Psi$ such that $F(\a)=\la\Psi|\a|\Psi\ra$.     The construction
is quite simple.
To make  $\Psi$  cyclic separating, vectors $\a\Psi$ are assumed to satisfy no relations ($\a\Psi\not=\b\Psi$ for $\a\not=\b$) and to
comprise a dense subspace $\K_0$ of $\K$.
  The inner product on $\K_0$ is defined to be $\la \a\Psi|\b\Psi\ra=F(\a^\dagger \b)$,  which in particular ensures
  that $\la\Psi|\a|\Psi\ra=\Tr_{\H}\,\rho\a$. All axioms of a Hilbert space are satisfied except completeness.  $\K$ is defined as the Hilbert space completion of $\K_0$.  Now $\A$ acts on one Hilbert space $\H$ with cyclic separating vector $\Omega$
(the vacuum) and on another Hilbert space $\K$ with cyclic separating vector $\Psi$.  So as in section \ref{statement}, we can
find an isometric embedding $T:\K\to\H$.  Then $T(\Psi)$ is the desired vector in $\H$ that is indistinguishable from $\rho$
for measurements in $\U$.

Research supported in part by NSF Grant PHY-1606531.    I thank N. Arkani-Hamed, B. Czech, C. Cordova, J. Cotler, X. Feng, D. Harlow, A. Jaffe, N. Lashkari, S. Rajagopal,
B. Schroer,
  B. Simon and especially R. Longo
for helpful comments and advice.


 \begin{appendix}

\section{More Holomorphy}\label{zelboxx}

\subsection{More On Subregions}\label{moresub}

Here (following \cite{Borchers}) we will prove a result relating the modular operators $\Delta_{\Psi;\,\U}$ and 
$\Delta_{\Psi;\,\t\U}$ for a pair of open
sets $\U$, $\t\U$ with $\t\U\subset \U$.  $\Psi$ is a vector that is cyclic separating for both algebras $\A_\U$ and $\A_{\t\U}$;
it is kept fixed in the following and will be omitted in the notation.    The result we will describe is useful in applications (for example, see
eqns. (6.7) and (6.8) in \cite{Faulkner}). 

From section \ref{theproof}, we know already that $\Delta_{\t\U}\geq \Delta_\U$, and from section \ref{practice}, it follows that
\be\label{pottor} \Delta_{\t\U}^\alpha\geq \Delta_\U^\alpha,~~~~ 0\leq \alpha\leq 1. \ee
From this, it follows that for any state $\chi$, and $0\leq \beta\leq 1/2$, we have \be\label{poff}\la\chi|\Delta_{\t\U}^{-\beta} \Delta_\U^{2\beta}
\Delta_{\t\U}^{-\beta}|\chi\ra=\la\Delta_{\t \U}^{-\beta}\chi|\Delta_\U^{2\beta}|\Delta_{\t\U}^{-\beta}\chi\ra\leq 
\la\Delta_{\t \U}^{-\beta}\chi|\Delta_{\t\U}^{2\beta}|\Delta_{\t\U}^{-\beta}\chi\ra=\la\chi|\chi\ra,\ee so
\be\label{ottor} \Delta_{\t\U}^{-\beta}\Delta_\U^{2\beta}\Delta_{\t\U}^{-\beta}\leq 1, ~~~ 0\leq \beta\leq 1/2.\ee
Since $X^\dagger X\leq 1$ implies $||X||\leq 1$, it follows that
\be\label{wottor} || \Delta_{\U}^\beta\Delta_{\t\U}^{-\beta}||\leq 1, ~~~ 0\leq \beta\leq 1/2. \ee
An imaginary shift in $\beta$ does not affect this bound, since the operators $\Delta_\U^{\i s}$, $\Delta_{\t\U}^{\i s}$, $s\in\R$
are unitary.   So 
\be\label{plottor} ||\Delta_{\U}^{-\i z}\Delta_{\t \U}^{\i z}||\leq 1 \ee
in the strip $1/2\geq \mathrm{Im}\,z\geq 0$.  This bound implies that the operator-valued function $\Delta_{\U}^{-\i z}\Delta_{\t\U}^{\i z}$
is holomorphic in that strip.

\subsection{More On Correlation Functions}\label{holoproof}

In section \ref{statement}, we needed to know that for $\x=\b^\dagger \a\in\A$,
\be\label{zame} F(z)=\la\Psi|\Delta_{\Psi|\Phi}^{\i z} \x\Delta_{\Psi|\Phi}^{1-\i z} |\Psi\ra=
\la\Psi|\Delta_{\Psi|\Phi}^{\i z} \x\Delta_{\Psi|\Phi}^{1/2-\i z} |\Delta_{\Psi|\Phi}^{1/2}\Psi\ra \ee
is holomorphic in the strip $0> \mathrm{Im}\,z>-1/2$ as well as continuous along the boundaries of the strip.  In fact,
we will prove that it is holomorphic in a larger strip\footnote{Similarly to eqn. (\ref{threp}), one would expect this
if one assumes a factorization $\H=\H_1\otimes \H_2$ of the Hilbert space.  In this appendix, we follow Araki's approach
to proving such statements without assuming a factorization. See \cite{ArakiFive}, section 3.} $0>\mathrm{Im}\,z>-1$ and again
continuous on the boundaries.

As we will see, it helps to consider first  the case that the state $\Delta_{\Psi|\Phi}^{1/2}\Psi$
is replaced by $\y\Psi$ for some $\y\in \A$.  So we consider the function
\be\label{wame}G(z)=\la\Psi|\Delta_{\Psi|\Phi}^{\i z} \x\Delta_{\Psi|\Phi}^{1/2-\i z}|\y\Psi\ra. \ee
Holomorphy in the strip is now trivial,  
because the condition
$0>\mathrm{Im }\,z>-1/2$ means that the exponents $\i z$ and $1/2-\i z$ in eqn. (\ref{wame})
both have real part between 0 and $1/2$, and consequently from section \ref{mod group}, we know that
both $\Delta_{\Psi|\Phi}^{1/2-\i z}|\y\Psi\ra$ and $\la\Psi|\Delta_{\Psi|\Phi}^{\i z}$ are holomorphic in this strip.

The norm of a state $\chi$ is $|\chi|=\sqrt{\la\chi|\chi\ra}$, and the norm $||\y||$ of a bounded operator $\y$
is the least upper bound of $|\y\chi|/|\chi|$ for any state $\chi$.   The following proof will depend on getting
an upper bound on $|G(z)|$ in the strip by a constant multiple of $|\y\Psi|$.   An immediate
upper bound is
\be\label{name} |G(z)|\leq |\Delta_{\Psi|\Phi}^{-\i \bar z}\Psi|\,||\x||\,|\Delta_{\Psi|\Phi}^{1/2-\i z}\y\Psi|.  \ee
If $z=s-\i\alpha$, with $s,\alpha\in\R$,  then the right hand side of eqn. (\ref{name}) only depends on $\alpha$, since
$\Delta_{\Psi|\Phi}^{\i s}$ is unitary.   For $s=0$, the function $G(z)$ is bounded on the compact set $0\leq \alpha\leq 1/2$
(for $\alpha$ in that range it is the inner product of two states that are well-defined  and bounded in Hilbert space according to
eqn. (\ref{woox})), 
so it is bounded in the whole strip $0\geq \mathrm{Im}\,z\geq -1/2$.  We need to improve this to get a bound by a multiple of
$|\y\Psi|$.

Let us look at the function $G(z)$ on the boundaries of the strip.  On the lower boundary $z=s-\i/2$, $\Delta^{1/2-\i z}$ 
is unitary.   Also on that boundary
$|\Delta_{\Psi|\Phi}^{-\i\bar z}\Psi|=|\Delta_{\Psi|\Phi}^{1/2}\Psi|<\infty$.
So on the lower boundary,  eqn. (\ref{name}) bounds $|G(z)|$ by a constant multiple of $|\y\Psi|$.  On the upper boundary
$z=s$, we write
\be\label{prof}|G(z)|=|\la \Delta_{\Psi|\Phi}^{1/2+\i s}\x^\dagger \Delta_{\Psi|\Phi}^{-\i s}\Psi|\y\Psi\ra|
\leq    |\Delta_{\Psi|\Phi}^{1/2} \Delta_{\Psi|\Phi}^{\i s }\x^\dagger \Delta_{\Psi|\Phi}^{-\i s}\Psi|\,|\y\Psi|.\ee
Reasoning similarly to  (\ref{prox}), this implies
\be\label{drof}|G(z)|\leq |\Delta_{\Psi|\Phi}^{\i s}\x\Delta_{\Psi|\Phi}^{-\i s}\Phi| \,|\y\Psi|. \ee
Because the operator $\Delta_{\Psi|\Phi}^{\i s}$ is unitary and $\la\Phi|\Phi\ra=1$, we get on the upper  boundary 
\be\label{turof} |G(z)|\leq ||\x|| \,|\y\Psi|. \ee
So there is a constant $C$, independent of $\y$ and $z$, such that on the boundaries of the strip,
\be\label{protof} |G(z)|\leq C|\y\Psi|. \ee

A holomorphic function, such as $G(z)$, that is bounded and holomorphic in a strip, 
and obeys a bound $|G(z)|\leq \hat C$ on the boundary of the strip,
obeys the same bound in the interior of the strip.  This statement is a special case of the Phragm\'{e}n-Lindel\"{o}f
principle, and can be proved as follows (we state the argument for our strip $0>\mathrm{Im}\, z>-1/2$).  
For $\epsilon>0$, the function $G_\epsilon(z)=\exp(-\epsilon z^2) G(z)$
satisfies $|G_\epsilon(z)|\leq \hat C \exp(\epsilon/4)$ on the boundary of the strip.  The function $G_\epsilon(z)$
vanishes rapidly for $\mathrm{Re}\,z\to\pm\infty$, so $|G_\epsilon(z)|$ achieves its maximum somewhere in the interior
of the strip or  its boundary.
By the maximum principle, this maximum is achieved somewhere on the boundary of the strip.   Therefore
the bound $|G_\epsilon(z)|\leq \hat C\exp(\epsilon/4)$ is satisfied throughout the strip.  As this is true for all 
$\epsilon$, we get $|G(z)|\leq \hat C$ throughout the strip.

 \begin{figure}
 \begin{center}
   \includegraphics[width=6.5in]{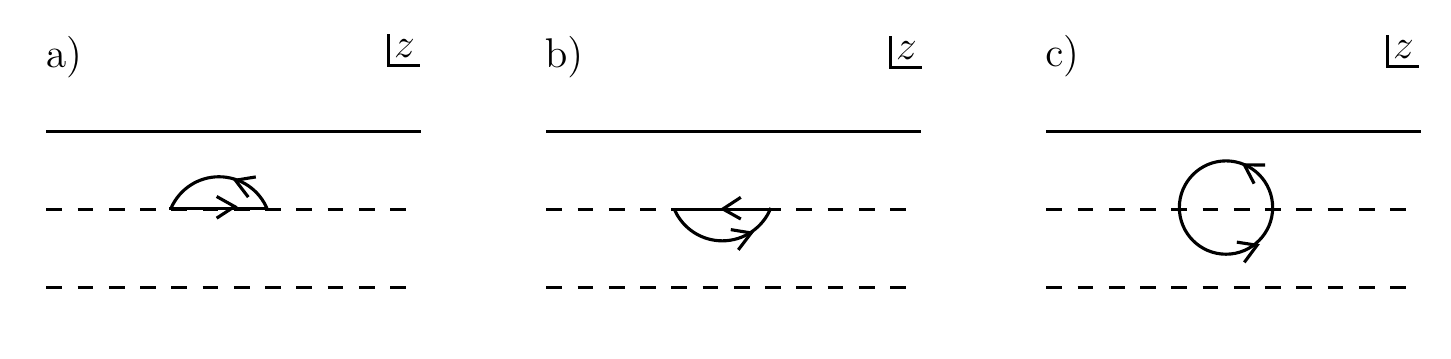}
 \end{center}
\caption{\small (a) If a function $F(z)$ is holomorphic in the strip $0>\mathrm{Im}\,z>-1/2$ and continuous
at the lower boundary of the strip, we can write a Cauchy integral formula with a contour that runs partly on the lower
boundary.  (b) If $F(z)$ is holomorphic for $-1/2>\mathrm{Im}\,z>-1$ and continuous on the upper
boundary of that strip, we can write a Cauchy integral formula with a contour that runs partly on the upper boundary.
(c) If $F(z)$ satisfies both conditions, we can combine the contours  from (a) and (b), choosing
them so that the part that runs on the line $\mathrm{Re}\,z=-1/2$ cancels.  The resulting Cauchy integral
formula shows that $F(z)$ is holomorphic on that line.  The argument sketched in fig. \ref{Fig1} of section \ref{proof}  is actually
the special case of this in which $F(z)$ vanishes in the lower strip.  \label{Fig6}}
\end{figure}

Going back to the original definition of $G(z)$ in eqn. (\ref{wame}), $G(z)$ can be interpreted as a linear functional
on the dense subset of $\H$ consisting of states $\y\Psi$, $\y\in \A$.  The validity of eqn.  (\ref{protof}) throughout the strip
says that this linear functional
is bounded.  
 A bounded linear functional on a dense subset of a Hilbert space $\H$ always extends to the
whole space, and remains bounded.  Moreover a bounded linear functional on a Hilbert space  $\H$ is always the
inner product with a state in $\H$.  So we learn that there is some $z$-dependent state $\chi(z)$ such that
\be\label{gulfo} G(z)=\la\chi(z)|\y\Psi\ra \ee
for all $\y\in \A$.   Moreover  $\la\chi(z)|$ is holomorphic in the strip since $G(z)$ is holomorphic in the
strip for all $\y$.    The fact that the linear functional in question extends over all of $\H$ means that for any  $\Upsilon\in\H$,
\be\label{nulb}\la\chi(z)|\Upsilon\ra\ee
is well-defined and holomorphic in the strip.

The original function $F(z)$ is then
\be\label{pokk} F(z)=\la \chi(z)|\Delta^{1/2}_{\Psi|\Phi}\Psi\ra.\ee
Here $\Delta_{\Psi|\Phi}^{1/2}\Psi$ is a Hilbert space state (as in eqn. (\ref{prox})), so this is a special case of (\ref{nulb}), and therefore  is holomorphic in the strip. 
Moreover the original definition (and bounds such as (\ref{woox}) that were used
along the way) make it clear that $F(z)$ has a continuous limit as one approaches the boundaries of the strip.

This is what we needed in section \ref{statement}, but actually the function $F(z)$ is holomorphic in a larger strip.
Writing
\be\label{zamez} F(z)=\la\Delta^{1/2}_{\Psi|\Phi}\Psi|\Delta_{\Psi|\Phi}^{-1/2+\i z} \x\Delta_{\Psi|\Phi}^{1-\i z} |\Psi\ra,\ee
we make an argument very similar to the above, but with the role of the bra and the ket exchanged.  Thus,
we begin by replacing $\Delta^{1/2}_{\Psi|\Phi}\Psi$ with $\y\Psi$ with  $\y\in\A$.  So we have to study
\be\label{wamez}H(z)=\la \y\Psi|\Delta_{\Psi|\Phi}^{-1/2+\i z} \x\Delta_{\Psi|\Phi}^{1-\i z} |\Psi\ra.\ee
We consider the function $H(z)$ in the strip $-1/2\geq \mathrm{Im}\,z\geq -1$.
An argument very similar to the above, reversing the role of the bra and the ket, shows that in this strip
$H(z)=\la \y\Psi|\Upsilon(z)\ra$, where $\Upsilon(z)$ is holomorphic in the strip.  Then
$F(z)=\la \Delta^{1/2}_{\Psi|\Phi}\Psi|\Upsilon(z)\ra$, and in this representation, holomorphy of $F(z)$ for
$-1/2>\mathrm{Re}\,z>-1$ is manifest.

We now have a function $F(z)$ that is holomorphic for $0>\mathrm{Im}\, z >-1/2$ and for $-1/2>\mathrm{Im}\,F(z)>-1$.
Moreover, this function is continuous on the line $\ell$ defined by $\mathrm{Im}\,z=-1/2$.   As sketched in fig. \ref{Fig6},
the Cauchy integral formula
can be used to show that $F(z)$ is actually holomorphic on the line $\ell$.     This fact about holomorphic functions of a single complex
variable has a less elementary analog, known as the Edge of the Wedge Theorem, for functions of several complex variables.  
For some of its applications in quantum field theory, see \cite{StW}.

\end{appendix}
\bibliographystyle{unsrt}

\begin{thebibliography}{99}
\bibitem{Nishioka}
T. Nishioka, ``Entanglement Entropy: Holography and Renormalization Group,'' 
arXiv:1801.10352.

\bibitem{RS}
H. Reeh and S. Schlieder, ``Bemerkungen zur Unita\"{a}r\"{a}quivalenz von Lorentzinvarienten Feldern,'' Nuovo Cimento {\bf 22} (1961) 1051.

\bibitem{BKLS}
L. Bombelli, R. K. Koul, J. Lee, and R. D. Sorkin, ``A Quantum Source of Entropy
For Black Holes,'' Phys. Rev. {\bf D34} (1986) 373-83.

\bibitem{Sred}
M. Srednicki, ``Entropy and Area,'' Phys. Rev. Lett. {\bf 71} (1993) 666-669, hep-th/9303048.

\bibitem{SU}
L. Susskind and J. Uglum, ``Black Hole Entropy in Canonical Quantum Gravity and Superstring Theory,''
Phys. Rev. {\bf D50} (1994) 2700-11.


\bibitem{McG}
M. McGuigan, ``Finite Black Hole Entropy and String Theory,'' Phys. Rev. {\bf D50} (1994) 5225-5231, arXiv:hep-th/9406201.

\bibitem{CW}
C. G. Callan, Jr. and F. Wilczek, ``On Geometric Entropy,'' Phys. Lett. {\bf B333} (1994) 55-61.

\bibitem{HLW}
C. Holzhey, F. Larsen, and F. Wilczek, ``Geometric and Renormalized Entropy in Conformal Field Theory,'' Nucl. Phys. {\bf B424} (1994)
443-67, arXiv:hep-th/9403108.

\bibitem{PapaRaju}
K. Papadodiamas and S. Raju, ``An Infalling Observer in AdS/CFT,''
JHEP {\bf 1310} (2013) 212, arXiv:1211.6767.

\bibitem{Faulkner}
S. Balakrishnan, T. Faulkner, Z. U. Khandker,
and H. Wang, ``A General Proof of the Quantum Null Energy Condition,''
arXiv:1706.09432.

\bibitem{Araki}
H. Araki, ``Relative Entropy of States of von Neumann Algebras,'' Publ. RIMS, Kyoto Univ.
{\bf 11} (1976) 809-33.

\bibitem{Araki2}
H. Araki, ``Inequalities in Von Neumann Algebras,'' Les rencontres physiciens-math\'{e}maticiens de Strasbourg, RCP25 {\bf 22} (1975) 1-25.


\bibitem{Petz}
D. Petz, ``Quasi-Entropies For Finite Quantum Systems,'' Rep. Math. Phys. {\bf 23} (1986) 57-65.

\bibitem{PetzNielsen}
M. A. Nielsen and D. Petz,  ``A Simple Proof of the Strong Subadditivity Inequality,'' Quantum Information and Computation {\bf 5} (2005)
507-13, arXiv:quant-ph/0408130.

\bibitem{LiebRuskai}
E. H. Lieb and M. B. Ruskai, ``Proof of the Strong Subadditivity of Quantum Mechanical Entropy,'' J. Math. Phys. {\bf 14} (1973) 
1938-41.

\bibitem{Lieb}
E. H. Lieb, ``Convex Trace Functions and the Wigner-Yanase-Dyson Conjecture,'' Adv. Math. {\bf 11} (1973) 267-88.

\bibitem{BiWi}J. Bisognano and E. H. Wichmann, ``On The Duality Condition For Quantum Fields,'' J. Math. Phys. {\bf 17} (1976)
303-21.

\bibitem{Unruh}
W. G. Unruh, ``Notes On Black-Hole Evaporation,'' Phys. Rev. {\bf D14} (1976) 870-892.


\bibitem{GibbonsHawking}
S. W. Hawking, ``Action Integrals And Partition Functions In Quantum Gravity,'' Phys. Rev. {\bf D15} (1977) 2752-6.

\bibitem{vonN}
J. von Neumann, ``On Infinite Direct Products,'' Comp. Math. {\bf 6} (1938) 1-77.

\bibitem{Powers}
R. T. Powers, ``Representations Of Uniformly Hyperfinite Algebras And Their Associated von Neumann
Rings,'' Ann. of Math. {\bf 86} (1967) 138-171.

\bibitem{ArakiWoods}
H. Araki and E. J. Woods, ``A Classification Of Factors,'' Publ. RIMS, Kyoto Univ. Ser. A. {\bf 3} (1968) 51-130.



\bibitem{Borchers}  H. J. Borchers, ``On Revolutionizing Quantum Field Theory With Tomita's Modular Theory,''
J. Math. Phys. {\bf 41} (2000) 3604-3673.  A more complete version of this article is available at \url{http://www.mat.univie.ac.at/~esiprpr/esi773.pdf}.

\bibitem{Haag}
R. Haag, {\it Local Quantum Physics} (Springer-Verlag, 1992).


\bibitem{SW}
S. J. Summers and R. Werner, ``Maximal Violation Of Bell's Inequalities Is Generic in Quantum Field Theory,'' Commun. Math. Phys. {\bf 110} (1987) 247-59.

\bibitem{NaT}
H. Narnhofer and W. Thirring, ``Entanglement, Bell Inequality, and All That,'' J. Math. Phys. {\bf 52} (2012) 095210.



\bibitem{HSa}
S. Hollands and K. Sanders, ``Entanglement Measures and Their Properties in Quantum Field Theory,'' arXiv:1702.04924.


\bibitem{Jones}
V. F. R. Jones, ``Von Neumann Algebras,'' available at \url{https://math.vanderbilt.edu/jonesvf/VONNEUMANNALGEBRAS2015/VonNeumann2015.pdf}.



\bibitem{StW}
R. F. Streater and A. S. Wightman, {\it PCT, Spin and Statistics, and All That} (W. A. Benjamin, 1964; paperback edition,
Princeton University Press, 2000).

\bibitem{HS}
R. Haag and B. Schroer, ``Postulates of Quantum Field Theory,'' J. Math. Phys. {\bf 3} (1962) 248.




\bibitem{Borchersold}H. J. Borchers,
``On the Converse of the Reeh-Schlieder Theorem,''
Commun. Math. Phys. {\bf 10} (1968) 269-93.

\bibitem{FRS}
K. Fredenhagen, K. H. Rehren, and B. Schroer, ``Superselection Sectors with Braid Group Statistics and Exchange Algebra, I. General
Theory,'' Commun. Math. Phys {\bf 125} (1989) 201-226.

\bibitem{SVW}
A. Strohmaier, R. Verch, and M. Wollenberg, ``Microlocal Analysis of Quantum Fields in Curved Spacetimes:
Analytic Wavefront Sets and Reeh-Schlieder Theorems,'' J. Math. Phys. {\bf 43} (2002) 5514-30, arXiv:math-ph/0202003.

\bibitem{GW}
C. G\'{e}rard and M. Wrochna, ``Analytic Hadamard States, Calder\'{o}n Projectors and Wick Rotation Near Analytic
Cauchy Surfaces,''  arXiv:1706.08942.

\bibitem{morrison}
I. A. Morrison, ``Boundary-to-Bulk Maps for AdS Causal Wedges and the Reeh-Schlieder
Property in Holography,'' arXiv:1403.3426.

\bibitem{KS}
K. Sanders, ``On The Reeh-Schlieder Property in Curved Spacetime,'' Commun. Math. Phys. {\bf 288} (2009) 271-85.

 \bibitem{EGJ} H. Epstein, V. Glaser and A. Jaffe, ``Nonpositivity of the Energy Density in Quantized Field
Theories,'' Nuovo Cim. {\bf 36} (1965) 10161022.

\bibitem{GL} D. Guido and R. Longo, ``An Algebraic Spin and Statistics Theorem,'' Commun. Math. Phys. {\bf 172} (1995) 517-33.

\bibitem{HG} R. Haag, ``Bemerkungen zum Nahmwirkungsprinzip in der Quantumphysik,'' Annalen der Physik Ser. 7 {\bf 11} (1963) 29-34.*-

\bibitem{LRT} P. Leyland, J. Roberts, D. Testard, ``Duality for Quantum Free Fields'' (Centre de Physique
ThŽorique, CNRS Marseille, 1978).

\bibitem{DL}
S. Doplicher and R. Longo, ``Standard and Split Inclusions of von Neumann Algebras,'' Invent. Math. {\bf 73} (1984) 493.


\bibitem{SchroerP} B. Schroer, ``Positivity and Causal Localization in Higher Spin Quantum Field Theories,'' 
arXiv:1712.02346.

\bibitem{HO} D. Harlow and H. Ooguri, to appear.

\bibitem{OtherBorchers}
H. J. Borchers, ``\"{U}ber die Vollst\"{a}ndigkeit lorentzinvarianter Felder in einer zeitartigen R\"{o}hre,'' Nuovo Cimento {\bf 19} (1961) 787.

\bibitem{OtherAraki}
H. Araki, ``A Generalization Of Borchers Theorem,'' Helv. Phys. Acta {\bf 36} (1963) 132-9.

\bibitem{Wightman}
A. S. Wightman, ``La Th\'{e}orie Quantique Locale et la Th\'{e}orie Quantique des Champs,'' Annales de l'I. H. P., section A {\bf 1} (1964)
403-420.

\bibitem{BorchersBook}
H.-J. Borchers,  { \it{Translation Group and Particle Representations in Quantum Field Theory}} (Springer-Verlag, 1996).

\bibitem{KL}S. Kullback and R. A. Leibler, ``On Information and Sufficiency,'' Ann. Math. Statistics {\bf 22} (1951) 79-86.

\bibitem{Umegaki}
H. Umegaki, ``Conditional Expectation In An Operator Algebra, IV (Entropy and Information),'' Kodai Math. Sem. Rep. {\bf 14} (1962) 59.

\bibitem{Casini}
H. Casini, ``Relative Entropy And The Bekenstein Bound,'' Class. Quant. Grav. {\bf 25} (2008) 205021, arXiv:0804.2182.

\bibitem{LongoXu}
R. Longo and F. Xu, ``Comment on the Bekenstein Bound,''
arXiv:1802.07184.


\bibitem{Wall}
A. Wall, ``A Proof of the Generalized Second Law for Rapidly Changing Fields and Arbitrary Horizon Slices,''
Phys. Rev. {\bf D85} (2012) 104049.


\bibitem{Uhlmann}
A. Uhlmann, ``Relative Entropy and the Wigner-Yanase-Dyson-Lieb Concavity In An Interpolation Theory,'' Commun. Math. Phys.
{\bf 54} (1977) 21-32.


\bibitem{Stone}
M. H. Stone, ``On Unbounded Operators In Hilbert Space,'' Jour. Ind. Math. Soc. {\bf 15} (1951) 155.

\bibitem{Halmos}
P. R. Halmos, ``Two Subspaces,'' Trans. Amer. Math. Soc. {\bf 144} (1969) 381-9.








\bibitem{BGMO}
A. Bernamonti, F. Galli, R. C. Myers, and J. Oppenheim, ``Holographic Second Laws Of Black Hole Thermodynamics,''
arXiv:1803.03633.


\bibitem{ReedSimon}
M. Reed and B. Simon, {\it Methods of Modern Mathematical Physics, I: Functional Analysis} (Academic Press, 1972).

\bibitem{Simon}
B. Simon, {\it Operator Theory: A Comprehensive Course in Analysis, Part 4} (American Mathematical Society, 2015).

\bibitem{LongoSImpleProof}
R. Longo, ``A Simple Proof Of The Existence Of Modular Automorphisms In Approximately Finite Dimensional Von Neumann Algebras,''
Pacific Journal of Mathematics {\bf 75}  (1978) 199-205.



\bibitem{ChaosBound}
J. Maldacena, S. H. Shenker, and D. Stanford, ``A Bound On Chaos,''
JHEP {\bf 1608} (2016) 106, arXiv:1503.01409.


\bibitem{ArakiFive}
H. Araki, ``Relative Hamiltonian for Faithful Normal States Of A Von Neumann Algebra,'' Publ. RIMS, Kyoto Univ. {\bf 9} (1973) 165-209.

\bibitem{NT}
H. Narnhofer and W. Thirring, ``From Relative Entropy to Entropy,'' Fizika {\bf 17} (1985) 257-65,
reprinted  in {\it Selected Papers Of Walter E. Thirring With Commentaries}
(American Mathematical Society, 1998).

\bibitem{GR}
S. Ghosh and S. Raju, ``Quantum Information Measures For Restricted Sets Of Observables,'' arXiv:1712.09365.

\bibitem{WY}
E. Wigner and M. M. Yanase, ``Information Content Of Distributions,'' Proc. Nat. Acad. Sci. USA (1963) 910-8.

\bibitem{Rindler}
W. Rindler, ``Kruskal Space and the Uniformly Accelerated Frame,'' Am. J. Phys. {\bf 34} (1966) 1174.


\bibitem{Hawking}
S. W. Hawking, ``Particle Creation By Black Holes,'' Commun. Math. Phys. {\bf 43} (1975) 199-220.



\bibitem{MoreGibbonsHawking}
G. W. Gibbons and S. W. Hawking, ``Cosmological Event Horizons, Thermodynamics, And Particle Creation,'' Phys. Rev. {\bf D15} (1977)
2738-51.

\bibitem{FHN}
R. Figari, R. Hoegh-Krohn, and C. R. Nappi, ``Interacting Relativistic
Boson Fields In The De Sitter Universe With Two Space-Time Dimensions,''
Commun. Math. Phys. {\bf 44} (1975) 265-78.






\bibitem{ArakiFree}
H. Araki, ``Type of von Neumann Algebras Associated to the Free Field,'' Prog. Theoret. Phys. {\bf 32} (1964) 956.

\bibitem{LongoKingston}
R. Longo, ``Algebraic And Modular Structure of Von Neumann Algebras of Physics,'' Proc. Symp. Pure Math. {\bf 38} (1982) part 2, 572.

\bibitem{Fredenhagen}
K. Fredenhagen, ``On The Modular Structure Of Local Algebras Of Observables,'' Commun. Math. Phys. {\bf 97} (1985) 79.

\bibitem{Roos}
H. Roos, ``Independence of Local Algebras in Quantum Field Theory,'' Commun. Math. Phys. {\bf 16} (1970) 238-46.

\bibitem{Buchholz}
D. Buchholz, ``Product States for Local Algebras,''  Commun. Math. Phys. {\bf 36} (1974) 287-304.


\bibitem{ArakiCone}
H. Araki, ``Some Properties Of Modular Conjugation Operator of Von Neumann Algebras and a Noncommutative Radon-Nikodym
Theorem With A Chain Rule,'' Pacific J. Math. {\bf 50} (1974) 309-54.

\bibitem{HaagKastler} 
R. Haag and D. Kastler, ``An Algebraic Approach To Quantum Field Theory,''
J. Math. Phys. {\bf 5} (1964) 848-61.


\end{thebibliography}

\end{document}